\documentclass[12pt]{article}
\pdfoutput=1

\usepackage{array} 
\usepackage{amssymb}
\usepackage{graphics,graphpap}
\usepackage{graphicx}
\usepackage{color}
\usepackage{graphicx}% Include figure files
\usepackage{dcolumn}% Align table columns on decimal point
\usepackage{epsfig}
\usepackage{epstopdf}
\usepackage{bm}% bold math
\usepackage{amsmath}
\usepackage{amsfonts}
\usepackage{textcomp}
\usepackage{bbm}
\usepackage{setspace}
\usepackage{url}
\usepackage{hyperref}

\renewcommand{\thefootnote}{\fnsymbol{footnote}}

\def\simge{\mathrel{%
   \rlap{\raise 0.511ex \hbox{$>$}}{\lower 0.511ex \hbox{$\sim$}}}}

\def\simle{\mathrel{
   \rlap{\raise 0.511ex \hbox{$<$}}{\lower 0.511ex \hbox{$\sim$}}}}

\def\s#1{\setbox0=\hbox{$#1$}%
\rlap{\ifdim\wd0>.7em\kern.22\wd0\else\kern.1\wd0\fi /}#1}

\newcommand{\emb}{\hookrightarrow}

\newcommand{\R}[1]{\text{\bf#1}}

\newcommand{\Rb}[1]{\bar{\text{\bf#1}}}

\newcommand{\I}{{\cal I}}
\newcommand{\Is}{\overline{{ \cal I}}}

\newcommand{\g}[1]{ #1}
\newcommand{\C}{\mathbb{C}}
\newcommand{\N}{\mathbb{N}}
\newcommand{\Z}{\mathbb{Z}}
\newcommand{\RR}{\mathbb{R}}

\newcommand{\Int}[3]{\I_{#1}[#2] = #3}
\newcommand{\Ints}[3]{\Is_{#1}[#2] = #3}
\newcommand{\Id}[1]{\mathbf{1}_{#1}}
\newcommand{\rep}[1]{{\cal R}(#1)}

\newcommand{\name}{\emph{SUtree}}
\begin{document}
%%%%%%%%%%%%%%%%%%%%%%%%%%%%%%%%%%%%%%%%%%%%%%%%%%%%%%%%%%%%%%%%%%%%%%%%%%%

\setcounter{footnote}{0}
\renewcommand{\thefootnote}{\arabic{footnote}}

\begin{titlepage}
\begin{flushright}\begin{tabular}{l}
CP3-Origins-2011-033 \\
DIAS-2011-25
\end{tabular}
\end{flushright}
\vskip1.5cm
\begin{center}
 {\Large \bf \boldmath Explicit and spontaneous breaking of $SU(3)$ into 
 its finite subgroups}

\vspace*{0.5cm}
{\sc Alexander Merle$^{\,a}$\footnote{\tt amerle@kth.se}  \& Roman Zwicky$^{\,b}$\footnote{\tt R.Zwicky@soton.ac.uk}}

\vspace*{0.5cm}
$^a${\normalsize \it KTH Royal Institute of Technology, School of Engineering Sciences,}\\
{\normalsize \it Department of Theoretical Physics, AlbaNova University Center,}\\
{\normalsize \it Roslagstullsbacken 21, 106 91 Stockholm, Sweden}\\[0.2cm]
$^b${\normalsize \it School of Physics \& Astronomy, University of Southampton,}\\
{\normalsize \it Highfield, Southampton SO17 1BJ, UK}
\vspace*{0.5cm}

{\large\bf Abstract:\\[8pt]} \parbox[t]{\textwidth}{
We investigate the breaking of $SU(3)$ into its subgroups from the viewpoints 
of explicit and spontaneous breaking. A one-to-one link between these two approaches 
is given by the complex spherical harmonics, which form a complete set of $SU(3)$-representation functions. An invariant of degrees $p$ and $q$ in complex conjugate variables 
corresponds to a singlet, or vacuum expectation value, in a $(p,q)$-representation of $SU(3)$. 
We review the formalism of the Molien function, 
which contains information on primary and secondary invariants. 
Generalizations of the Molien function to the tensor generating functions 
are discussed.   The latter allows all branching rules to be deduced. 
We have computed all primary and secondary invariants
for all proper finite subgroups of order smaller than $512$, 
for the entire series of groups $\Delta(3n^2)$, $\Delta(6n^2)$, and for all  crystallographic groups.  
Examples of sufficient conditions for breaking into a subgroup are worked out for
the entire $T_{n[a]}$-, $\Delta(3n^2)$-, $\Delta(6n^2)$-series and for all crystallographic groups $\Sigma(X)$. The corresponding invariants provide an alternative definition 
of these groups.
A Mathematica package, \name,  is provided which allows the extraction of  the invariants, Molien and generating functions, syzygies, VEVs, branching rules, character tables, matrix $(p,q)_{SU(3)}$-representations, Kronecker products, etc.\ for the groups
discussed above.}

\vfill

\end{center}
\end{titlepage}

\setcounter{footnote}{0}

\tableofcontents

\newpage

\section{Introduction }

The aim of this work is to study the breaking of a group $G$, referred to as the  \emph{mastergroup}, into 
one of its subgroups in the frameworks of explicit  and spontaneous symmetry breaking. As the master group we have in mind $G = SU(3)$ and as subgroups proper finite subgroups thereof, denoted by ${\cal F}_3$, are considered. 
The methods that we use are general but some of the results, such as the necessary and sufficient 
conditions for $SU(3) \to {\cal F}_3$, are specific to $SU(3)$. The focus on the latter is motivated by the hope that patterns in the flavour sector  
of the Standard Model (SM) are linked with such a symmetry. This hope was fueled ever since the tribimaximal mixing matrix for the lepton sector has been proposed~\cite{tribi}, resulting in many model studies based on discrete subgroups of $SU(3)$~\cite{FlavourModels}.
Discrete Abelian \cite{Buchmuller:2008uq}  and non-Abelian \cite{Kobayashi:2006wq} symmetries further arise in string theory.

Let us begin with a group theoretic introduction. A group can be defined either algebraically, e.g.\ by giving relations between its generators, 
or through a specific (faithful) representation. For example, the well-known group $O(3)$
can be defined from its fundamental representation, which corresponds to a rotation 
in a $3$-dimensional space, as follows:
\begin{equation}
\label{eq:tbook}
O(3) = \{ O \in M_3(\RR) | O^T O = 1 \} \;,
\end{equation}
where $M_3(\RR)$ denotes the set of $3 \times 3$ matrices over $\RR$. 
Equivalently, if we regard $O\in M_3(\RR)$ as acting on a  three-dimensional (representation) space $\RR^3$ via the matrix-vector multiplication $\vec{x}=(x,y,z) \mapsto O \vec{x}$, then $O(3)$ can be defined as the set of matrices leaving the polynomial
\begin{equation}
\label{eq:P2}
P_2 =  x^2 + y^2 + z^2 = \vec{x}^T \vec{x}
\end{equation}
invariant.  The condition \eqref{eq:tbook} and the invariance of \eqref{eq:P2} are 
equivalent, which is easily verified.
Thus the polynomial $P_2$ defines 
the group $O(3)$. Note that $P_2$ will be denoted by $\I_2[SO(3)]$ later on.
The realisation of this idea to finite groups is the main goal of this work. 
More precisely, we shall be concerned with finding a minimal number of invariants
that enforce a faithful irreducible representation (irrep) of a group and thus
can be seen as an alternative definition of the group under consideration.

The so-called \emph{Molien function} provides a powerful and simple tool to obtain
the number of algebraically independent and dependent invariants of a group. 
An important and subtle question, to be discussed, is which invariants are necessary and sufficient to define a group as there are groups which have common invariants.
This question will lead to the investigation of maximal subgroups.
For example, the symmetries of the  permutation groups on 4 elements, $S_4$ and $A_4  \subset S_4$, both leave the polynomial $P_4 = x^4 + y^4 + z^4$ invariant. 
Imposing $P_4$ in addition to \eqref{eq:P2} leads to  $S_4$ and not $A_4$, since the latter is a subgroup of the former.
In this work we shall provide a pragmatic solution to the problem and thus find the necessary 
conditions to break from $SU(3)$ to one of its finite subgroups via \emph{invariant polynomials}, called \emph{explicit breaking}.
Moreover we shall show how the language of invariant polynomials can be translated into 
the language of \emph{vacuum expectation values} (VEVs), called \emph{spontaneous symmetry breaking} (SSB) in the physics literature. This is achieved by 
the so-called \emph{complex spherical harmonics}, which are the generalization of the spherical harmonics from $SO(3)$ to $SU(3)$.

The paper is organized as follows: In Sec.~\ref{sec:mainideas} we discuss the main conceptual ideas and tools.
In Sec.~\ref{sec:SU(3)database} we introduce the list of $SU(3)$ subgroups that we are going to study and which are implemented in our database \name, as given in Tab.~\ref{tab:GroupList}. We also present all invariants of the $\Delta$-groups, by which we mean the countable series
$\Delta(3n^2)$ and $\Delta(6n^2)$ here and thereafter. In Sec.~\ref{sec:breaking} we discuss
the necessary and sufficient conditions for breaking into a large class of $SU(3)$ subgroups. Some examples are given for illustration. 
In Sec.~\ref{sec:embedding} we show that our results are independent on 
the embedding up to trivial transformations.
In Sec.~\ref{sec:conclusions} we conclude and give an outlook. 
Useful details and topics are discussed in various appendices.
In particular the tensor generating functions and branching rules are discussed 
in App.~\ref{app:generating}. 

This is a paper targeted at a physicist audience.
The physics background, such as spontaneous symmetry breaking, etc.\ is not explained
in any sufficient detail, and a language with reference to a physics background is used at times. 
Modulo this issue the text should be readable for mathematicians as well.
Some familiarity with basic finite group theory beyond the facts mentioned in App.~\ref{app:facts}, such as character tables, etc.\ is assumed.

\section{Main conceptual ideas and tools}
\label{sec:mainideas}

In this section we discuss the main conceptual ideas and tools of this paper with 
the example of the breaking of $SO(3) \to S_4$.\footnote{The generalization of these ideas 
is partly obvious and the specific implementation to $SU(3)$ will be discussed in Sec.~\ref{sec:breaking}.} The main topics are the explicit breaking~\ref{sec:explicit}, 
the breaking via VEVs~\ref{sec:VEV}, 
the  connection of the latter two~\ref{sec:connection}, 
how to obtain all algebraically independent invariants and thus VEVs~\ref{sec:Molien}, 
and the question of maximal subgroups~\ref{sec:sufficient}, the latter being the most subtle point to handle 
in practice.
The discussion  mostly follows  the language of Lagrangian field theory; occasionally 
the mathematical perspective is added for clarity.

\subsection{Explicit breaking }
\label{sec:explicit} 

The discussion in this subsection partly overlaps with the introduction. 
Suppose $(x,y,z)\in \mathbb{R}^3$ is the space upon which the fundamental representation 
of $SO(3)$, denoted by $\R{3}$, acts.\footnote{Note that $SO(3)$ is a subgroup 
of $SU(3)$. In case we were to consider the breaking $SU(3) \to S_4$ we would have to impose the polynomial $P_2$ from Eq.~\eqref{eq:P2} as additional invariant. When two 
or more invariants have to be imposed, this can be understood as a sequential breaking, e.g., $SU(3) \to SO(3) \to S_4$.}  
For the breaking into $S_4$, $SO(3) \to S_4$, it is  
 is sufficient
to demand invariance under the following polynomial,\footnote{It is crucial here that the master groupis $SO(3)$ and not $SU(3)$, as otherwise $SU(3) \to \Delta(6\cdot 4^2)
\supset \Delta(6 \cdot 2^2) \simeq S_4$
with the invariant mentioned above, as we shall see in Sec.~\ref{sec:breaking}.} 
\begin{alignat}{2}
\label{eq:example}
& {\cal I}_4[S_4] = x^4 + y^4 + z^4\;,     \qquad                   & &   SO(3) \to S_4 \;.
\end{alignat}
N.B.: We have left aside how to find invariant polynomials to Sec.~\ref{sec:Molien} and 
the more subtle question of the choice of polynomials to Sec.~\ref{sec:sufficient}.

How is this phrased in the language of Lagrangian field theory? 
We would think of $SO(3)$ or $SU(3)$ as internal symmetries\footnote{As opposed to 
a space-time symmetries.} of a field $(\varphi_1, \varphi_2,\varphi_3)$ within the representation space. 
The \emph{explicit breaking} in Lagrangian language is accomplished 
by adding a polynomial of the invariant~\eqref{eq:example},\footnote{In the case where 
 we restrict ourselves to renormalizable terms, $f$ ought to be linear.}
\begin{equation}
{\cal L}_{S_4} = {\cal L}_{SO(3)}(\varphi_1^2  + \varphi_2^2 + \varphi_3^2 )  + f (\varphi_1^4  + \varphi_2^4 + \varphi_3^4 ) \; ,
\label{eq:explicit}
\end{equation}
to the original $SO(3)$ invariant Lagrangian ${\cal L}_{SO(3)}$, where $f$ is a polynomial function.
The new term can be regarded as an addition to the potential.
The term explicit breaking has to be contrasted with the term spontaneous breaking, to be discussed below, which is more indirect.

\subsection{Breaking via vacuum expectation values (VEVs)}
\label{sec:VEV}

If we were to consider  $SO(3)$ in its fundamental representation $\R{3}$ 
and single out one direction, then the symmetry  breaks down to $SO(2)$. 
This ought to be obvious from a spatial drawing. 
How can the breaking of $SO(3)$ to, say, $S_4$ or any group different from $SO(2)$ occur in this language? This happens when higher, not fundamental, representations are considered.\footnote{In Ref.~\cite{Adulpravitchai:2009kd} a few 
small representations were considered for $SU(2)$ and $SU(3)$, and it was found that they cannot break to any non-Abelian group except for $D_2'$, which is the double cover of the the dihedral group $D_2$.}

We shall formulate this idea first in mathematical language without referring to $SO(3)$ or $SU(3)$.  
Let us choose a   certain (faithful) representation of the master group $G$, 
denoted by ${\cal R} = {\cal R}(\g{G})$, acting on the representation space $V$ with $
 d_G  =  \dim(V) > 3$. 
Further we single  out a certain representation vector $v \in V$ and collect 
the following elements,
\begin{equation}
\label{eq:Hv}
H = \{ g \in G \,|\, {\cal R}(g) v = v  \} \; .
\end{equation}
It is readily verified that $H$ constitutes a representation of a proper subgroup of $G$. 
The group $H$ is called \emph{stabilizer}, \emph{isotropy group}, or \emph{little group},
depending on the area of research.

In the language of Lagrangian field theory one would add a  potential 
to the initial kinetic term:
\begin{equation}
\label{eq:explicit}
{\cal L}_H = {\cal L}_{SO(3)}^{\rm kinetic}  - U(\varphi_1, ..,\varphi_{d_G}) \;,
\end{equation}
which is ${\cal R}(G)$-invariant but whose extremum $v$, obtained by
\begin{equation}
\label{eq:extremum}
\frac{\partial}{\partial \varphi_i} U(\varphi_1, ..,\varphi_{d_G})  = 0 \;, \; i = 1, .., d_G \quad \Leftrightarrow  \quad
(\varphi_1, ... ,\varphi_{d_G} ) = v \;,
\end{equation}
is ${\cal R}(H)$-invariant but not ${\cal R}(G)$-invariant. As the notation suggests, $v$ in Eq.~\eqref{eq:extremum} corresponds to the representative $v$ in Eq.~\eqref{eq:Hv}. 
In the physics literature this phenomenon is called 
\emph{spontaneous symmetry breaking} and $v$ is referred to as a \emph{VEV}.

\subsection{From invariants to VEVs and back }
\label{sec:connection}

It is  natural to ask of whether, given $\I_4[S_4]$ in Eq.~\eqref{eq:example}, one can determine the corresponding $v$ that leads to  $H = S_4$ in Eq.~\eqref{eq:Hv} and vice versa. 
In other words: Is there a link between explicit breaking and spontaneous breaking? \footnote{This question was raised but deferred to later work in Ref.~\cite{turkey}.}

The link is readily established by noting that certain polynomial functions furnish a 
representation of the group.
In the case of $SO(3)$ this is usually given by the spherical harmonics $Y_{l,m}$.
The latter correspond to a complete set of representation functions of $SO(3)$ for $l = 0,1,2,...$,
with representation dimensions $2l+1$ as $m$ ranges from $m = -l$ to $m=+l$ in integer steps.
 
Expanding the invariant polynomial in spherical harmonics and using the orthogonality relations, one obtains:
\begin{equation}
\label{eq:Y4m}
{\cal I}[S_4] = x^4 + y^4 + z^4   = c \left(Y_{4,-4} + \sqrt{\frac{14}{5}} Y_{4, 0} +   Y_{4,4}\right) \;,
\end{equation}
where $c$ is an irrelevant constant depending on the normalization of the spherical harmonics.
This means that in an $l=4$ representation, which is $9$-dimensional, a direction
\begin{equation}
\label{eq:vev}
v \sim (1,0,0,0, \sqrt{\frac{14}{5}},0,0,0,1)
\end{equation}
breaks  $SO(3) \to S_4$.   In the language  of branching rules this reads,
\begin{equation}
\label{eq:branch_mini}
\R{9}_{SO(3)}|_{S_4} \to \R{1}_{S_4} + ... \;,
\end{equation}
where the dots stand for higher representations. Note that it is the VEV $v$~\eqref{eq:vev} 
which corresponds to the trivial irrep under the restriction to $S_4$. 
How to obtain the other irreps in Eq.~\eqref{eq:branch_mini}, which are the ones of interest for model building, is described in App.~\ref{app:branching}.
 
It is the goal of this paper to generalize this to $SU(3)$. A few explicit examples can be found in Sec.~\ref{sec:examples}; the complex spherical harmonics are discussed in App.~\ref{app:csh}.

\subsection{Invariants of a representation -- the Molien function }
\label{sec:Molien}

Given a certain finite group $H$, is it possible to 
obtain all polynomial algebraic invariants? 
The answer is affirmative through the so-called \emph{Molien function}~\cite{Molien}, more generally known as the 
\emph{generating function}~\cite{Patera:1978qx}.
The Molien function is defined, for finite groups, as follows:
\begin{equation}
\label{eq:Molien}
M_{{\cal R}(H)}(P) \equiv \frac{1}{|{\cal R}(H)|} \sum_{h \in {\cal R}(H)} \frac{1}{\det(\Id{}-P\, h)} = \sum_{m \geq 0} h_m  P^m \; ,
\end{equation}
where $P$ is a real number, ${\cal R}(H)$ is a representation of $H$, and $|{\cal R}(H)|$
denotes the number of elements in that representation. Thus the Molien function 
is the average of the inverses of the characteristic polynomials over the group.
The \emph{Molien theorem} states that
the positive integer numbers $h_m$ correspond to the numbers of invariants $\I_{m}$ of
degree $m$ that leave the subgroup ${\cal R}(H)$ invariant, see e.g.~\cite{sturmfels} 
or~\cite{Meyer} for a discussion within $SO(3)$.\footnote{
The concept of the Molien function finds its generalization in the generating function.
For a generic review on this powerful subject we refer the reader to Ref.~\cite{Patera:1978qx} and references therein.  In App.~\ref{app:generating} the generalization 
from counting invariants to counting covariants is presented. As previously mentioned, the branching rules can be obtained 
in this framework as well.}
Since any (polynomial) function of invariants is also an invariant, the question of minimality imposes itself. Thus, what are the algebraically independent invariants and how 
do the dependencies between the others work out?
 
It turns out 
that, for an $n$-dimensional representation,  
there are exactly $n$ algebraically independent invariants~\cite{noether}, the so-called  \emph{fundamental} or \emph{primary} invariants~\cite{burnside}. Furthermore there are the \emph{secondary} invariants, denoted by $\Is$ as opposed to $\I$, which are \emph{not} algebraically independent.
Relations of them and primary and 
 secondary invariants are as follows
\cite{sturmfels}:
\begin{equation}
\Is_{n_i}^{\;2} = f_0(\I_{m_1},\I_{m_2},\I_{m_3}) + \sum_j f_1^{(j)}(\I_{m_1},\I_{m_2},\I_{m_3}) \cdot \Is_{n_j}\; ,
\label{eq:syzygy}
\end{equation}
where $f_0$ and $f^{(j)}_1$ are (polynomial) functions that depend only on the primary invariants, as indicated. 
Relations, as the one in Eq.~\eqref{eq:syzygy}, are called \emph{syzygies} in the mathematical literature.  Note that, once Eq.~\eqref{eq:syzygy} is verified, we can be sure that 
we have found a valid set of primary and secondary invariants. 

As a matter of fact, given a set $\{ \I_{m_1}, \I_{m_2},\I_{m_3}, \Is_{n_i}, .. \}$ of primary and secondary invariants, the Molien function  can be
written as~\cite{sturmfels}:
\begin{equation}
\{ \I_{m_1}, \I_{m_2},\I_{m_3}, \Is_{n_i}, .. \}  \quad \Rightarrow \quad   M_{\g{H}(\R{3})}(P) = \frac{1 + \sum_i a_{n_i} P^{n_i}}{(1-P^{m_1})(1-P^{m_2})(1-P^{m_3})} \; .
\label{eq:Molien_ex}
\end{equation}
Here we have specialized to a $3$-dimensional representation but the generalization 
should be obvious.
The three  primary invariants are of degrees $m_1$, $m_2$, and $m_3$, respectively. 
Further to that there are  $1+\sum_i a_{n_i}$  secondary invariants one of which  
 is the trivial invariant and there 
are $a_{n_i}$ invariants of degree $n_i$.
Note that the syzygies in~\eqref{eq:syzygy} are consistent with the fact that 
secondary invariants, associated with $P^{n_i}$, 
do not appear to any other power than one.

The representation of the Molien function~\eqref{eq:Molien_ex} is not unique and this 
is why the logical arrow only goes from left to right and not the other way around.
 In practice
invariants can be found by following the three step procedure below:
\begin{enumerate}
\item A form of the Molien function as in Eq.~\eqref{eq:Molien_ex} is guessed. \\
{\small One should also check that it verifies the proposition in Eq.~\eqref{eq:proposition}.}
\item The corresponding invariants are generated (to be discussed below) and the algebraic independence of the primary invariants is verified. \\
{\small Algebraic independence of potential primary invariants can be checked with 
the Jacobian criterion, Eq.~\eqref{eq:Jac_Del3}.}\\
{\small In the case where primary and/or secondary invariants are degenerate in degree, complications may arise, c.f.\ App.~\ref{app:degenerate_same}.}
\item The syzygies from Eq.~\eqref{eq:syzygy} are verified.
\end{enumerate}
If the latter step fails one has to return to the non-uniqueness of steps one and two.
Issues about the non-uniqueness of the form of the Molien function and strategies on how
to deal with cases when the degrees of the polynomials are degenerate are discussed 
in App.~\ref{app:subtle}.
One  powerful fact, c.f.\ proposition~2.3.6.\ in Ref.~\cite{sturmfels}, which helps with point one is 
that the number of secondary invariants equals,
\begin{equation}
\label{eq:proposition}
\text{number of secondary invariants} \equiv 1 + \sum_i a_{n_i}  = \frac{m_1\cdot m_2 \cdot m_3}{|H|} \;,
\end{equation}
with an obvious generalization to an $n$-dimensional representation.

Given the information on the degrees of the invariants from Eq.~\eqref{eq:Molien_ex}, how 
can the invariants be constructed? This is rather straightforward, modulo ambiguities in 
form of degeneracies, by symmetrization of trial polynomials. 
We  observe that for any (polynomial) function $f(x, y, z)$, an invariant ${\cal I}(x,y,z)$ can be obtained as follows,
\begin{equation}
\label{eq:genI}
{\cal I}(x,y, z) =  \frac{1}{|{\cal R}(H)|} \sum_{h \in  {\cal R}(H) } f(h \circ x, h \circ y , h \circ z) \;,
\end{equation}
where here and thereafter $\circ$ denotes the action of the group on an element of 
the representation space. Verification of the invariance in \eqref{eq:genI} is immediate and left to the reader.
This operation is known as the \emph{Reynolds operator} in the mathematical literature, see e.g.~\cite{sturmfels}. Note that the form of the invariants is dependent on the embedding, e.g.\ on similarity transformations as discussed in Sec.~\ref{sec:embedding}.
In practice this means that the invariants can be obtained by taking a suitable ansatz 
for the function $f(x,y,z)$. For our purposes the most convenient trial functions are:
\begin{equation}
f(x,y,z) =  x^n y^m z^{4-n-m} \;, \qquad m,n \geq 0 \;.
\end{equation}
N.B.: For most trial functions this invariant is going to be zero, which is the trivial invariant.
An example on how to obtain the invariants, with $S_4$, is discussed in App.~\ref{app:exampleS4_reynolds}.

It is now time to return to our example $S_4$ and execute the three step procedure outlined 
previously.  The degrees of the invariants are such that no problems of the kind mentioned in point two occur.

Step 1:  By computing~\eqref{eq:Molien} 
and looking for poles we guess that the Molien function as in 
Eq.~\eqref{eq:Molien_ex} takes the following form,
\begin{equation}
\label{eq:MolienS4}
M_{S_4}(P) =  \frac{1+P^9}{(1-P^2)(1-P^4)(1-P^6)}\;,
\end{equation}
which satisfies proposition \eqref{eq:proposition}.

Step 2: By using the Reynolds operator 
the primary,
\begin{equation} 
\label{eq:IS4}
\Int{2}{S_4}{x^2 + y^2 + z^2} \;,\quad
 \Int{6}{S_4}{(xyz)^2} \;,\quad
 \Int{4}{S_4}{x^4 + y^4 + z^4}\;,
\end{equation}
and secondary,
\begin{equation}
\label{eq:IS4s}
\Ints{9}{S_4}{xyz(x^2 - y^2)(y^2-z^2)(z^2-x^2)}\;,
\end{equation}
invariants are readily computed, and the algebraic independence of the former can be shown easily by, e.g., the Jacobian criterion~\cite{Jacobi}.
Note that the first primary invariant is merely the statement that $S_4$ is a subgroup of $SO(3)$.

Step 3: The syzygy~\eqref{eq:syzygy} is verified to be
\begin{eqnarray}
&& \overline{\I}_9^2  = \I_2^{\,4} \I_4 \I_6  -\frac{1}{4} \I_2^{\,6}  \I_6   - \frac{5}{4}  \I_2^{\,2} \I_4^{\,2}  \I_6
+ \frac{1}{2}   \I_4^{\,3}  \I_6 + 5  \I_2^{\,3}  \I_6^{\,2} - 9  \I_2 \I_4  \I_6^{\,2} 
- 27   \I_6^{\,3} \;,
\label{eq:I9sq}
\end{eqnarray}
where we have omitted the $[S_4]$ on the invariants. By verifying the syzygy, we have completed the program and shown that~\eqref{eq:MolienS4} is indeed the Molien function as in Eq.~\eqref{eq:Molien_ex}.

\begin{figure}[h]
  \centering
  \includegraphics[width=3.0in, trim=0.85mm 0.85mm 0.85mm 0.85mm, clip]{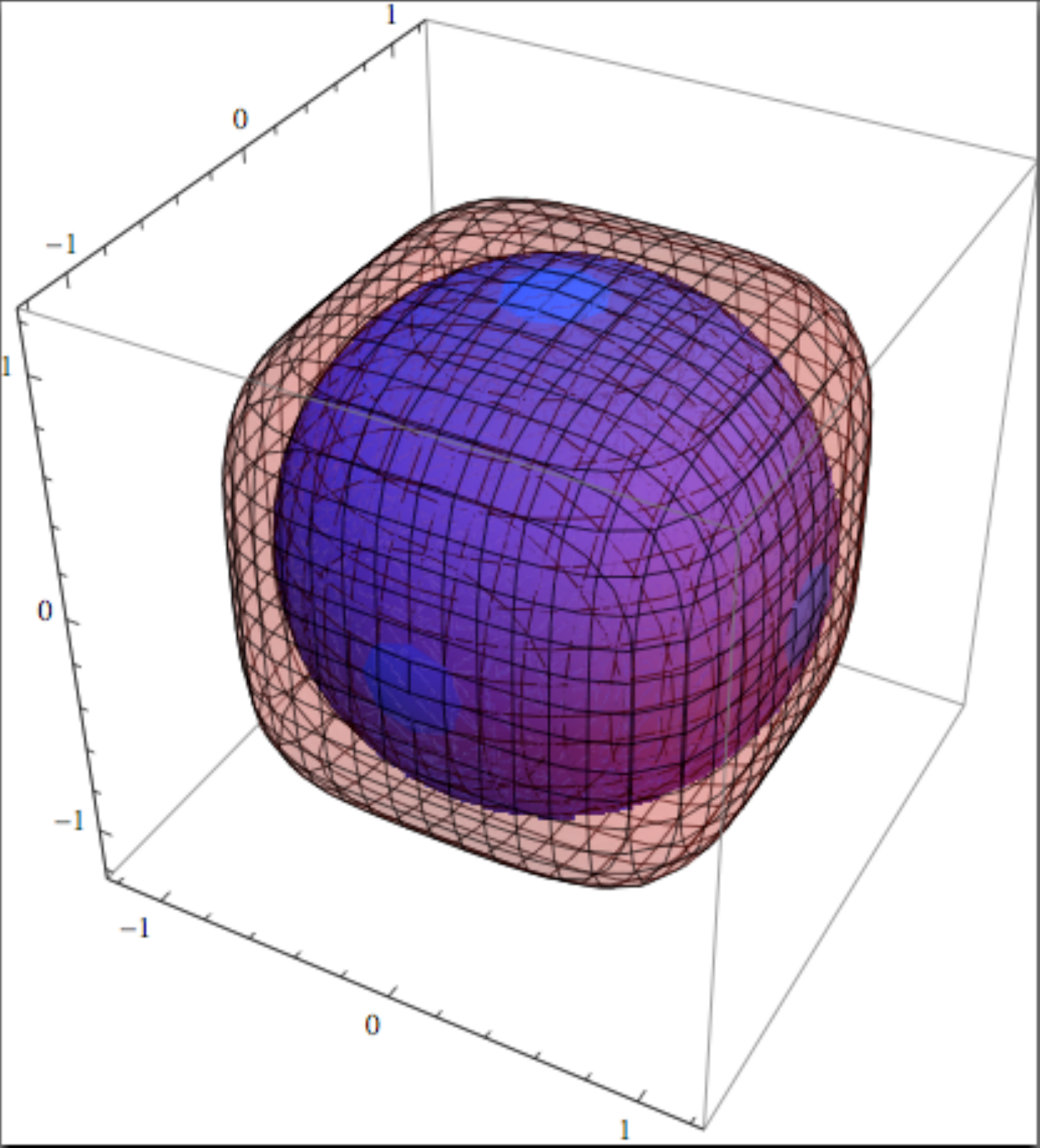}
  \includegraphics[width=3.0in, trim=0mm 0.85mm 0mm 0mm, clip]{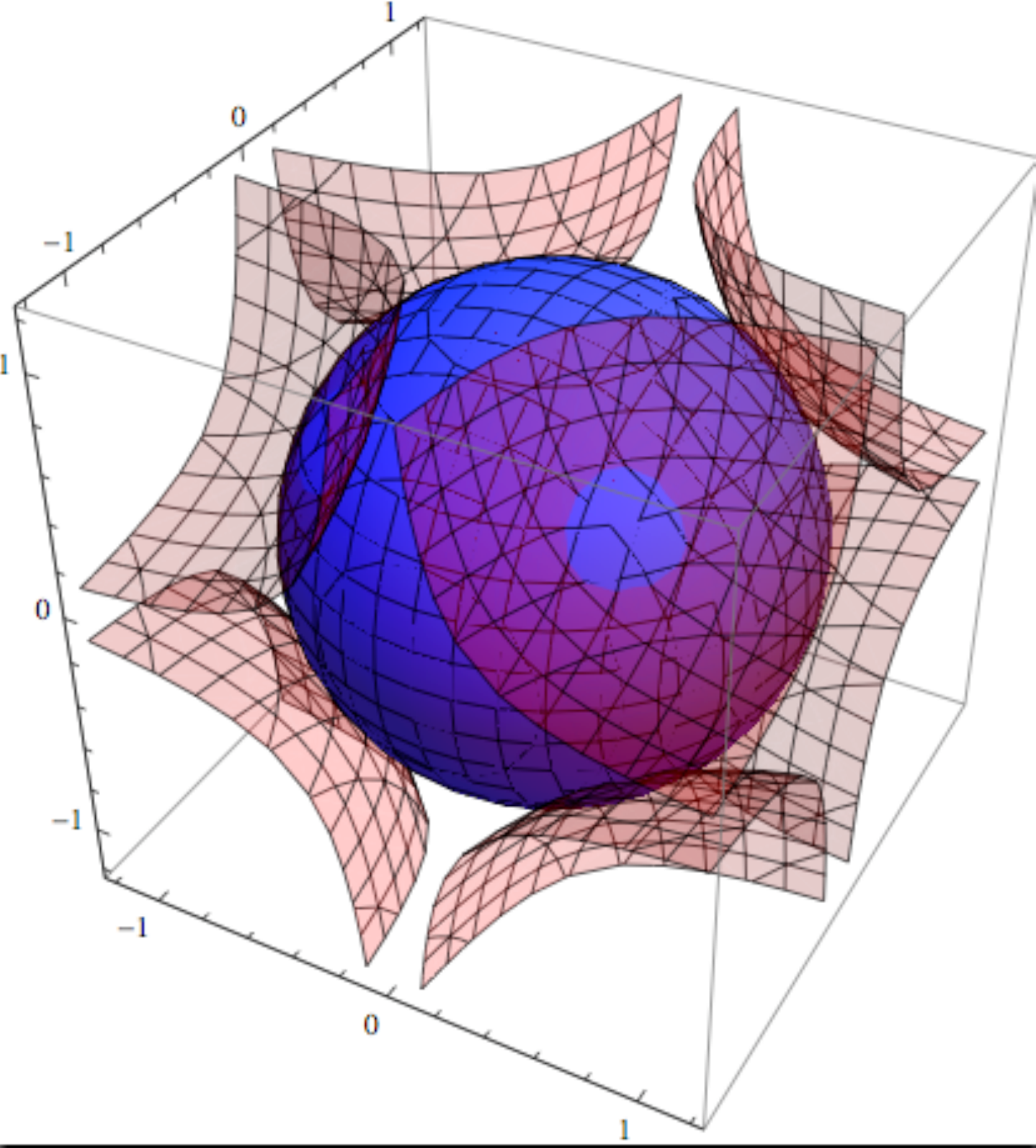} 
 \caption{\small This figure shows the 3-sphere (blue/dark gray) corresponding to $SO(3)$ 
 and (left) the equipotential surfaces (light brown/gray) corresponding to the invariant $\I_4[S_4] = x^4 + y^4 + z^4$ and (right) the same for $\I_6[S_4] = (xyz)^2$. 
 The two invariants, as shall be argued later in Sec.~\ref{sec:breaking}, are sufficient 
 to break $SO(3) \to S_4$. The polyhedric symmetries of the hexahedron (cube) and octahedron, 
 which are dual to each other under interchange of faces and edges, beautifully reveal themselves in this plot. On the left the intersection of the sphere and 
 the $\I_4$-invariant corresponds to the six faces or six edges of the hexahedron 
 and octahedron, respectively, whereas on the right the analoguous intersection corresponds 
 to the eight edges and eight faces of the hexahedron and octahedron, respectively. 
 This is why $S_4$ is, at times, called the \emph{hexahedron} or \emph{octahedron group}. }
\label{fig:S4graphics}
\end{figure}

This completes the exemplification of the Molien function and primary and secondary invariants for $S_4$.  In Fig.~\ref{fig:S4graphics}  the geometric nature of
these invariants is revealed in plots.
In Sec.~\ref{sec:Del36} we obtain all primary and secondary invariants
including the syzygies for the entire $\Delta$-series. In App.~\ref{app:generating} 
the generalization from invariants to covariants, by which we mean tensor objects,
is discussed by going from the Molien function to the tensor generating function.
From the tensor generating function of a group the branching rules can be deduced, as 
implemented in our package \name.

\subsection{Criteria for breaking into subgroups - maximal subgroups}
\label{sec:sufficient}

In the previous discussion we have simply assumed that, by imposing the invariant~\eqref{eq:example}, the group breaks from $SO(3) \to S_4$. How can we be sure of that? 
What are the necessary and sufficient criteria?

Consider a group $H$ and an unordered list of invariants 
$\{\I[H]_1, \I[H]_2,..\}$ associated to it,
as well as the corresponding vectors $\{v[H]_1,v[H]_2, ..\}$ constructed as in   Eqs.~\eqref{eq:Y4m} and~\eqref{eq:vev}. 
The certain fact is that 
the group $H$ leaves $\I[H]_j$ and $v[H]_j$ invariant by  construction.
However, there can be other groups $H'$ which leave 
them invariant as well. 
It is a fact~\cite{Michel} that in this case the group must break into the 
largest group, which we shall denote by $\overline H \subset G$. Basically, there are three distinct relations between $H$, $H'$, and $\overline H$:
\begin{itemize}
\item subgroup:  $H \subset ..  H' .. \subset \overline H$
\item supergroup: $H' \subset .. H .. \subset \overline H$
\item no such relation: $H \subset .. \overline H ..  \supset H'$
\end{itemize}

As an  example of the second case we mention that 
$A_4$ and $S_4$  
both  leave $\I_4$ from Eq.~\eqref{eq:IS4} invariant, but since  $A_4 \subset S_4$ 
the master group $SO(3)$ breaks into $S_4$ with $\I_4$. Note, we have assumed that there does
not exist a group $H'$ with $S_4 \subset H' \subset G$ which leaves $\I_4$ invariant.

Thus finding the sufficient criteria for breaking into a subgroup is a subtle issue. 
This problem can be handled once one knows the ``tree'' of subgroups from the group $G$.
Ideally we would therefore like to know when given two groups of which one is  a subgroup 
of the other, $H \subset G$, whether $H$ is a maximal subgroup or whether there is another group $H'$ in between, $H \subset H' \subset G$. In the case where $G$ is a finite group this question
can be settled by a computer algorithm using corollary 1.5.A in~\cite{permutationgroups}. In the case where $G$ is not finite, e.g.\ continuous, the question cannot  be answered in general.  In connection with Lie groups a general discussion can be found in~\cite{O'Raifeartaigh:1986vq}, and many examples are given in~\cite{nice}.
In the context of $SO(3)$ the so-called \emph{Michel 
criterion} was known for some time~\cite{Michel}, but counterexamples
have been found~\cite{counter}.
In Sec.~\ref{sec:maxSU(3)} we shall discuss strategies to cope with the proper finite 
 subgroups of $SU(3)$.

\section{$\boldsymbol{SU(3)}$ database}
\label{sec:SU(3)database}

The  $SU(3)$ subgroups have been classified almost 100 years ago 
\cite{Milleretal}. For a review of the contributions thereafter we refer the reader to 
the introduction of Ref.~\cite{Ludl:2011gn}.
The generators of the  proper finite $SU(3)$ subgroups are
 given in Tab.~\ref{tab:SU3_SubGen}, taken from Ref.~\cite{Ludl:2010bj}. 
The question to what extent these results are embedding dependent 
is discussed in Sec.~\ref{sec:embedding}. 
By proper we mean  the groups which are not subgroups of $SU(2)$.
The invariants of the $SO(3)$ and $SU(2)$ subgroups have been discussed 
extensively in the literature in regard to applications in crystallography. 
There are the crystallographic types $A_4$, $S_4$, and $A_5$, 
the dihedral groups $D_n \simeq \Z_n \rtimes \Z_2$, and the cyclic groups $\Z_n$. For a discussion 
of the invariants of these groups we refer the reader to the extensive review~\cite{MichelPR}.

In our database \name, we are going to restrict ourselves 
to the proper finite $SU(3)$ subgroups of order smaller than $512$,\footnote{As is nicely illustrated in~\cite{Ludl:2010bj}, above the order of 512 the number of 
groups becomes rather large and barely manageable.} supplemented 
by the three crystallographic subgroups of $SO(3)$ mentioned above and by the 
two crystallographic groups $\Sigma(216\phi)$ and $\Sigma(360\phi)$ whose orders
are $3 \cdot 216$ and $3 \cdot 360$, respectively, and exceed $512$.
A complete list is given in Tab.~\ref{tab:GroupList} in App.~\ref{app:GenGroup} with our code (group number), GAP code, and alternative names. 
The classification basically falls into two types: The groups $\Sigma(X)$, the so-called 
\emph{crystallographic groups}, and the countable series of $C$- and $D$-groups of which 
$\Delta(6n^2)$, $\Delta(3n^2)$, and $T_{n[a]}$ are special cases as can be inferred from 
Tab.~\ref{tab:SU3_SubGen}.  This table differs from the one in~\cite{Ludl:2010bj} by writing 
the generators $M,J,P,Q$ in terms of $C$- and $D$-type generators and by $T_n \to T_{n[a]}$.
The latter is necessary as for certain $n$ there exist several solutions for $a$, e.g.\ for $n=\{91,133\}$ in Tab.~\ref{tab:GroupList}.
To some degree the division into crystallographic and non-crystallographic groups is arbitrary, as $A_4 \sim \Delta(12)$ and $S_4 \sim \Delta(24)$.
We have worked out the subgroup structure or ``subgroup tree" of this entire list 
with the generator basis and GAP,  and 
it is given in App.~\ref{app:group_tree} in Fig.~\ref{fig:su3tree} and Tab.~\ref{tab:group_tree}, where
further remarks on this process can be found.
 We do not include groups which are of the type 
${\cal F}_3 \times \Z_n$, where $\Z_n$ denotes the cyclic group of order $n$.  
These groups can  be reconstructed by using theorem~II.2~\cite{Ludl:2010bj}, which we quote 
in App.~\ref{app:facts} for the reader's convenience.\footnote{$A_4 \times \Z_3$ 
has become popular in model building recently~\cite{Ludl:2010bj}. 
Note that $\Delta(27) \times \Z_3$ has no faithful $3$-dimensional irrep since the center of $\Delta(27)$ is $\Z_3$ and thus does not fall into the category of the theorem~II.2. 
If this was not the case the criteria for breaking into 
$\Delta(3 (3n)^2)$ groups given in Sec.~\ref{sec:breaking} would need further refinement.} An investigation of subgroups of order smaller than $100$ including 
all groups with direct products can be found in Ref.~\cite{Parattu:2010cy}.
More subtle is the question of $U(3)$ subgroups versus $SU(3)$ subgroups. 
The classification of the former has not been completed.\footnote{In~\cite{Ludl:2010bj}
the $U(3)$ subgroups of order smaller than 512 were considered. Moreover this reference
uncovers further series of $U(3)$ subgroups in generalizing the $\Delta$-series. 
Ref.~\cite{Ishimori:2010au} investigates a number of finite $U(3)$ subgroups.}
In particular there is more to it than ${\cal F}_{3} \times \Z_n$, as 
$U(3) \simeq SU(3) \times U(1)$ might suggest,
e.g.\ $S_2(4) \simeq 
A_4 \rtimes \Z_4$~\cite{Ludl:2010bj}.  In this case the $3$-dimensional irreps are obtained
by multiplying a certain generator by $\pm i$. This means that among the invariants quoted in~\eqref{eq:IS4}   and~\eqref{eq:IS4s} only $\I_4$ will remain an invariant.

For the sake of completeness let us mention the topological structures of the non-crystallographic 
groups:\footnote{
It should be kept in mind that the semidirect product is only complete
once the group-homomorphism is given. Thus the notation above does not yet determine the group.}
\begin{alignat}{2}
T_n &: ~~\Z_n \rtimes \Z_3  \qquad  &  \qquad  ~\cite{Bovier:1980ga}\;,  \nonumber    \\[0.1cm]
\Delta(3n^2) &: ~~(\Z_n \times \Z_n) \rtimes \Z_3  & \qquad ~\cite{FFK}\;, \nonumber \\[0.1cm]
\Delta(6n^2) &: ~(\Z_n \times \Z_n) \rtimes S_3  & \qquad ~\cite{FFK}\;,
\end{alignat}
where $\rtimes$ stands for the \emph{semidirect product}. 
The $\Delta$-groups, sometimes called \emph{trihedral groups}, are to be seen as direct generalizations of 
the dihedral groups $D_n \subset SO(3)$, which have the topological structure 
$D_n \sim \Z_n \rtimes \Z_2$.
The general $C$- and $D$-groups have been shown to have the following topological structures:
\begin{alignat}{2}
C(n,a,b) & : ~~(\Z_t \times \Z_u) \rtimes \Z_3  \qquad  &  
\qquad ~\cite{Ludl:2011gn}\;,  \nonumber    \\[0.1cm]
 D(n,a,b,d,r,s) &: ~~  (\Z_t \times \Z_u) \rtimes S_3   & 
 \qquad ~\cite{GL11} \;.
\end{alignat}
An algorithm but no explicit formulae for $t$ and $u$ were given.
For $C$-groups the following is true: $t \leq u \leq n$, which is consistent with  
the statement in Eq.~\eqref{eq:CDsubgroups} in Sec.~\ref{sec:break_noncrystallographics}.

\begin{table}
\begin{center}
\begin{tabular}{|l| l| l| l | }
\hline
Group & Generators   $C$-,$D$-type     &  $\Sigma(X)$-type & MBD \\\hline \hline
$C(n,a,b)$ & $E$, $F(n,a,b)$ & & $C$ \\ 
$D(n,a,b;d,r,s)$ & $E$, $F(n,a,b)$, $G(d,r,s)$ & &  $D$ \\
\hline
$\Delta(3n^2)=C(n,0,1),$ $n\ge 2$ & $E$, $F(n,0,1)$ & &  $\in C$\\
$\Delta(6n^2)=D(n,0,1;2,1,1)$, $n\ge 2$ & $E$, $F(n,0,1)$, $G(2,1,1)$ & & $\in D$ \\
$T_{n[a]}=C(n,1,a)$, $(1+a+a^2)= n \Z$ & $E$, $F(n,1,a)$ & &  $\in C$ \\ \hline \hline
$\Sigma(60)=A_5 = I=Y$ & $E$, $F(2,0,1)$ & $H$   & $H$ \\
$\Sigma(168)  = PSL(2,7)$ & $E$, $M \equiv F(7,1,2)$ &  $N$ & $J$ \\
$\Sigma(36\phi)$ & $E$, $J \equiv F(3,0,1) $ & $K$ & $E$ \\
$\Sigma(72\phi)$ & $E$, $J \equiv F(3,0,1) $ &  $K$, $L$ & $F$\\
$\Sigma(216\phi)$ & $E$, $J= F(3,0,1) $, $P \equiv F(9,2,2)$ &  $K$  & $G$\\
$\Sigma(360\phi)$ & $E$, $F(2,0,1)$,  $Q \equiv G(6,3,5)$ & $H$ & $I$ \\
\hline
\end{tabular}
\caption{\label{tab:SU3_SubGen} \small  Types of finite subgroups of $SU(3)$ which are not subgroups 
of $SU(2)$ with the exceptions of $A_4 \simeq \Delta(12)$, $S_4 \simeq \Delta(24)$, 
and $A_5 \simeq \Sigma(60)$. Explicit generators are given in App.~\ref{app:generators}. 
For $T_{n[a]}$: $n = 3 p$ or $n=p$ with $p$ equal to the product of primes of the form $3 \N +1$~\cite{response,Bovier:1980ga}. Some further remarks on these groups and the 
fact that neither $a$ nor $n$ determine each other can be found in Sec.~\ref{sec:break_noncrystallographics}. This is why we have extended the notation 
from $T_n$ to $T_{n[a]}$. The column MBD corresponds to the classification used in 
\cite{Milleretal} and the letters should not be confused with the generators. The $A$ and $B$ types correspond to direct products of Abelian factors and are not of the type considered here.}
\end{center}
\end{table}

%%%%%%%%%%%%%%%%%%%%%%%%%%%%%%%%%%%%%%%%%%%%%%%%%%%%%%%%%%%%%%%%%%%%%%%%%%%
\subsection{\label{sec:SUtree}Our database: \name }
%%%%%%%%%%%%%%%%%%%%%%%%%%%%%%%%%%%%%%%%%%%%%%%%%%%%%%%%%%%%%%%%%%%%%%%%%%%

We have developed an accompanying software package called \name~with this paper. After having downloaded the software from
\begin{center}
\url{http://theophys.kth.se/~amerle/SUtree/SUtree.html}
\end{center}
and having extracted the files, the easiest way to go is to open the example file
\begin{center}
{\tt ExampleNotebookSUtree.nb}
\end{center}
with Mathematica. In there, all features of the database are explained and exemplified.
A short dialogue example can be found in Sec.~\ref{sec:examples}.

A short summary is given here: Using the group numbers, GAP numbers, or group names, as given in Tab.~\ref{tab:GroupList}, one can efficiently refer to all of the 61 groups in our list.\footnote{Note that some of our results are only true for the representation used in Tab.~\ref{tab:GroupList}, c.f.\ Sec.~\ref{sec:embedding}.} All the group elements are stored numerically in the database. We have used these matrices to calculate the primary invariants, secondary invariants,  Molien functions, and tensor generating functions for all 61 groups in our database. Note that, depending on the group, the expressions in particular for the secondary invariants can be relatively lengthy [e.g.\ the one of $\Sigma(360\phi)$], and the number of secondary invariants can be quite large, too (e.g.\ $T_{163[58]}$ has 116 secondary invariants). In addition, we have also calculated all the corresponding syzygies for the secondary invariants.\footnote{This is true except for the syzygies of the four secondary invariants of $T_{163[58]}$ of highest degree, as well as for the syzygy of $\Sigma(360\phi)$. Furthermore, the syzygy of $A_5$ was obtained  numerically only.} Furthermore, the database contains a routine to calculate any $(l)_{SO(3)}$- or $(p,q)_{SU(3)}$-basis (cf.\ App.~\ref{app:pqBasis}), and to translate all invariant polynomials into VEVs and vice versa. In addition, the character tables are given for all groups, as well as the (tensor-) generating functions,\footnote{This is true for all groups except for $T_{163[58]}$ and for $T_{169[22]}$, for which we did not succeed in finding generating functions, due to the large numbers of conjugacy classes.} from which the branching rules and Kronecker products can be derived with \name.

The functions of \name~can be used to verify many of the following calculations. In some cases, as will be shown below for the $\Delta(3n^2)$ and $\Delta(6n^2)$ groups, it is even possible to use our program to guess general results that can then be proven a posteriori.

%%%%%%%%%%%%%%%%%%%%%%%%%%%%%%%%%%%%%%%%%%%%%%%%%%%%%%%%%%%%%%%%%%%%%%%%%%%
\subsection{\label{sec:Del36}Invariants of $\Delta(3n^2)$ and $\Delta(6n^2)$ }
%%%%%%%%%%%%%%%%%%%%%%%%%%%%%%%%%%%%%%%%%%%%%%%%%%%%%%%%%%%%%%%%%%%%%%%%%%%

In this section we compute the primary and secondary invariants of the groups $\Delta(3n^2)$
and $\Delta(6n^2)$, valid for any $n$.
From Tab.~\ref{tab:SU3_SubGen} and Eq.~\eqref{eq:Gens} we infer that,
\begin{equation}
 E=\begin{pmatrix}
 0 & 1 & 0\\
 0 & 0 & 1\\
 1 & 0 & 0
 \end{pmatrix}\;, \qquad 
 F=F(n,0,1)=\begin{pmatrix}
 1 & 0 & 0\\
 0 & \eta & 0\\
 0 & 0 & \eta^{-1}
 \end{pmatrix} \;, \qquad 
  G=G(2,1,1)=\begin{pmatrix}
 -1 & 0 & 0\\
 0 & 0 & -1 \\
 0 & -1 & 0
 \end{pmatrix}\;,
 \label{eq:Gen_Del3}
\end{equation}
and that $\{E,F\}$ and $\{E,F,G\}$ generate the groups $\Delta(3n^2)$ and 
$\Delta(6n^2)$, respectively, with  $\eta=e^{2\pi i /n}$. 
We are now going through the three step procedure of Sec.~\ref{sec:Molien}.

Step 1: By computing the Molien function~\eqref{eq:Molien} 
for a few cases with lower $n$ the following Molien functions~\eqref{eq:Molien_ex} suggest themselves,
\begin{alignat}{2}
& M_{\Delta (3 n^2)}(P) &=& \frac{1+P^{3n}}{(1-P^3) (1-P^n) (1-P^{2 n})}\;, \nonumber \\[0.1cm]
& M_{\Delta (6 n^2)}(P) &=&
\left\{   \begin{array}{ll} 
         \frac{1+P^{3n+3}}{(1-P^6) (1-P^n) (1-P^{2 n})}  &  n \text{ even}\;,   \\[0.1cm]
         \frac{1+P^{n+3}+P^{3n}+P^{4n+3}}{(1-P^6) (1-P^{2 n}) (1-P^{2 n})} &  n \text{ odd}\;,
          \end{array} \right.
          \label{eq:Mol_Del36}
\end{alignat}
which verified proposition \eqref{eq:proposition}.

Step 2: By computing a few invariants for $n$ without problems of degeneracies, c.f.\ App.~\ref{app:degeneracies}, primary and secondary invariants 
have been obtained as shown in  Tab.~\ref{tab:Delta(3/6n2)}.

The algebraic independence of the primary invariants can be verified using the Jacobian criterion~\cite{Jacobi}.
For, e.g., the $\Delta(3n^2)$ primary invariants we get:
\begin{equation}
 {\rm det} \begin{pmatrix}
 \partial_x \I_3 & \partial_x \I_n & \partial_x \I_{2n}\\
 \partial_y \I_3 & \partial_y \I_n & \partial_y \I_{2n}\\
 \partial_z \I_3 & \partial_z \I_n & \partial_z \I_{2n}
 \end{pmatrix}  = 2 n^2 (x^n - y^n) (y^n - z^n) (z^n - x^n) \neq 0\;, 
 \label{eq:Jac_Del3}
\end{equation}
for general  $x,y,z$.

Step 3: The syzygies are given in that table as well. This completes the analysis and proves 
that~\eqref{eq:Mol_Del36} are Molien functions in the form of Eq.~\eqref{eq:Molien_ex} with the interpretation of primary and secondary invariants.

Let us end this section by pointing out that it is rather remarkable that the invariants 
of $\Delta(3n^2)$ and $\Delta(6n^2)$ are expressible in such a simple manner 
for general $n$. In particular the length of the syzygies does not depend on $n$ 
and this is the reason why we were able to compute all the data. 

In fact, 
$\Delta(3n^2)$ and $\Delta(6n^2)_{n \in 2\N}$ can be seen as the extensions of
the $SO(3)$ subgroups $A_4$ and $S_4$, where the Euclidian distance~\eqref{eq:P2}
is generalized from a power of $2$ to $n$:
\begin{alignat}{2}
\label{eq:c1}
& A_4;S_4:     &\to&    \qquad \Delta(3n^2);\Delta(6n^2)_{n \in 2\N}: \nonumber \\
&   x^2 + y^2 + z^2 \qquad   &\to& \qquad   x^n + y^n + z^n\;.
\end{alignat}
It is like the $\Delta(3n^2),\Delta(6n^2)_{n \in 2\N}$ are the $A_4,S_4$ of a space where distances are measured with $x^n + y^n + z^n$.  The series $\Delta(6n^2)_{n \in 2\N+1}$
can be seen as the following extension:
\begin{alignat}{2}
\label{eq:c2}
& S_3:     &\to&    \qquad \Delta(6n^2)_{n \in 2\N+1}: \nonumber \\
&   x y + y z + z x \ \qquad   &\to& \qquad   x^n y^n + y^n z^n  + z^n x^n \;,
\end{alignat}
since $S_3 \simeq \Delta(6 \cdot 1^2)$. Note that $S_3$ is not a proper finite $SU(3)$ subgroup since it does not contain a faithful $3$-dimensional irrep which, already, follows 
from the dimensionality theorem (c.f.\ App.~\ref{app:facts}), $|S_3| = 6  < 1^2+3^2$.
\begin{table}
\begin{center}
\begin{tabular}{|l|l|l|}\hline
Group & Type & Invariants \\ \hline\hline
$\Delta(3n^2)$ & primary & $\I_3 = x y z$,\\
                        &               & $\I_n = x^n + y^n + z^n$ \\
                        &               & $\I_{2n} = x^{2 n}  + y^{2 n}  + z^{2 n} $ \\
                          & secondary  & $\I_{3n} = x^{3 n}  + y^{3 n}  + z^{3 n}$ \\ \hline
                          & syzygy & $\Is_{3n}^2  =  9 \I_3^{2 n} + 9 \I_3^n \I_n \I_{2 n} + \frac{9}{4} \I_n^2 \I_{2 n}^2 - 3 \I_3^n \I_n^3 - \frac{3}{2} \I_n^4 \I_{2n} + \frac{1}{4} \I_n^6$ \\
     \hline \hline       
     $\Delta(6n^2)$   & primary & $\I_6 =(x y z)^2$,\\
        even $n$                &               & $\I_n = x^n + y^n + z^n$ \\
                        &               & $\I_{2n} = x^{2 n}  + y^{2 n}  + z^{2 n} $ \\
                          & secondary  & $\Is_{3n+3} = x y z (x^n - y^n) (y^n - z^n) (z^n - x^n)$ \\ \hline
                                                   & syzygy & $\Is_{3n+3}^2 = \I_6 \left[ \frac{1}{2} \I_{2 n}^3 - 27 \I_6^n - 9 \I_6^{n/2} \I_n \I_{2 n} + 5 \I_6^{n/2} \I_n^3 + \I_n^4 \I_{2 n} - \frac{1}{4} \I_n^6 - \frac{5}{4} \I_n^2 \I_{2 n}^2 \right]$ \\
     \hline \hline 
        $\Delta(6n^2)$   & primary & $\I_6 =(x y z)^2$,\\
        odd $n$                &               & $\I_{2n}' = x^n y^n + y^n z^n + z^n x^n$ \\
                        &               & $\I_{2n} = x^{2 n}  + y^{2 n}  + z^{2 n} $ \\
                     & secondary  & $\Is_{n+3} =  x y z (x^n + y^n + z^n)$ \\
             &              & $\Is_{3n} = (x^n - y^n) (y^n - z^n) (z^n - x^n)$ \\
              &              & $\Is_{4n+3}= x y z \left[ (x^{3n} y^n - y^{3n} x^n) + (y^{3n} z^n - z^{3n} y^n) + (z^{3n} x^n - x^{3n} z^n) \right]$ \\ \hline
               & syzygies &  $\Is_{n+3}^2 =  \I_6 \left( \I_{2 n} + 2 \I'_{2n} \right)$ \\
                &  &         $    \Is_{3n}^2 =  \I_{2 n} (\I'_{2 n})^2 - 2 (\I'_{2 n})^3 - 4 \I_6^{(n-1)/2} \Is_{n+3} \I_{2 n} + 10 \I_6^{(n-1)/2} 
\Is_{n+3} \I'_{2 n} - 27 \I_6^n$ \\
             &  &         $    \Is_{4n+3}^2 = \I_6 \Is_{3 n}^2 \left( \I_{2 n} + 2 \I'_{2 n} \right)$ \\
     \hline         
\end{tabular}
\end{center}
\caption{\small Primary and secondary invariants of $\Delta(3n^2)$ and $\Delta(6n^2)$ and 
the corresponding syzygies. They have been guessed by explicit calculation of the first few cases and then proven to be correct a posteriori by using the generators from Eq.~\eqref{eq:Gen_Del3}. Let us stress once more that the choice of primary invariants is in certain cases like
the choice of a basis. Note that, for even $n$ in $\Delta(6n^2)$, another choice for $\I_{2n}$
is $ \I'_{2n} = x^{n} y^{n} + y^{n} z^{n} + z^{n} x^{n}$, which is symmetric as well.
It is readily verified that $2 \I'_{2n} =  \I_n^2 - \I_{2n}$. }
\label{tab:Delta(3/6n2)}
\end{table}

%%%%%%%%%%%%%%%%%%%%%%%%%%%%%%%%%%%%%%%%%%%%%%%%%%%%%%%%%%%%%%%%%%%%%%%%%%%
\section{Breaking of $\boldsymbol{SU(3) \to \Sigma(X), \Delta(6n^2),  \Delta(3n^2), T_{n[a]} }$ }
%%%%%%%%%%%%%%%%%%%%%%%%%%%%%%%%%%%%%%%%%%%%%%%%%%%%%%%%%%%%%%%%%%%%%%%%%%%
\label{sec:breaking}

In this section we provide example solutions to the problem, discussed in general 
terms in Sec.~\ref{sec:sufficient}, of selecting the invariants for a specific group $H$ that break $SU(3) \to H$.
In Tab.~\ref{tab:group_tree} we list the groups in our database with subgroup references 
from where the subgroup tree can be derived, as in Fig.~\ref{fig:su3tree}. 
In the database only  $\{S_4, A_5\}$ and $\{ \Sigma(168),\Sigma(216\phi),\Sigma(360\phi)\}$ are maximal 
subgroups of $SO(3)$ and $SU(3)$, respectively. For all other cases there exists a group 
$H'$ such that $H \subsetneq H' \subsetneq SU(3)$. The task then becomes to show that 
the invariants selected for $H$ are not invariants of any $H'$ as well. 
The group $H'$ could be either of finite, continuous, or of mixed type.\footnote{Note that discrete subgroups of $SU(3)$ which are not finite, such as $SU(3)$ elements with rational entries, are of no interest here since they would never leave invariant the kind of polynomials we are considering.} In Sec.~\ref{sec:maxSU(3)} potential groups $H'$ of continuous-type are discussed. In Secs.~\ref{sec:crystal_break} and~\ref{sec:break_noncrystallographics}  we give examples of sufficient invariants for $T_{n[a]}$, $\Delta(3n^2)$, $\Delta(6n^2)$, and all $\Sigma(X)$. 
The subtle question of why it is legitimate to work with the explicit generators,
as given in Tab.~\ref{tab:SU3_SubGen}, is discussed in Sec.~\ref{sec:3irreps}.
We have not attempted to find sufficient conditions for generic   
$C$- and $D$-groups. Possibly  more work is needed on the structure of these groups.\footnote{Some effort has been undertaken recently in Refs.~\cite{Ludl_diplom,DMFV} 
and especially in Ref.~\cite{Ludl:2011gn}.}

\subsection{Continuous subgroups of  $SU(3)$ }
\label{sec:maxSU(3)}

The continuous subgroups of $SU(3)$ are $SO(3)$ and $U(2) = SU(2) \times U(1)$, 
and subgroups thereof.  
We observe that  all groups in our 
list contain the generator $E$ (cf.\ Tab.~\ref{tab:SU3_SubGen} or  App.~\ref{app:GenGroup}), which corresponds to a cyclic permutation of the three variables
$\{x,y,z\}$. Let us first discuss the group $SO(3)$. The finite subgroups of $SO(3)$ are
the well-known $A_4$, $S_4$, and $A_5$ for which we have all the data, and the dihedral groups
which are not invariant under a cyclic permutation  since 
they correspond to the symmetry of a molecule with one distinguished axis. 
The subgroup $U(2)$  is not invariant under cyclic permutations as the embedding $U(2) =  SU(2) \times U(1) \emb SU(3)$,\footnote{The group $SU(2)$ can be embedded in such a way that $\R{3}_{SU(3)} \to \R{3}_{SU(2)}$, but then it is the same 
as $\R{3}_{SO(3)}$ which we have already discussed.} denoted by the symbol $\emb$, singles out a direction and is therefore eliminated for the same reason.
For groups of mixed type, only $U(1) \times U(1) \rtimes \Z_3;S_3$
are known~\cite{Milleretal,FFK}, and they can be understood as the formal limits $n \to \infty$ 
 of $\Delta(3n^2);\Delta(6n^2)$. The latter are implicitly included in our discussion through 
 the $\Delta$-groups.
In summary the cyclic symmetry of our groups forbids any groups of 
continuous or mixed type.\footnote{Note that this line of reasoning is general and much simpler 
than algebraic methods, which have to be applied case by case, see Ref.~\cite{Luhn:2011ip}.
The cyclicity is evident in the language of invariant polynomials as opposed to the language of VEVs.}

\subsection{Breaking to crystallographic groups $\Sigma(X)$}
\label{sec:crystal_break}

\begin{figure}[h]
  \centering
  \includegraphics[width=5.0in]{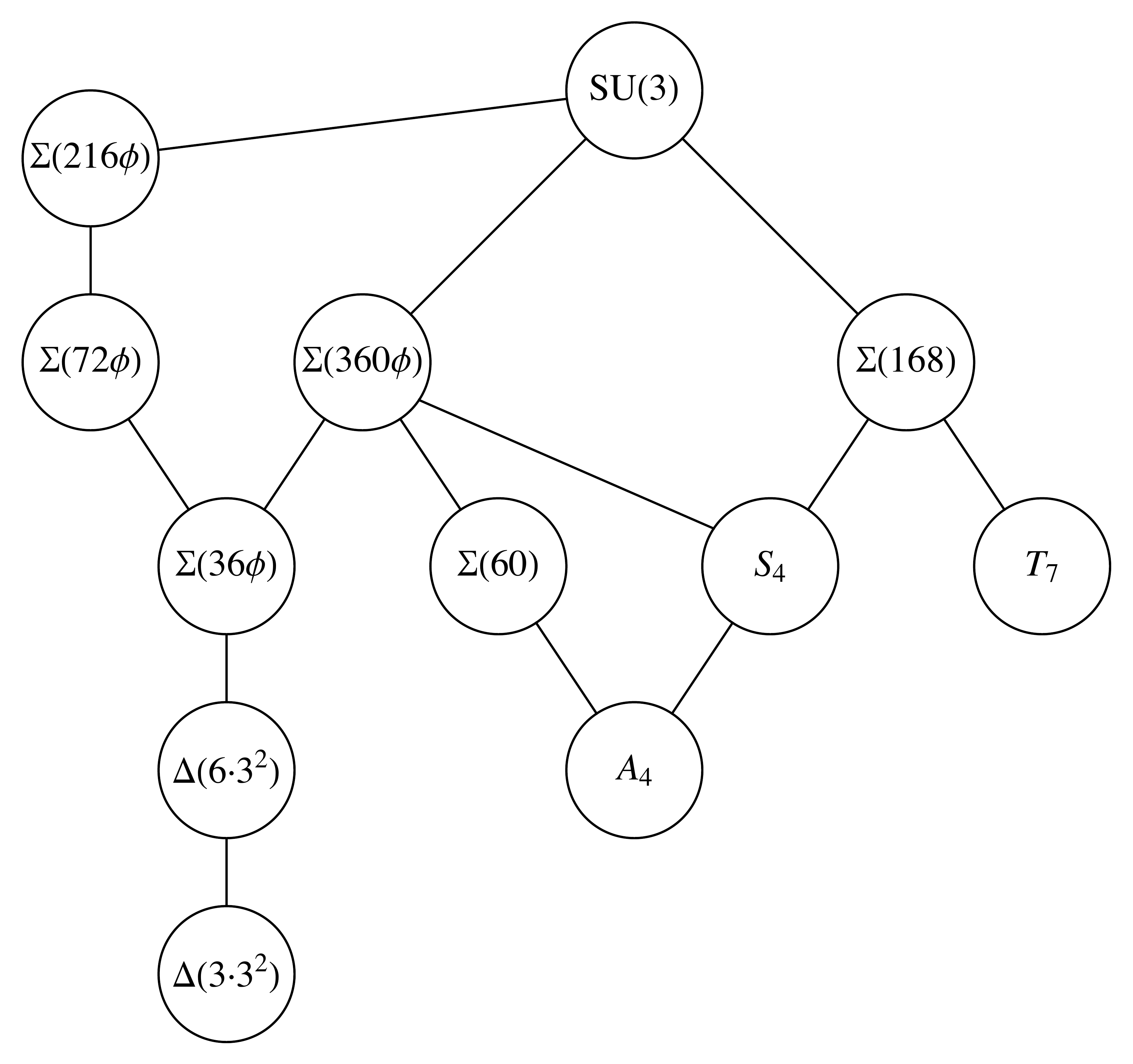} 
 \caption{\small The subgroup tree of the crystallographic groups. Note this is only a partial tree.  The entire tree, within our database, is shown in Fig.~\ref{fig:su3tree}.}
\label{fig:minitree}
\end{figure}

The partial  subgroup tree in the crystallographic sector is shown in Fig.~\ref{fig:minitree},
from where we infer that  $\Sigma(216\phi)$, $\Sigma(360\phi)$, and $\Sigma(168)$ are maximal 
subgroups of $SU(3)$. This can be seen as follows: First they are maximal in the chain 
of crystallographic groups. Second, they cannot be subgroups of the  $C$- and $D$-type
groups since the latter contain irreps of dimensions not higher than six~\cite{DMFV}, 
whereas $\Sigma(216\phi)$, $\Sigma(360\phi)$, and $\Sigma(168)$ all contain irreps 
of dimensions larger than six. 

Using our Mathematica package \name, the common invariants of the subgroups can be 
identified, and thus breaking from $SU(3)$ into these subgroups can be worked out.
Let us list the Molien functions
and the lowest invariants that break $SU(3) \to \Sigma(X)$:\footnote{A subtle point is that 
the subgroup relation $\Sigma(36\phi) \subset \Sigma(360\phi)$ is not apparent from 
its generators. Thus one has to be cautious when comparing invariants. In order to verify that the invariant of degree six proposed for the breaking 
$SU(3) \to \Sigma(36\phi)$ is correct, one has to use the basis transformation given 
in App.~\ref{app:group_tree}. We have verified that, in that basis, the invariant discussed is not left invariant by  $\Sigma(360\phi)$.}

\begin{tabular}{|l|l|l|}
\hline
Group  &  Molien function  & Invariant of lowest degree that breaks $SU(3) \to \Sigma(X)$ \\[0.1cm] \hline
$\Sigma(60)$ & $\frac{1+ P^{15}}{(1-P^{2})(1-P^{6})(1-P^{10})}$ & 
$ (\phi_0^2 x^2 -y^2 ) (\phi_0^2  z^2-x^2)(\phi_0^2 y^2-z^2)$ \\[0.1cm]
$\Sigma(36\phi)$ & \mbox{$\frac{1+ P^{9}+  P^{12}+ P^{21}}{(1-P^{6})^2(1-P^{12})}$} &  
$(x^6 + 2 x^3 y^3  -6 x^4 y z + \text{cy.}) -18 x^2 y^2 z^2$ \\[0.1cm]
$\Sigma(168)$ & $\frac{1+ P^{21}}{(1-P^{4})(1-P^{6})(1-P^{14})}$ & $x^3 z + z^3 y + y^3 x$ \\[0.1cm]
$\Sigma(72\phi)$ & $\frac{1+ P^{12}+P^{24}}{(1-P^{6})(1-P^{9})(1-P^{12})}$  & 
$x^6 + y^6 + z^6 - 10 x^3 y^3 - 10 y^3 z^3 - 10 z^3 x^3$  \\[0.1cm]
$\Sigma(216\phi)$ & $\frac{1+ P^{18}+P^{36}}{(1-P^{9})(1-P^{12})(1-P^{18})}$ &  
$x^6 (y^3-z^3) + y^6 (z^3-x^3) + z^6 (x^3-y^3)$ \\[0.1cm]
$\Sigma(360\phi)$ & $\frac{1+ P^{45}}{(1-P^{6})(1-P^{12})(1-P^{30})}$ &  
$x^6 + y^6+z^6 + a x^2 y^2 z^2 + b_+  \left(x^4 y^2+ \text{cy.}  \right) + b_- \left(x^4 z^2 +  \text{cy.} \right) $ \\ \hline
\end{tabular}\\

In the table, we have used:
\begin{equation}
\phi_0 \equiv \frac{1+\sqrt{5}}{2} \;, \; a=3 \left(5-i \sqrt{15}\right)  \;, \; b_\pm = \frac{3}{8} \left[5 \mp 3\sqrt{5} + i \left( \sqrt{15} \pm 5 \sqrt{3} \right)\right]\;,
\end{equation}
where ``cy.'' stands for cyclic permutations in the variables $x$, $y$, and $z$.
Let us add that the Molien function for $\Sigma(216\phi)$ differs from 
the one in~\cite{Milleretal}, but it is the same as in~\cite{analytic},
where the ones for $\Sigma(360\phi)$ and $\Sigma(168)$ were also presented. The reader should be able to find invariants of higher degrees that achieve the same.
A subtle point to be stressed is that not all subgroups relations are apparent
from the generators as given in Tab.~\ref{tab:SU3_SubGen}. Thus, when comparing invariants or checking their invariance with respect to supergroups, one has to account for this fact by similarity transformations, c.f.\ App.~\ref{app:group_tree},
as we did for the case $\Sigma(36\phi) \subset \Sigma(360\phi)$ as described in an earlier
footnote in this section.

%%%%%%%%%%%%%%%%%%%%%%%%%%%%%%%%%%%%
\subsection{Breaking to $C$- and $D$-groups ($\Delta(6n^2)$, $\Delta(3n^2)$, and $T_{n[a]}$)}
\label{sec:break_noncrystallographics}
%%%%%%%%%%%%%%%%%%%%%%%%%%%%%%%%%%%%

\begin{figure}[h]
  \centering
  \includegraphics[angle=0,width=4.0in]{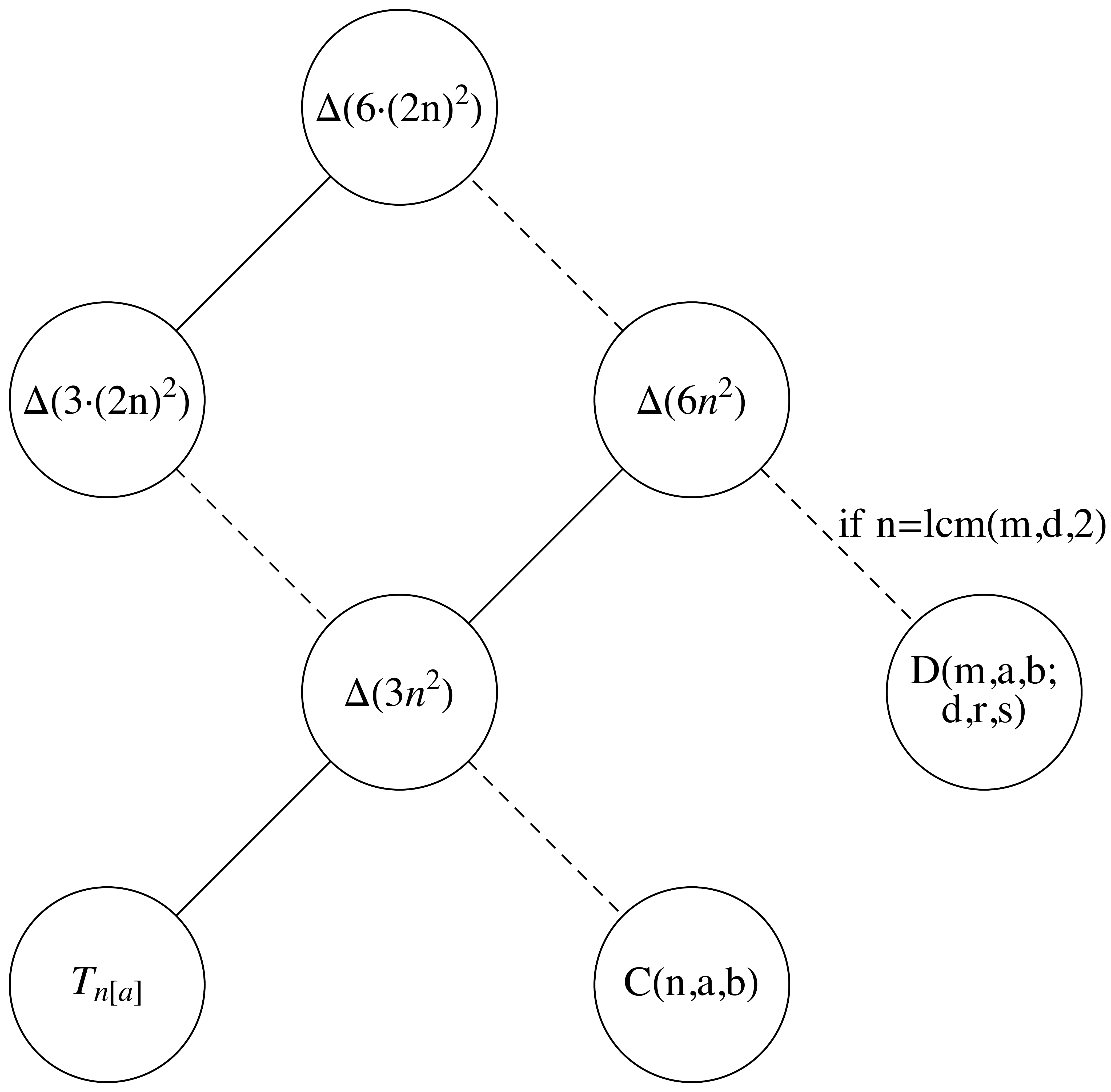} 
   \caption{\small Tree of subgroups in the $C$- and $D$-sector as given 
   in Eq.~\eqref{eq:CDsubgroups}. The dashed lines allude 
 to the fact that there could be other subgroups in between, whereas the solid lines 
 are maximal subgroup relations.}
\label{fig:CDtree}
\end{figure}

\begin{figure}[h]
  \centering
  \hspace{-2.5cm}
  \includegraphics[angle=0,width=16cm]{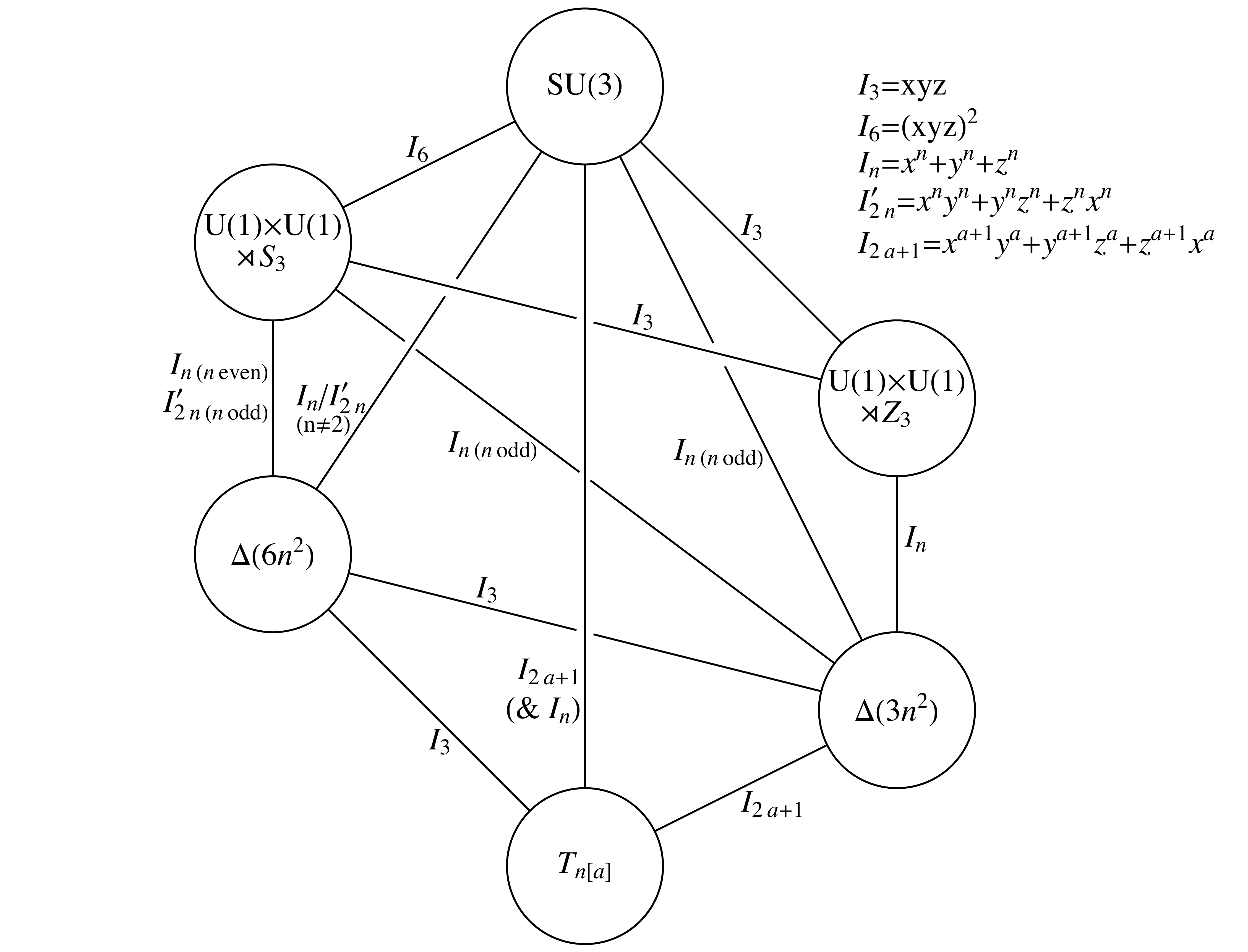} 
   \caption{\small Summary of of breaking patterns for the $T_{n[a]}$-, $\Delta(3n^2)$-, and 
   $\Delta(6n^2)$-groups. The groups $U(1) \times U(1) \rtimes \Z_3$ [$S_3$] are understood to be the formal limits of $n \to \infty$ of $\Delta(3n^2)$ [$\Delta(6n^2)$].}
\label{fig:Deltatree}
\end{figure}

Before discussing the groups $\Delta(3n^2)$ and $\Delta(6n^2)$ in more detail let
us discuss some generalities about subgroup structures,
\begin{eqnarray}
\label{eq:CDsubgroups}
\Delta(3n^2)   &\stackrel{m}{\subset}& \Delta(6n^2)      \;, \nonumber \\[0.1cm]
\Delta(3n^2)   &\subset& \Delta(3 (2n)^2) \;, \nonumber \\[0.1cm]
\Delta(6n^2)   &\subset& \Delta(6 (2n)^2) \;, \nonumber \\[0.1cm]
C(n,a,b) &\subseteq& \Delta(3n^2)  \;,  \qquad \text{e.g. } T_{n[a]} \stackrel{m}{\subset} \Delta(3n^2)\;, \nonumber \\[0.1cm]
D(m,a,b;d,r,s) &\subseteq& \Delta(6 n^2) \;, \qquad  n={\rm lcm}(m,d,2)  \;,
\end{eqnarray}
depicted in Fig.~\ref{fig:CDtree}. The acronym ``lcm'' stands for lowest common multiple and the symbol $\stackrel{m}{\subset}$ for maximal subgroup. In the cases at hand this follows by virtue of Lagrange's theorem, c.f.\ App.~\ref{app:facts}.
The first three statements are obvious from the generators. The fourth one comes about 
by realizing that any $C(n,a,b)$ corresponds to a $\Delta(3n^2)$~\cite{Ludl:2011gn}. Crucially some of those
representations are not faithful so that $C(n,a,b)$, depending on $a$ and $b$, can be a proper subgroup of $\Delta(3n^2)$. For the fifth statement we refer the reader to~\cite{DMFV}. The groups $\Delta(3n^2)$, $\Delta(6n^2)$, and $T_{n[a]}$ are discussed case by case, and illustrated in Fig.~\ref{fig:Deltatree}.

%%%%%%%%%%%%%%%%%%%%%%%%%%%%%%%%%%%%%%%%%%%%%%%%%%%%%%%%%%%%%%%%%%%%%%%%%%%%%%%
\subsubsection*{\label{sec:break_Del6}Breaking to $\boldsymbol{\Delta(6 n^2)}$ }
%%%%%%%%%%%%%%%%%%%%%%%%%%%%%%%%%%%%%%%%%%%%%%%%%%%%%%%%%%%%%%%%%%%%%%%%%%%%%%%

We propose that $SU(3) \to \Delta(6n^2)$ by imposing 
\begin{alignat}{3}
\label{eq:breakD6}
& \I_{2n}'[\Delta(6n^2)]\; &=& \;  x^n y^n + y^n z^n + z^n x^n\;, \qquad  & & n \text{ odd},  \nonumber   \\[0.1cm]
& \I_n[\Delta(6n^2)]\; &=&  \;x^n + y^n + z^n\;, \qquad  & & n \neq 2   \text{ and even}.
\end{alignat}
Note in the case where $n=2$, which corresponds to $S_4 = \Delta(6 \!\cdot \! 2^2)$, $SU(3) \to SO(3)$ and thus a further invariant, say 
$\I_4[S_4]$ or $\I_6[S_4]$~\eqref{eq:IS4}, has to be imposed. 
We have checked that none of the crystallographic generators in Tab.~\ref{tab:SU3_SubGen} leaves either $ \I_n$ or $\I_{2n}'$ invariant. 
It remains to show that the action of $F(m,a,b)$ and $G(d,r,s)$ for generic parameters 
$\{m,a,b,d,r,s\}$ together with the constraint of~\eqref{eq:breakD6} being invariant implies
that they are contained within $\Delta(6n^2)$.
\begin{itemize}
\item \emph{$F(m,a,b)$}:
Let us assume that  $F(m,a,b)$ exists which leaves $\I_n$~\eqref{eq:breakD6} invariant.
Then the following ought to be true:
\begin{equation}
 \eta^{a n}=1\ \ {\rm and}\ \ \eta^{b n}=1\;,
 \label{eq:inv_conds_1}
\end{equation}
where $\eta=e^{2\pi i/m}$. 
Writing $\theta \in \{ a, b \}$, it  follows from Eq.~\eqref{eq:inv_conds_1} that
\begin{equation}
 \eta^{\theta n}=e^{2 \pi i \theta n /m} \Leftrightarrow \frac{\theta n}{m} \equiv k_\theta \in \mathbb{N}_0\;.
  \label{eq:inv_conds_2}
\end{equation}
This allows us to rewrite the initial generator as
\begin{eqnarray}
  F(m,a,b) &=&  {\rm diag}\left( e^{2 \pi i a/m}, e^{2 \pi i b/m}, e^{-2 \pi i (a+b)/m} \right)\nonumber \\
 & \stackrel{\eqref{eq:inv_conds_2}}{=}& {\rm diag}\left( e^{2 \pi i k_a/n}, e^{2 \pi i k_b/n}, e^{-2 \pi i (k_a+k_b)/n} \right) = F(n, k_a, k_b)\;.
 \label{eq:inv_conds_3}
\end{eqnarray}
Thus we have traded the $m$ for $n$ by $(a,b) \to (k_a,k_b)$. There is no special need to be specific
about the latter two as $(k_a, k_b)= (0,1)$, and a few equivalences, generate $\Delta(3n^2)$ and second 
the other choices lead to smaller groups as stated in Eq.~\eqref{eq:CDsubgroups}.
Since the breaking, however, will always lead to the \emph{largest} group to which one could possibly break, this observation completes the argument.

\item \emph{$G(m,a,b)$}:
The investigation of $G(d,r,s)$ calls for a distinction of odd and even $n$:
\begin{itemize}
\item \emph{$n$ even:} The action of $G(d,r,s)$ leaves $\I_n$ invariant if and only if
$(rn/d,sn/d) \in \Z^2$, where we have used that for even $n$ the phase factor, $(-1)^n=1$, is unity.
Thus we may write $(r/d,s/d) = (R/n,S/n)$ with $(R,S) \in \Z^2$, and 
therefore $G(d,r,s) \to G(n,R,S)$.
\item \emph{$n$ odd:} The very same action on $\I_{2n}'$ lead to the conclusion 
that $(rn/d,sn/d) \in (2\Z+1)^2$, which by the same argumentation leads to
$(r/d,s/d) = (R/n,S/n)$ with $(R,S) \in \Z^2$ and 
therefore $G(d,r,s) \to G(n,R,S)$ as above.
\end{itemize}
Making the observation that $F(n,a,b) G(2,1,1) = G(n,-a,-b)$ we can infer that 
$G(n,R,S) \in \Delta(6n^2)$, since the latter is generated by 
$\{ E,F(n,0,1), G(2,1,1)\}$. 
In order to appreciate the last step it should be added that $\{ E,F(n,a,b), G(2,1,1)\}$ can only be a subgroup 
of $\Delta(6n^2)$.
 \end{itemize}

%%%%%%%%%%%%%%%%%%%%%%%%%%%%%%%%%%%%%%%%%%%%%%%%%%%%%%%%%%%%%%%%%%%%%%%%%%%%%%%
\subsubsection*{\label{sec:break_Del3}Breaking to $\boldsymbol{\Delta(3 n^2)}$ }
%%%%%%%%%%%%%%%%%%%%%%%%%%%%%%%%%%%%%%%%%%%%%%%%%%%%%%%%%%%%%%%%%%%%%%%%%%%%%%%

We propose that $SU(3) \to \Delta(3n^2)$ by imposing 
\begin{alignat}{3}
\label{eq:breakD3}
& \I_{2n}'[\Delta(3n^2)]\; &=& \;  \;x^n + y^n + z^n\;, \qquad  & & n \text{ odd,}   \nonumber   \\[0.1cm]
 & \I_n[\Delta(3n^2)]\; &=&  \;x^n + y^n + z^n\;, \qquad \I_3[\Delta(3n^2)] = x y z\;,  \qquad & & n \text{ even.}  
\end{alignat}
The results follow, rather directly, from the analysis of $\Delta(6n^2)$ 
in the previous subsection. We will not repeat all arguments in detail.
\begin{itemize}
\item \emph{$n$ odd}: $\I_n$ is not a $\Delta(6n^2)$ invariant because 
of the generator $G(2,1,1)$. Moreover, specifically a generic $G(d,r,s)$ does not leave 
$\I_n$ invariant for odd $n$ because of the minus sign in $(-\delta^{-r-s})$.
Thus $\I_n$ breaks $SU(3)$ to $\Delta(3n^2)$ for odd 
$n$.\footnote{The case 
in~\cite{Luhn:2011ip} for $\Delta(27)$, $n =3$, 
can be seen as a special case of our finding. The VEV found in that reference 
ought to translate into $\I_n$.}
\item \emph{$n$ even}: Since $\Delta(3n^2)$ is a maximal subgroup of 
$\Delta(6n^2)$ it suffices to find one invariant, e.g.\ $\I_3[\Delta(3n^2)]$, 
of $\Delta(3n^2)$ in order to break from $\Delta(6n^2)$ to $\Delta(3n^2)$. 
Imposing the two invariants from Eq.~\eqref{eq:breakD3} can be seen as a sequential 
breaking $SU(3) \stackrel{\I_n}{\to} \Delta(6n^2) \stackrel{\I_3}{\to} \Delta(3n^2)$.
\end{itemize}

%%%%%%%%%%%%%%%%%%%%%%%%%%%%%%%%%%%%%%%%%%%%%%%%%%%%%%%%%%%%%%%%%%%%%%%%%%%%%%%
\subsubsection*{\label{sec:break_Tn}Breaking to $\boldsymbol{T_{n[a]} }$}
%%%%%%%%%%%%%%%%%%%%%%%%%%%%%%%%%%%%%%%%%%%%%%%%%%%%%%%%%%%%%%%%%%%%%%%%%%%%%%%

We propose that $SU(3) \to T_{n[a]}$ for\footnote{The fact that a $(3,0)=\R{10}$  was found 
to break $SU(3) \to T_{7[2]}$~\cite{Luhn:2011ip} can be seen as a special case of the analysis. The VEV found in that reference ought to translate into the invariant $\I_{2a+1}$.}
\begin{alignat}{3}
\label{eq:breakTna}
& \I_{2a+1}[T_{n[a]}] \; &=&  \ x^{a+1} y^{a} + y^{a+1} z^{a} + z^{a+1} x^{a}\;, \qquad 
& & a^2 + a + 1 = 1 \cdot n\;,   \nonumber \\
& \I_{2a+1}[T_{n[a]}] \; &\;,&  \;\;\; \I_n[T_{n[a]}] = x^n + y^n + z^n\;,   \qquad 
& & a^2 + a + 1 = m \cdot n \;, \quad  m \in \N + 1  \;.
\end{alignat}
Let us discuss the first case first.
The generators of crystallographic type listed in Tab.~\ref{tab:SU3_SubGen} 
do not leave $ \I_{2a+1}[T_{n[a]}]$ invariant. Idem for the generator $G(d,r,s)$ as it exchanges $y$ and $z$ but not $x$. Second, considering a generator 
$F(\eta,\alpha,\beta)$, we get three equations which add up to zero. So we effectively have two conditions:
\begin{eqnarray}
\label{eq:sowas}
(a+1)\alpha + a \beta = 0 \mod \eta \;, \quad (a+1)\beta + a (-\alpha-\beta) = 0 \mod \eta \;.
\end{eqnarray}
Considering $\alpha = \alpha(a)$ and $\beta = \beta(a)$, and differentiating both equations 
with respect to $a$ we get a set of first order coupled homogeneous differential equations whose
solution is unique and given by $(\alpha(a),\beta(a)) = (1,a)$. Reinserting this solution into~\eqref{eq:sowas} we get $a^2 + a +1 = 0\mod \eta$ and, 
using the condition $a^2 + a + 1 = n$, we get $\eta = n$ 
if $F(\eta,1,a)$ is not to be a subgroup of $F(n,1,a)$. This completes the argument.

In the second case we have $a^2 + a + 1 = mn$ and we cannot conclude $n = \eta$.
Imposing $ \I_{2a+1}[T_{n[a]}]$ alone in this case will break $SU(3) \to 
C(n\cdot m ,1,a) \supset T_{n[a]}$.\footnote{In fact, for $T_{91[16]}$ and $T_{133[30]}$, $m$ is $3$ and $7$, respectively, 
and thus $C(91\cdot 3,1,16) \simeq T_{273[16]}$ and 
$C(133\cdot 7,1,30) \simeq T_{931[30]}$ are indeed of the $T_{n[a]}$-series.} This can be remedied by imposing the additional invariant $\I_n[T_{n[a]}]$ as proposed above.
 
 With respect to the classification of the $T_{n[a]}$-series we note that that neither 
 $n$ determines $a$ nor does $a$ determine $n$. Thus the double label seems appropriate.

%%%%%%%%%%%%%%%%%%%%%

 \subsection{Examples}
 \label{sec:examples}
 A few explicit examples can be found below within the basis quoted at the end of this section. All these examples can also be found in the example notebook of \name.
\begin{itemize}
 \item From $(4,0)_{SU(3)} \to \Sigma(168)$:  $(4,0) \simeq \R{15}'$
 \begin{eqnarray}
 \I_4[\Sigma(168)]  &=& x^3 z + z^3 y + y^3 x =  (-\sqrt{6})v[\Sigma(168)]_{4,0} \cdot 
 {\cal B}_{(4,0)}\;, \nonumber  \\
  v[\Sigma(168)]_{4,0} &=& (0,0,0,1,0,1,0,0,0,0,0,0,-1,0,0)  \;.
 \end{eqnarray}
 \item From $(4,0)_{SU(3)} \to \Delta(96)$:  
 \begin{eqnarray}
 \I_4[\Delta(96)] &=&  x^4 + y^4 + z^4   =   2 \sqrt{6} v[\Delta(96)]_{4,0} \cdot 
 {\cal B}_{(4,0)}\;, \nonumber  \\
  v[\Delta(96)]_{4,0} &=& (1,0,0,1,0,0,0,0,0,0,0,0,0,0,1)  \;.
 \end{eqnarray}
  \item From $(1,1)_{SU(3)} \to S_3$:  $(1,1) \simeq \R{8}$ \\
  The irreps of $S_3$ 
 are $\{\R{1},\R{1}',\R{2}\}$ and generators $\{ E, G(2,1,1) \}$, Eq.~\eqref{eq:Gen_Del3}. 
Note that we have used the fact that $S_3 = \Delta(6\cdot 1^2)$. 
From the generators we can infer that $\R{3}_{SU(3)} \to \R{1}'+ \R{2}$.
The branching rule  is computed using the methods of 
App.~\ref{app:generating}:
 \begin{equation}
 (1,1)_{SU(3)}|_{S_3} \to (\R{1} + \R{1}'  +  3 \cdot \R{2} )_{S_3} \;.
 \end{equation}
 Thus there is one single invariant in that representation. 
 The invariant is easily guessed,\footnote{As particle physicists we might want to replace $(x,y,z) \to (u,d,s)$ and think in terms of meson states organized by the $SU(3)$-flavour symmetry of the eightfold way.}
 \begin{eqnarray}
 \I[S_3]_{1,1} &=&  x  y^* + y  x^* + z^* y + y^* z + x^* y + x  z^*  = v[S_3]_{1,1} \cdot 
 {\cal B}_{(1,1)}\;, \nonumber  \\
  v[S_3]_{1,1} &=& (1,-1,0,1,0,-1,-1,-1)  \;,
 \end{eqnarray}
 as it corresponds to the Weyl-symmetry of the root diagram, which is $S_n$ for $SU(n)$.
 
 \item Example dialogue in the Mathematica package \name:  
\begin{center}
\begin{tabular}{rl}\\
{\scriptsize In[1]:=} & {\tt \hspace{-0.5cm} SetDirectory[\textrm{''}...(your directory).../SUtree{\bf \_}v1p0/\textrm{''}];}\\
&\\
{\scriptsize In[2]:=} & {\tt \hspace{-0.5cm} \$RecursionLimit=260;}\\
&{\tt \hspace{-0.5cm} <<SUtree.m}\\
&\\
{\scriptsize In[3]:= } & {\tt \hspace{-0.5cm} BranchingSU3[\{3,0\}, \textrm{''}A$_{\tt 4}$\textrm{''}];}\\
{\scriptsize Out[3]= } & {\tt \hspace{-0.5cm} \{\{3, 0\}, 10, \{1, 1\}, \{3, 1\}, \{3, 1\}, \{3, 1\}\}}
\end{tabular}
\end{center}
1.~The directory has to be set to the path where the package and its data directory reside. 2.~The package is loaded  via ``{\tt <<SUtree.m}" and the recursion limit is enlarged. 
 3.~The branching rule for $(3,0)_{SU(3)} \to (\R{1}_1 + 3 \cdot  \R{3}_1)_{A_4}$ is  obtained.
 More details about the output can be learned by typing ``{\tt ?BranchingSU3}" into 
 the Mathematica dialogue. Here we shall just add that the second entry in the list corresponds 
 to the dimension of the irrep $(3,0)$ and that ${\tt \{3,1\}}$ corresponds to the first 3-dimensional 
 irrep in the character table.
 \end{itemize}

The explicit bases used above are derived from~\eqref{eq:SU(3)_GF}.
The ordering is such that $rst$ is interpreted as a number with constraints~\eqref{eq:rst}, e.g.\ $(001,002,...010,011, ...100, ...)$. The bases are given by:
 \begin{eqnarray}
{\cal B}_{(1,1)} &=&  \left\{x z^*,-y z^*,\frac{x x^*+y y^*-2 z z^*}{\sqrt{6}},x y^*,\frac{x x^*-y y^*}{\sqrt{2}},-x^* y,-y^* z,-x^* z\right\}, \nonumber \\[0.1cm]
{\cal B}_{(4,0)}  &=&  \left\{\frac{x^4}{2 \sqrt{6}},-\frac{x^3 y}{\sqrt{6}},\frac{x^2 y^2}{2},-\frac{x y^3}{\sqrt{6}},\frac{y^4}{2 \sqrt{6}},-\frac{x^3 z}{\sqrt{6}},\frac{x^2 y z}{\sqrt{2}},-\frac{x y^2 z}{\sqrt{2}},\frac{y^3 z}{\sqrt{6}},\frac{x^2 z^2}{2},\right. \nonumber \\
& &\left. -\frac{x y z^2}{\sqrt{2}},\frac{y^2 z^2}{2},-\frac{x z^3}{\sqrt{6}},\frac{y z^3}{\sqrt{6}},\frac{z^4}{2 \sqrt{6}} \right\}.
 \end{eqnarray}

 %%%%%%%%%%%%%%%%%%%%%%%%%%%%%%%%%%%%%%%%%%%%%%%%%%%%%%%%%%%%%%%%%%%%%%%%%%%%%%%
\section{The question of the embedding $\boldsymbol{\R{3}_{{\cal F}_3} \emb \R{3}_{SU(3)} }$}
%%%%%%%%%%%%%%%%%%%%%%%%%%%%%%%%%%%%%%%%%%%%%%%%%%%%%%%%%%%%%%%%%%%%%%%%%%%%%%%
\label{sec:embedding}

In our analysis we have chosen a particular embedding,
\begin{equation}
\label{eq:3to3}
\R{3}_{{\cal F}_3} \emb \R{3}_{SU(3)}  \; ,
\end{equation}
namely the one given in Tab.~\ref{tab:SU3_SubGen}. 
It is therefore a legitimate question 
whether our results are dependent on  it. 
We shall discuss this issue from the viewpoint of explicit breaking and not 
from the viewpoint of VEVs. Since the two are equivalent this is sufficient.

Generically  an embedding for groups, denoted by $H \emb G$, is an (injective) map from 
$H$ to $G$ that preserves the group structure.
One distinguishes embeddings 
up to similarity transformations \ref{sec:similar} and those who do not
fall into this class \ref{sec:non-similar}. The former case resembles 
the choice of a coordinate system and  the latter corresponds to inequivalent irreps.
In the case where the irrep is of the same dimension as the group it is embedded in, 
as in Eq.~\eqref{eq:3to3}, this corresponds to different irreps in the character table.
In Sec.~\ref{sec:Molien_embedding} we discuss 
the impact of the embedding on the Molien function and on the invariants.  
In Sec.~\ref{sec:3irreps} it is analyzed whether the inequivalent 
$3$-dimensional faithful irreps of the $\Sigma(X)$-, $\Delta$-, and $T_{n[a]}$-groups can be distinguished
with respect to each other.

Before embarking on these topics, we would like to add a few more comments
in connection with larger groups and embedding into larger groups:
\begin{itemize}
\item
In this work we have restrained ourselves to  $3$-dimensional 
(irreducible) representations in view of the three generations of particles in the lepton 
and quark sector of the SM. If there was a fourth generation, which is possible, then 
we would be studying something like:
\begin{equation}
 \R{4}_{A_5}  \emb \R{4}_{SU(4)}     \quad   \text{   instead of   } \quad 
 \R{3}_{A_5}  \emb \R{3}_{SU(3)}  \;,
\end{equation}
for example.  Finite  $SU(4)$ subgroups have been studied in Ref.~\cite{Hanany:1999sp}.
\item
For model building it is interesting to consider $\R{3}_{{\cal F}_3} \emb X$ with 
$|X| > 3$. For example the chain,
\begin{equation}
\label{eq:SU(6)}
 \R{3}_{A_5} + \R{3}'_{A_5} \emb \R{6}_{S_5}   \emb \R{6}_{SU(6)}  \;,
\end{equation}
could very well be part of an interesting model. The embedding theory of this kind
is well developed for Lie groups, where inequivalent emebddings are characterized 
by an embedding index~\cite{conformal} (or~\cite{Ramond} for an alternative discussion).
An example often discussed in books~\cite{Ramond,Cornwell} is $SU(2) \emb SU(3)$,
as quoted in Sec.~\ref{sec:maxSU(3)}.  
Finding all embeddings is equivalent to finding all branching rules. For finite groups no complete theory is known to our knowledge. 

\item
A possibility, frequently used in model building, is to introduce
several fields carrying different irreps of $SU(3)$ or of one of its subgroups.
For this setting the embedding up to similarity transformations does matter.
This phenomenon is known under the name of
\emph{vacuum alignment} and is briefly  outlined in App.~\ref{app:flavons}.
\end{itemize}

\subsection{Embedding up to similarity transformations }
\label{sec:similar}

Given a certain representation ${\cal R}(h)$ of $H$, which we shall denote
for the sake of brevity by $h$ only, the similarity transformation,
\begin{equation}
\label{eq:similar}
h' = A h A^{-1} \;, \qquad \text{ where $A$ is an invertible matrix,}
\end{equation}
provides another  representation of the group. 
Note that $h$ and $h'$ are unitary representations if and only if  
$A$ is a unitary matrix, e.g.~\cite{Cornwell}.

Importantly the transformation~\eqref{eq:similar} does 
\emph{not} correspond to an inequivalent irrep. In the finite case the character and therefore the character table is left invariant. A point we would like to emphasize is that under~\eqref{eq:similar} 
the invariants transform unless $A \in {\cal R}(H)$.

\subsubsection*{Specht's theorem - criteria for unitary equivalence}

It is an important practical question, given a set of matrices 
$h$ and $h'$, of whether they are unitary equivalent, $h' = U h U^\dagger$. 
The criteria are given 
by \emph{Specht's theorem}~\cite{Specht}, which gives sufficient conditions.
For three dimensions they amount to: 
\begin{eqnarray}
& & {\rm tr}[h] = {\rm tr}[h']  \;,\; {\rm tr}[h^2] = {\rm tr}[h'^2]  \;,\;
{\rm tr}[h h^\dagger] = {\rm tr}[h' h'^\dagger] \;,\; {\rm tr}[h^3] = {\rm tr}[h'^3]
 \;,\;  \\[0.1cm] 
& & {\rm tr}[h^2 h^\dagger] = {\rm tr}[h'^2 h'^\dagger]  \;,\; 
  {\rm tr}[h^2 (h^2)^\dagger] = {\rm tr}[h'^2 (h'^2)^\dagger] 
 \;,\;  {\rm tr}[h^2 (h^2)^\dagger h h^\dagger ] = {\rm tr}[h'^2 (h'^2)^\dagger h' h'^\dagger]\;.  \nonumber
  \end{eqnarray}

\subsection{Inequivalent embeddings }
\label{sec:non-similar}

As stated above, for the embedding type~\eqref{eq:3to3}, inequivalent embeddings correspond to different $3$-dimensional irreps of the group.
 An example is given by the two representations $\R{3}_1$ and $\R{3}_2$ of
 $A_5$, see e.g.~\cite{Ludl_diplom}.

\subsection{Molien function and embeddings }
\label{sec:Molien_embedding}

We begin by observing that the Molien functions 
and the invariants of two complex conjugate representations $\R{3}$ and $\Rb{3}$ are related 
to each other as:
\begin{equation}
\label{eq:conjugate}
M_{\Rb{3}}(P)  = M_{\R{3}}(P) \;, \quad \I[\Rb{3}] = \I[\R{3}]^* \;,
\end{equation}
where the symbol $\phantom{}^*$ denotes complex conjugation here and thereafter.
This directly follows from~\eqref{eq:Molien} and~\eqref{eq:genI}.
Note that the Molien function on the right-hand side (RHS) in the equation above is not
complex conjugated for the very reason that it is real by virtue of Molien's theorem.

Let us denote the set of matrices of a representation by $\{\R{3}\}$, sometimes called the \emph{image}, as opposed to $\R{3}$ for just the representation itself.
In the case where two inequivalent representations, say $\R{3}$ and 
$\R{3}'$, have the same  image, 
\begin{equation}
\label{eq:image}
\{\R{3}\} = \{\R{3}'\} \;,  
\end{equation}
the Molien functions and the invariants are identical.\footnote{The fact that two inequivalent representations
have the same representation matrices might be a bit of a surprise at first thought.  
A simple
example is $\Z_3$ which has three irreps, the identity $\R{1}$ and two complex conjugate 
pairs $\R{1}'$ and $\Rb{1}'$ which are generated by 
$A = \exp(2\pi i\cdot 1/3)$ and 
$A^* = \exp(2\pi i \cdot 2/3)$, respectively. Yet there is no inner automorphism that maps one irrep to the other. When embedded into the dihedral group $D_3 \simeq \Z_3 \rtimes \Z_2$
in a block diagonal way, $\R{2}_{D_3}|_{\Z_3} = {\rm diag}( \R{1}'_{\Z_3},\Rb{1}'_{\Z_3})$, then the inner automorphism linking the two irreps is given by the Pauli matrix $\sigma_2 \in \R{2}_{D_3} $,  
 $\sigma_2  {\rm diag}( \R{1}'_{\Z_3},\Rb{1}'_{\Z_3}) \sigma_2^{-1}  = 
 {\rm diag}( \Rb{1}'_{\Z_3},\R{1}'_{\Z_3})$.}
They have the same Molien function and also the same invariants as 
is obvious from Eqs.~\eqref{eq:Molien} and~\eqref{eq:genI}.

A particular, but not infrequent, case is when
\begin{equation}
\label{eq:same_image}
\R{h} \in {\cal R}(H)   \Rightarrow \R{h}^* \in {\cal R}(H)   
\end{equation}
applies, the complex conjugate representation has the same image, Eq.~\eqref{eq:image}.
It goes without saying that this is trivial and not useful if the representation is real.
Note that, if an invariant is not real, then~\eqref{eq:conjugate} and the observations 
above imply that the complex conjugates do not have the same image. The converse is not true. The results above are summarized in Tab.~\ref{tab:Molien}.

\begin{table}
\begin{center}
\begin{tabular}{l|l|l|l}
       & $\R{3}$ vs.\ $\Rb{3}$ &  $\{\R{3}\} = \{\R{3}'\}$   & $h' = A h A^{-1} \;,\; h \in H$  \\\hline
Molien function &  identical   & identical  & identical   \\ 
Invariants         &   complex conjugate     & identical  & change unless $A \in H$  \\
\end{tabular}
\caption{\small Summary of transformation properties of Molien function and the invariants, as discussed in the text, 
with respect to the relation as given in the first row. Most of these properties 
are easily inferred from the definitions~\eqref{eq:Molien} and~\eqref{eq:genI}. 
In what regards the third case it is noted that any element can be conjugated 
by a separate matrix $A_h$ and the Molien function is still left invariant. 
}
\label{tab:Molien}
\end{center}
\end{table}
Crucially, if two irreps $\R{3}$ and $\R{3}'$  have the same image, then the fact that they have 
the same invariants, see Eq.~\eqref{eq:image}, means that there is no way, in our framework, 
to distinguish  $ \R{3} \emb \R{3}_{SU(3)}$   from $\R{3}' \emb \R{3}_{SU(3)} $. This 
apparent ambiguity corresponds to the arbitrariness of  labeling of the irreps $\R{3}$ 
and $\R{3}'$. Associating $\R{3}_{SU(3)}  \to \R{3}$, for example $\R{3}'$ can be generated from
tensor products of the latter, since $\R{3}_{SU(3)}$ is the fundamental irrep of $SU(3)$ from which all other irreps are generated.

In connection with this observation we would like to add two remarks:
First, if two irreps have the same image this ought to imply 
that the Kronecker products of $\R{3}$ and $\R{3}'$ are identical 
under the interchange of irreps of the same order.
One can verify this for the example of $\R{3}$ and $\R{3}'$ for $A_5 = \Sigma(60)$ 
\cite{Ludl_diplom,Everett:2008et}.
Second, if we consider a higher dimensional case such as 
$\R{6}_{SU(6)} \to \R{6}_{S_5} \to \R{3}_{A_5} + \R{3}'_{A_5}$, see~\eqref{eq:SU(6)}, then the two irreps 
can be distinguished. This can be seen or described as follows: One can choose
an embedding of $S_5$ such that under $A_5$ the two irreps are block diagonal,
\begin{equation}
\R{6}_{S_5}|_{A_5}  = 
 \left( \begin{array}{cc}
	\R{3}_{A_5} & 0 \\
	0 & \R{3}'_{A_5}
	\end{array} \right),
\end{equation}
\noindent and associate the six-dimensional representation space by the variables
$\{ x_1,..,x_6\}$. Assuming an invariant $\I_{A_5}(x_1,x_2,x_3)$ breaks $\R{6}_{S_5} \to \R{3}_{A_5} + 3 \cdot \R{1}_{A_5}$, then same invariant $\I_{A_5}(x_4,x_5,x_6)$ breaks 
$\R{6}_{S_5} \to \R{3}'_{A_5} + 3 \! \cdot \!\R{1}_{A_5}$. 
In order to determine this invariant one ought to look at all embeddings 
$\R{3}_X + \R{3}_X \emb \R{6}_{S_5}$, and then go through the same reasoning as in Secs.~\ref{sec:sufficient} and~\ref{sec:breaking}, respectively.

\subsection{The $3$-dimensional irreps of $\Sigma(X)$, $T_{n[a]}$, $\Delta(3n^2)$, and $\Delta(6n^2)$ }
\label{sec:3irreps}

\subsubsection*{Equivalent embeddings}

It is conceivable that a similarity transformation \eqref{eq:similar} on 
the list of groups in
Tab.~\ref{tab:SU3_SubGen} would lead to an embedding that leaves
say \eqref{eq:breakD6} invariant and is a supergroup of $\Delta(6n^2)$. It would thus invalidate the condition in Eq.~\eqref{eq:breakD6}. We shall see below that, due to Schur's Lemma, c.f.\ App.~\ref{app:facts}, this is not the case.

Consider the conditions \eqref{eq:breakD6} and \eqref{eq:breakD3} 
for $\Delta(6n^2)$ and $\Delta(3n^2)$: These polynomials 
imply that  $E$ and $F(n,0,1)$ are part of the groups that leave them invariant. 
Since $E$ and $F(n,0,1)$ generate a $\Delta(3n^2)$-irrep 
of dimension three, by virtue of Schur's Lemma, there does not exist a matrix, 
other than a multiple of the identity, that commutes with $E$ and $F(n,0,1)$.
Therefore we were right to consider, for instance, $G(d,r,s)$ only and not some $A G(d,r,s) A^{-1}$ 
in the previous sections. 
The same argument holds for  $T_{n[a]}$ \eqref{eq:breakTna} with 
$F(n,0,1)$ replaced by $F(n,1,a)$. Similar arguments validate the chains
$\Sigma(36 \phi) \subset \Sigma(72\phi) \subset \Sigma(216\phi)$ and 
$\Sigma(60) \subset \Sigma(360\phi)$, since the supergroups differ from 
the subgroups by one generator only.

\subsubsection*{Inequivalent embeddings}

In this section, we are interested in whether inequivalent embeddings 
of the $3$-dimensional irreps of the $\Sigma(X)$-, $\Delta$-, and $T_{n[a]}$-groups give 
rise to distinct invariants and are thus distinguishable in 
$\R{3}_{SU(3)} \to \R{3}_{\Sigma(X);\Delta}$.
An invaluable source for this endeavour is the diploma thesis
of Patrick Ludl~\cite{Ludl_diplom}, which we shall use frequently below.
The main results are summarized in Tab.~\ref{tab:3embedding}.

\begin{table}
\begin{center}
\begin{tabular}{c|c|c|c|c|c}
Group & number $\R{3}$   & not faithful &  not in $SU(3)$ & same image & remain  \\\hline
$\Sigma(60)$ &  2 & 0 & 0 & 1 & 1 \\
$\Sigma(36\phi)$      & (4,4)  & 0 & (3,3) & 1  & 1 \\
$\Sigma(168) $        &  (1,1) & 0 & 0 & 1 & 1 \\
$\Sigma(72\phi)$      & (4,4)  & 0  & (3,3) & 0 & (1,1) \\
$\Sigma(216\phi)$ &  (4,4) +1  & 1 &  (3,3) & 1 & 1 \\
$\Sigma(360\phi)$ &   (2,2) & 0 & 0 & (1,1) &  (1,1) \\ \hline
$\Delta(3n^2),\; n \;  \s{\!\!\in} 3 \Z$ & $\frac{n^2-1}{3}, \;\R{3}_{a,b}$  & 
${\rm gcd}(a[b],n) > 1$ & 0 & all faithful & 1   \\
$\Delta(3n^2),\; n  \in  3 \Z$ & $\frac{n^2-3}{3}, \;\R{3}_{a,b}$  & idem & 0 & idem & 1   \\
$\Delta(6n^2)$   & $2(n-1), \;\R{3}_a$ & $n/a \in 2,..\frac{n}{2}$ & half of them & idem & 1  
\end{tabular}
\caption{\small $(n,n)$ stands for $n$ pairs of complex conjugate representations. 
The subtraction of the third, fourth, and fifth columns from the second column results
in the  last column.
The irreps which are not in $SU(3)$ do not satisfy the unit
determinant criteria; they are irreps of $U(3)$ rather than $SU(3)$.
The only non-faithful irrep is $\R{3}_{\Sigma_{216\phi}} \simeq \R{3}_{A_4}$.
The same image criteria is discussed around Eq.~\eqref{eq:image}.
The acronym ``gcd'' stands for greatest common divisor, and $a[b]$ stands for $a$ and/or $b$. Since $T_{n[a]} \subset \Delta(3n^2)$ and the latter has only $3$-dimensional irreps 
the $T_{n[a]}$ $3$-dimensional irreps form a subset of the latter.
}
\label{tab:3embedding}
\end{center}
\end{table}

We will not go through all the points but just mention a few facts.
For the groups $\Sigma(36\phi,72\phi,216\phi)$ several irreps are not
in $SU(3)$ since they are obtained from the irrep $\R{3}_{\Sigma(36\phi,72\phi,216\phi)} \emb \R{3}_{SU(3)}$ by multiplying a certain generator by $-1$, $i$, or $-i$, which violates the
determinant condition for $SU(3)$~\cite{Ludl_diplom}. The $\Sigma(X)$ irreps are faithful with the exception 
of $\R{3}_{\Sigma_{216\phi}} \simeq \R{3}_{A_4}$~\cite{Ludl_diplom}, as 
mentioned in Tab.~\ref{tab:3embedding}.
For the $\Delta$-groups the conditions for groups to be non-faithful are given 
in Tab.~\ref{tab:3embedding} as well. A faithful irrep is always 
provided by $(a,b) = (1,0)$, which corresponds to $F(n,1,0)$ and is the one
used throughout this paper, e.g.\ in Tab.~\ref{tab:SU3_SubGen}. 
The important point is though that all faithful irreps have got the same image, 
and thus the same invariants.  The same image of irreps
is determined by criterion~\eqref{eq:same_image} in the cases of 
$\Sigma(36 \phi)$, $\Sigma(168)$, $\Sigma(216\phi)$, and $\Sigma(360\phi)$.

In conclusion we have not missed anything by restricting ourselves to 
a particular embedding in Tab.~\ref{tab:SU3_SubGen}. 
For the case of complex conjugate pairs one has to choose the complex
conjugate invariant in order to distinguish the two cases. However, 
a $\R{3}$ and a $\Rb{3}$ are not really different in the same way as anti-matter 
is not really different from matter.

\section{Epilogue }
\label{sec:conclusions}

In this work we have been studying the breaking of $SU(3)$ into its  proper finite subgroups 
${\cal F}_3$, from the viewpoints of explicit breaking and SSB.
These two approaches are linked 
by the complex spherical harmonics, the representation functions of $SU(3)$, as explained 
in Sec.~\ref{sec:connection} for $SO(3)$ and illustrated for $SU(3)$ in Sec.~\ref{sec:examples}.

In the explicit breaking approach a field $\phi$ transforming 
under the fundamental irrep $\R{3} = (1,0)$ is considered.
The crucial question is which term(s) have to be added to an $SU(3)$-invariant Lagrangian in order to break to ${\cal F}_3$: 
\begin{equation}
\label{eq:explicit_epi}
{\cal L}_{SU(3) \to {\cal F}_3} = {\cal L}_{SU(3)}(\phi_1,\phi_2,\phi_3) +
 {\cal L}_{{\cal F}_3}(\phi_1,\phi_2,\phi_3) \; .
\end{equation}
In retrospect of Sec.~\ref{sec:breaking} we may say that such terms, with the exception 
of a few small groups like $A_4$, $T_{7[2]}$, and $\Sigma(168)$,  
lead to potentials which are not  renormalizable by power counting, 
as their polynomial degrees exceed four.\footnote{In four space-time dimensions a term in the Lagrangian is powercounting renormalizable if its mass dimension is equal to or below four. A scalar field
has mass dimension one in four space-time dimensions.
For instance to enforce $SU(3) \to \Delta(75)$, an explicit term 
$\delta {\cal L} = \frac{c}{\Lambda} ( \phi_1^5 + \phi_2^5 + \phi_3^5) $ would serve the
purpose according to Fig.~\ref{fig:Deltatree}. Restricting oneself to terms up to 
dimension four with symmetry 
$\Delta(75)$, only $\delta {\cal L} =  c' \Lambda  \phi_1 \phi_2 \phi_3$ would remain but 
would lead, according to Fig.~\ref{fig:Deltatree}, to an accidentally larger symmetry 
$SU(3) \to U(1) \times U(1) \rtimes \Z_3 \left[ \supset \Delta(75) \right]$, reminiscent of the baryon number 
conservation in the renormalizable SM.}
  
In the approach of SSB, a field $\tilde \phi$ in an irrep $(p,q)$ of $SU(3)$ is considered. The association of a VEV to this field, singling out a direction, breaks the symmetry:
\begin{equation}
  \label{eq:branch_epi}
(p,q)_{SU(3)} \to |\R{(p,q)}|_{{\cal F}_3}  =   \R{1}_{{\cal F}_3} + ...
\end{equation}
The full relation, including the omitted terms, is called the \emph{branching rule}.
 In the case at hand the branching rule  necessarily contains 
 the trivial irrep, as indicated. The branching rules can be computed with our program 
 \name~by the formalism of the generating functions.  This is outlined in App.~\ref{app:generating} and exemplified in Sec.~\ref{sec:examples} for our package \name.  

It is straightforward to find structures of invariant polynomials by virtue 
of the Reynolds operator~\eqref{eq:genI}, and thus  VEVs 
which leave the group structure ${\cal F}_3$ invariant. They are linked 
by the complex spherical harmonics, and their degrees and dimensions are related as follows:
\begin{equation}
\label{eq:link_epi}
(p,q) = ({\rm deg}_{\phi_i} \I[{\cal F}_3],{\rm deg}_{\phi_i^*} \I[{\cal F}_3]) 
 \quad \leftrightarrow  \;
\quad v[{\cal F}_3] \in \C^{|(p,q)|}\;,
\end{equation}
where $|(p,q)| = \frac{1}{2} (p+1)(q+1)(p+q+2)$ is the dimension of the $(p,q)$-irrep, 
and for the sake of clarity, $(p,q) = (4,5)$ if for example $\I[{\cal F}_3] = \phi_1^2 \phi_2^2 \phi_1^* (\phi_3^*)^4 $.

The non-trivial issue is to find \emph{sufficient} conditions, since a supergroup 
always shares common invariants with its subgroups. 
We have provided solutions for all crystallographic groups $\Sigma(X)$ and for the 
series of trihedral groups $T_{n[a]}$, $\Delta(3n^2)$, and $\Delta(6n^2)$ in Sec.~\ref{sec:breaking} for representations of the $(p,0)$-type. We were able to do so 
by having at our disposal the explicit generators   of the proper finite $SU(3)$ subgroups 
and showing that the results are independent of  the particular embedding 
Tab.~\ref{tab:SU3_SubGen}. We wish to emphasize once more that 
the criterion  for breaking into faithful irrep can be seen as an alternative definition 
of the group. This is close, but  not identical, to the original classification of SU(3) subgroups\cite{Milleretal}.
The reason we are restricted
to the $(p,0)$-type is that we have not considered the case, in explicit breaking, 
where the complex conjugate field 
$\phi^*$, transforming  as $\Rb{3} = (0,1)$,  is added to the Lagrangian in Eq.~\eqref{eq:explicit_epi}. For the case of SSB the limitation to $(p,0)$ is not very restrictive, 
as $(0,q)$-fields and other $(p',0)$-fields can easily be accommodated into the potential~\eqref{eq:explicit}.\footnote{For explicit breaking, an inclusion 
of $\phi^*$ might be necessary depending on the charges of the field. 
For particle physics model-building the SSB approach is more important, as it is the model-builders' goal to explain symmetry patterns dynamically rather than to work in a framework 
where the symmetry is broken explicitly.}
Nevertheless the  $(p,q)$-case  is more generic
and  doable with 
this formalism through the tensor generating function. We leave such a possibility to future work.

Further to that we have computed all primary and secondary invariants, and thus the
syzygies of the entire $\Delta$-series.
This has led to the intuitive geometric interpretation \eqref{eq:c1}
that the $\Delta$-groups are generalizations of $A_4$ and $S_4$ under
a deformation of the Euclidian metric. We have computed the same data for
the remaining groups in the database as given in Tab.~\ref{tab:GroupList}.
This information is stored in the  software package and database \name.
Further to that Molien functions, tensor generating functions,
branching rules, translations from invariants to VEVs and back, character tables, Kronecker products, and further things can be found in the example notebook.

Let us end by emphasizing an interesting nuance: Whereas there is 
a one-to-one link between the degree of explicit terms and the dimension
of the irrep   in the SSB scenario~\eqref{eq:link_epi} for $SU(3) \to {\cal F}_3$, 
we are not aware of a relation to the form of the potential $U(\phi_i)$ 
enforcing SSB, in particular to the degrees of terms needed.
As the explicit terms tend to be non-renormalizable, as discussed above, 
it is thus an interesting question of whether they could be renormalizable in 
the SSB approach.
Low dimensional irreps which lead to power counting renormalizable potentials have
been analyzed in~\cite{Luhn:2011ip,Etesi:1997jv,turkey2,Berger:2009tt}.
Possibly one or the other counterexample already exists in the literature.\\

\noindent \emph{Note added:} Shortly after this paper was finished, the preprint~\cite{Holthausen:2011vd} on discrete groups appeared, which is more directed to the practical aspects used in model building. That paper is accompanied by the software package {\tt Discrete} and it is a very useful addendum to our work.

\section*{Acknowledgments}

We are grateful to Maximilian Albrecht, Claudia Hagedorn, Gareth Jones, Ron King, Patrick Ludl, Christoph Luhn, and Tim Morris
for useful discussions and/or comments on the manuscript. We are indebted to Thomas Fischbacher for collaboration in the early stages of the project and for discussions on embedding theory. RZ is grateful to Thomas
Mannel for his sincere interest in the subject.
The work of AM is supported by the G\"oran Gustafsson foundation.
RZ gratefully acknowledges the support of an advanced STFC fellowship.

\appendix

\section{Mini group theory compendium}
\label{app:facts}
\setcounter{equation}{0}
\renewcommand{\theequation}{A.\arabic{equation}}

In this appendix, for the reader's convenience, we
state a few definitions, facts, and theorems (frequently) used throughout our work.

\begin{itemize}
\item {\bf Branching rule:}  Let $\R{g}$ be an irrep of $G$ and $\R{h}_i$ be irreps of $H$ 
where $H \subset G$. Then the restriction of $G$ to $H$ leads to
\begin{equation}
\R{g}|_H \to \sum_i a_{h_i} \R{h}_i \;.
\end{equation}
The positive number $a_{h_i}$ counts how many times the irrep $\R{h}_i$ is contained in $\R{g}$.
\item {\bf Dimensionality theorem:} The order of a group is equal to the sum of 
squares of the dimensions of all its irreps,
\begin{equation}
 |H| = \sum_i^{\text{irreps}} | {\cal R}_i(H)|^2 \;.
\end{equation}
\item {\bf Center of a group}: The center $C$ of a group $G$ is the set of elements that commute with all group elements, $C:= \{ g'\in G : \forall g\in G: g g' = g' g \}$.
\item A version of {\bf Schur's lemma}: If ${\cal R}(G)$ is a $d$-dimensional irrep of $G$ and $A {\cal R}(G) = {\cal R}(G) A$ for some matrix $A$, then $A$ can only be a multiple of the $d$-dimensional identity matrix.
\item  {\bf Lagrange's theorem:} Let $H$ be a subgroup of the finite group $G$. Then $|G|/|H|$ is an integer.
\item {\bf Semidirect product}: The semidirect product $G \rtimes H\equiv G \rtimes_\phi H$ between two groups $G$ and $H$ is defined as the operation mapping $(g_1,h_1)$ and $(g_2,h_2)$, with $g_{1,2}\in G$ and $h_{1,2}\in H$, onto $(g_1\phi_{h_2}(g_2), h_1 h_2)$, where $\phi_{h_2}$ is a homomorphic mapping $H\to G$.
\item {\bf Theorem II.2}~\cite{Ludl:2010bj}:  Let $G$ be a finite group 
with $m$-dimensional faithful irrep and $c$ the order of the center, 
then $G \times \Z_n$ has an $m$-dimensional faithful irrep $\Leftrightarrow$
${\rm gcd}(n,c) = 1$. 
(The acronym ``gcd'' stands for the greatest common divisor.)
\item {\bf Multiplicity}: Writing the Kronecker product of two irreps as 
 $${\cal R}_1(G) \times 
{\cal R}_2(G) = n_{12}^3 {\cal R}_3(G) + ...\; , $$
the positive number $n_{12}^3$ is the multiplicity. 
It is   computed via the
scalar product $n_{12}^3 = \langle {\cal R}_1 {\cal R}_2, {\cal R}_3 \rangle$, 
where $\langle {\cal R}_i,{\cal R}_j \rangle \equiv |G|^{-1} \sum_{g \in G} 
\chi_i[g] \chi_j[g]^*$ with character 
$\chi_i[g] = 
{\rm tr}[{\cal R}_i(g)]$.
\end{itemize}

\paragraph{Notation:}
\begin{itemize}
\item $\simeq$ isomorphic
\item $\emb$ group embedding 
\end{itemize}

To this end let us mention that representations of a group $H$ 
are generically denoted by ${\cal R}(H)$, but when a very specific group 
is considered often the dimension of the representation is boldfaced, as 
in $\R{3}$, which is not unambiguous and often results in writing a second 
$3$-dimensional irrep by $\R{3}'$ for instance. 
In the cases of $SO(3)$ and $SU(3)$ it is common to refer to an irrep by 
$(l)$ and $(p,q)$, respectively. The latter are unambiguous and partly discussed 
in App.~\ref{app:csh}.
We switch between these notations
throughout this work always adopting to the most convenient one.

%%%%%%%%%%%%%%%%%%%%%%%%%%%%%%%%%%%%%%%%%%%%%%%%%%%%%%%%%%%%%%%%%%%%%%%%%%%%%%%
\section{The group database}
%%%%%%%%%%%%%%%%%%%%%%%%%%%%%%%%%%%%%%%%%%%%%%%%%%%%%%%%%%%%%%%%%%%%%%%%%%%%%%%
\label{app:GenGroup}
\setcounter{equation}{0}
\renewcommand{\theequation}{B.\arabic{equation}}

In this appendix additional useful information can be found on 
the group database which is listed in Tab.~\ref{tab:GroupList} 
and described in the main Sec.~\ref{sec:SU(3)database}.

\begin{table}
\centering
\begin{tabular}{lr}

\begin{minipage}{8cm}
\begin{center}
\begin{tabular}{|l|l|l|l|}\hline
No. & $\mbox{\textlbrackdbl}g,j\mbox{\textrbrackdbl}$ & $c$ & Names\\ \hline
01 & $\mbox{\textlbrackdbl} 12, 3 \mbox{\textrbrackdbl}$ & 1 & $\Delta(3\cdot 2^2)$, $A_4$, $T$\\
02 & $\mbox{\textlbrackdbl} 21, 1 \mbox{\textrbrackdbl}$ & 1 & $C(7,1,2)$, $T_{7[2]}$\\
03 & $\mbox{\textlbrackdbl} 24, 12 \mbox{\textrbrackdbl}$ & 1 & $\Delta(6\cdot 2^2)$, $S_4$, $O$\\
04 & $\mbox{\textlbrackdbl} 27, 3 \mbox{\textrbrackdbl}$ & 3 & $\Delta(3\cdot 3^2)$\\
05 & $\mbox{\textlbrackdbl} 39, 1 \mbox{\textrbrackdbl}$ & 1 & $C(13,1,3)$, $T_{13[3]}$\\
06 & $\mbox{\textlbrackdbl} 48, 3 \mbox{\textrbrackdbl}$ & 1 & $\Delta(3\cdot 4^2)$\\
07 & $\mbox{\textlbrackdbl} 54, 8 \mbox{\textrbrackdbl}$ & 3 & $\Delta(6\cdot 3^2)$\\
08 & $\mbox{\textlbrackdbl} 57, 1 \mbox{\textrbrackdbl}$ & 1 & $C(19,1,7)$, $T_{19[7]}$\\
09 & $\mbox{\textlbrackdbl} 60, 5 \mbox{\textrbrackdbl}$ & 1 & $A_5$, $\Sigma(60)$, $I$, $Y$\\
10 & $\mbox{\textlbrackdbl} 75, 2 \mbox{\textrbrackdbl}$ & 1 & $\Delta(3\cdot 5^2)$\\
11 & $\mbox{\textlbrackdbl} 81, 9 \mbox{\textrbrackdbl}$ & 3 & $C(9,1,1)$\\
12 & $\mbox{\textlbrackdbl} 84, 11 \mbox{\textrbrackdbl}$ & 1 & $C(14,1,2)$\\
13 & $\mbox{\textlbrackdbl} 93, 1 \mbox{\textrbrackdbl}$ & 1 & $C(31,1,5)$, $T_{31[5]}$\\
14 & $\mbox{\textlbrackdbl} 96, 64 \mbox{\textrbrackdbl}$ & 1 & $\Delta(6\cdot 4^2)$\\
15 & $\mbox{\textlbrackdbl} 108, 15 \mbox{\textrbrackdbl}$ & 3 & $\Sigma(36\phi)$\\
16 & $\mbox{\textlbrackdbl} 108, 22 \mbox{\textrbrackdbl}$ & 3 & $\Delta(3\cdot 6^2)$\\
17 & $\mbox{\textlbrackdbl} 111, 1 \mbox{\textrbrackdbl}$ & 1 & $C(37,1,10)$, $T_{37[10]}$\\
18 & $\mbox{\textlbrackdbl} 129, 1 \mbox{\textrbrackdbl}$ & 1 & $C(43,1,6)$, $T_{43[6]}$\\
19 & $\mbox{\textlbrackdbl} 147, 1 \mbox{\textrbrackdbl}$ & 1 & $C(49,10,6)$\\
20 & $\mbox{\textlbrackdbl} 147, 5 \mbox{\textrbrackdbl}$ & 1 & $\Delta(3\cdot 7^2)$\\
21 & $\mbox{\textlbrackdbl} 150, 5 \mbox{\textrbrackdbl}$ & 1 & $\Delta(6\cdot 5^2)$\\
22 & $\mbox{\textlbrackdbl} 156, 14 \mbox{\textrbrackdbl}$ & 1 & $C(26,1,3)$\\
23 & $\mbox{\textlbrackdbl} 162, 14 \mbox{\textrbrackdbl}$ & 3 & $D(9,1,1;2,1,1)$\\
24 & $\mbox{\textlbrackdbl} 168, 42 \mbox{\textrbrackdbl}$ & 1 & $PSL(2,7)$, $\Sigma(168)$\\
25 & $\mbox{\textlbrackdbl} 183, 1 \mbox{\textrbrackdbl}$ & 1 & $C(61,1,13)$, $T_{61[13]}$\\
26 & $\mbox{\textlbrackdbl} 189, 8 \mbox{\textrbrackdbl}$ & 3 & $C(21,1,2)$\\
27 & $\mbox{\textlbrackdbl} 192, 3 \mbox{\textrbrackdbl}$ & 1 & $\Delta(3\cdot 8^2)$\\
28 & $\mbox{\textlbrackdbl} 201, 1 \mbox{\textrbrackdbl}$ & 1 & $C(67,1,29)$, $T_{67[29]}$\\
29 & $\mbox{\textlbrackdbl} 216, 88 \mbox{\textrbrackdbl}$ & 3 & $\Sigma(72\phi)$\\
30 & $\mbox{\textlbrackdbl} 216, 95 \mbox{\textrbrackdbl}$ & 3 & $\Delta(6\cdot 6^2)$\\
\hline
\end{tabular}
\end{center}
\end{minipage}

&

\begin{minipage}{8cm}
\begin{center}
\begin{tabular}{|l|l|l|l|}\hline
31 & $\mbox{\textlbrackdbl} 219, 1 \mbox{\textrbrackdbl}$ & 1 & $C(73,1,8)$, $T_{73[8]}$\\
32 & $\mbox{\textlbrackdbl} 228, 11 \mbox{\textrbrackdbl}$ & 1 & $C(38,1,7)$\\
33 & $\mbox{\textlbrackdbl} 237, 1 \mbox{\textrbrackdbl}$ & 1 & $C(79,1,23)$, $T_{79[23]}$\\
34 & $\mbox{\textlbrackdbl} 243, 26 \mbox{\textrbrackdbl}$ & 3 & $\Delta(3\cdot 9^2)$\\
35 & $\mbox{\textlbrackdbl} 273, 3 \mbox{\textrbrackdbl}$ & 1 & $C(91,1,16)$, $T_{91[16]}$\\
36 & $\mbox{\textlbrackdbl} 273, 4 \mbox{\textrbrackdbl}$ & 1 & $C(91,1,9)$, $T_{91[9]}$\\
37 & $\mbox{\textlbrackdbl} 291, 1 \mbox{\textrbrackdbl}$ & 1 & $C(97,1,35)$, $T_{97[35]}$\\
38 & $\mbox{\textlbrackdbl} 294, 7 \mbox{\textrbrackdbl}$ & 1 & $\Delta(6\cdot 7^2)$\\
39 & $\mbox{\textlbrackdbl} 300, 43 \mbox{\textrbrackdbl}$ & 1 & $\Delta(3\cdot 10^2)$\\
40 & $\mbox{\textlbrackdbl} 309, 1 \mbox{\textrbrackdbl}$ & 1 & $C(103,1,46)$, $T_{103[46]}$\\
41 & $\mbox{\textlbrackdbl} 324, 50 \mbox{\textrbrackdbl}$ & 3 & $C(18,1,1)$\\
42 & $\mbox{\textlbrackdbl} 327, 1 \mbox{\textrbrackdbl}$ & 1 & $C(109,1,45)$, $T_{109[45]}$\\
43 & $\mbox{\textlbrackdbl} 336, 57 \mbox{\textrbrackdbl}$ & 1 & $C(28,1,2)$\\
44 & $\mbox{\textlbrackdbl} 351, 8 \mbox{\textrbrackdbl}$ & 3 & $C(39,1,3)$\\
45 & $\mbox{\textlbrackdbl} 363, 2 \mbox{\textrbrackdbl}$ & 1 & $\Delta(3\cdot 11^2)$\\
46 & $\mbox{\textlbrackdbl} 372, 11 \mbox{\textrbrackdbl}$ & 1 & $C(62,1,5)$\\
47 & $\mbox{\textlbrackdbl} 381, 1 \mbox{\textrbrackdbl}$ & 1 & $C(127,1,19)$, $T_{127[19]}$\\
48 & $\mbox{\textlbrackdbl} 384, 568 \mbox{\textrbrackdbl}$ & 1 & $\Delta(6\cdot 8^2)$\\
49 & $\mbox{\textlbrackdbl} 399, 3 \mbox{\textrbrackdbl}$ & 1 & $C(133,1,11)$, $T_{133[11]}$\\
50 & $\mbox{\textlbrackdbl} 399, 4 \mbox{\textrbrackdbl}$ & 1 & $C(133,1,30)$, $T_{133[30]}$\\
51 & $\mbox{\textlbrackdbl} 417, 1 \mbox{\textrbrackdbl}$ & 1 & $C(139,1,42)$, $T_{139[42]}$\\
52 & $\mbox{\textlbrackdbl} 432, 103 \mbox{\textrbrackdbl}$ & 3 & $\Delta(3\cdot 12^2)$\\
53 & $\mbox{\textlbrackdbl} 444, 14 \mbox{\textrbrackdbl}$ & 1 & $C(74,1,10)$\\
54 & $\mbox{\textlbrackdbl} 453, 1 \mbox{\textrbrackdbl}$ & 1 & $C(151,1,32)$, $T_{151[32]}$\\
55 & $\mbox{\textlbrackdbl} 471, 1 \mbox{\textrbrackdbl}$ & 1 & $C(157,1,12)$, $T_{157[12]}$\\
56 & $\mbox{\textlbrackdbl} 486, 61 \mbox{\textrbrackdbl}$ & 3 & $\Delta(6\cdot 9^2)$\\
57 & $\mbox{\textlbrackdbl} 489, 1 \mbox{\textrbrackdbl}$ & 1 & $C(163,1,58)$, $T_{163[58]}$\\
58 & $\mbox{\textlbrackdbl} 507, 1 \mbox{\textrbrackdbl}$ & 1 & $C(169,1,22)$, $T_{169[22]}$\\
59 & $\mbox{\textlbrackdbl} 507, 5 \mbox{\textrbrackdbl}$ & 1 & $\Delta(3\cdot 13^2)$\\ 
60 & $\mbox{\textlbrackdbl} 648, 532 \mbox{\textrbrackdbl}$ & 3 & $\Sigma(216\phi)$\\ 
61 & $\mbox{\textlbrackdbl} 1080,260 \mbox{\textrbrackdbl}$ & 3 & $\Sigma(360\phi)$\\ \hline
\end{tabular}
\end{center}
\end{minipage}

\end{tabular}
\caption{\label{tab:GroupList} \small The groups contained in our database, together with their group numbers and GAP numbers ~\cite{GAP,SmallGroups}, while $c={\rm ord}(C)$ is the order of the center of the respective group, which can only be $1$ or $3$ by theorem II.2 stated in App.~\ref{app:facts}. Note that, in some cases, it might not work out to describe the $T_{n[a]}$ groups by the number $n$ only, as different choices for the second parameter $a$ might be possible, due to the definition of these groups as $C(n,1,a)$ with $a^2+a+1 = 0\ {\rm mod}\ n$.}
\end{table}

\subsection{Generators }
\label{app:generators}

The generators needed for the groups in 
Tab.~\ref{tab:SU3_SubGen} are given by~\cite{Ludl:2010bj}:
\begin{eqnarray}
&& E = \left( \begin{array}{ccc}
	0 & 1 & 0 \\ 0 & 0 & 1 \\ 1 & 0 & 0
	\end{array} \right), \ \ 
	F(n,a,b) = \left( \begin{array}{ccc}
	\eta^a & 0 & 0 \\ 0 & \eta^b & 0 \\ 0 & 0 & \eta^{-a-b}
	\end{array} \right),\ \  
	G(d,r,s) = \left( \begin{array}{ccc}
	\delta^r & 0 & 0 \\ 0 & 0 & \delta^s \\ 0 & -\delta^{-r-s} & 0
	\end{array} \right)\;,\nonumber\\
&& H = \frac{1}{2} \left( \begin{array}{ccc}
	-1 & \mu_- & \mu_+ \\ \mu_- & \mu_+ & -1 \\ \mu_+ & -1 & \mu_-
	\end{array} \right),\ \  
	J = \left( \begin{array}{ccc}
	1 & 0 & 0 \\ 0 & \omega & 0 \\ 0 & 0 & \omega^2 
	\end{array} \right),\ \ 
	K= 
	\frac{1}{\sqrt{3}\,i} \left( \begin{array}{ccc}
	1 & 1 & 1 \\ 
	1 & \omega & \omega^2 \\ 
	1 & \omega^2 & \omega
	\end{array} \right)\;,\nonumber\\
&& L = 
	\frac{1}{\sqrt{3}\,i} \left( \begin{array}{ccc}
	1 & 1 & \omega^2 \\ 
	1 & \omega & \omega \\ 
	\omega & 1 & \omega
	\end{array} \right),\ \ 
	M= \left( \begin{array}{ccc}
	\beta & 0 & 0 \\ 0 & \beta^2 & 0 \\ 0 & 0 & \beta^4 
	\end{array} \right),\ \ 
	N= 
	\frac{i}{\sqrt{7}} \left( \begin{array}{ccc}
	\beta^4 - \beta^3 & \beta^2 - \beta^5 & \beta   - \beta^6 \\ 
	\beta^2 - \beta^5 & \beta   - \beta^6 & \beta^4 - \beta^3 \\ 
	\beta   - \beta^6 & \beta^4 - \beta^3 & \beta^2 - \beta^5 
	\end{array} \right),\nonumber \\
&&	P= 
	\left( \begin{array}{ccc}
	\epsilon & 0 & 0 \\ 
	0 & \epsilon & 0 \\ 
	0 & 0 & \epsilon\omega
	\end{array} \right),\ \ Q=\left( \begin{array}{ccc}
	-1 & 0 & 0 \\ 
	0 & 0 & -\omega \\ 
	0 & -\omega^2 & 0
	\end{array} \right).
	\label{eq:Gens}
\end{eqnarray}
Here, we have used the abbreviations
\begin{equation}
 \eta\equiv e^{2\pi i/n},\quad \delta\equiv e^{2\pi i/d},\quad \mu_{\pm} \equiv \frac{1}{2} \left( -1 \pm \sqrt{5} \right),\quad \omega\equiv e^{2\pi i/3},\quad\beta\equiv e^{2\pi i/7},\quad \epsilon\equiv e^{4\pi i/9}.\nonumber
\end{equation}
Further to that note that the generators  $J,M,P,Q$ can be expressed as follows:
\begin{equation}
\label{eq:JMPQ}
J=F(3,0,1) \;, \quad 
M=F(7,1,2) \;, \quad 
P=F(9,2,2) \;, \quad 
Q=G(6,3,5) \;.
\end{equation}
The orders of the generators, $X^o = \Id{}$, are:
\begin{center}
\begin{tabular}{c||c|c|c|c|c|c|c|c|c|c|c}
Generator $X$  & $E$ & $F(n,a,b)$ & $G(d,r,s)$ & $H$ & $J$ & $K$ & $L$ & $M$ & $N$ & $P$ & $Q$ \\ \hline
$o$ & $3$ & $\frac{n}{{\rm gcd}(n,a,b)}$ & $\frac{d}{{\rm gcd}(d,r,s)}$ & $2$ & $3$ & $4$ & $4$ & $7$ & $2$ & $9$ & $2$ 
\end{tabular}
\end{center}
The orders must divide the order of the group by virtue of Lagrange's theorem, as they 
generate the subgroup $\Z_o \in {\cal F}_3$. The acronym ``gcd'' stands for greatest common divisor.

\begin{table}
\centering
\begin{tabular}{l  |  c |c |c |c |c |c |c |c |c  }
$\mathcal{G} $ & $g_1$ & $g_2$ &$g_3$ &$g_4$ &$g_5$ &$g_6$ &$g_7$ &$g_8$ \\ \hline 
$E$   & 0 & $\frac{4}{3\sqrt{3}}$ & 0 & 0 & $-\frac{4}{3\sqrt{3}}$ & 0 & $\frac{4}{3\sqrt{3}}$ & 0  \\
$F$ & 0& 0& $\frac{2(a-b)}{n} $ & 0& 0& 0& 0&  $\frac{2 \sqrt{3}(a+b)}{n} $  \\
$G$ &  0& 0& $\frac{3}{d} r $ & 0& 0& $\sin\left( \frac{\pi (r + 2s)}{d}\right )$ & $\cos\left( \frac{\pi (r + 2s)}{d}\right )$&  $\frac{ \sqrt{3} }{n} r  $  \\
$H$  & $\frac{1+\sqrt{5}}{4}$ & 0 & $\frac{1+\sqrt{5}}{8}$ & $-\frac{1}{1+\sqrt{5}}$ & 0 &  $\frac{1}{2}$ & 0 & $\frac{1-3 \sqrt{5}}{8 \sqrt{3}}$ \\
$K$  & $-\frac{1}{\sqrt{3}}$ & 0 & $-\frac{\sqrt{3}}{4}$ & $-\frac{1}{\sqrt{3}}$ & 0 & $\frac{1}{2\sqrt{3}}$ & 0 & $-\frac{1}{4}$  \\
$L$ & $-\frac{1}{\sqrt{3}}$ & 0 & $-\frac{\sqrt{3}}{4}$ & $\frac{1}{2\sqrt{3}}$ & $-\frac{1}{2}$ & $-\frac{1}{4\sqrt{3}}$ & $\frac{1}{4}$ & $-\frac{1}{4}$ \\
$N$ & $-\frac{1}{\sqrt{3}}$ & 0 & $-\frac{\sqrt{3}}{4}$ & $-\frac{1}{\sqrt{3}}$ & 0 & $\frac{1}{2 \sqrt{3}}$ & 0 & $-\frac{1}{4}$
\end{tabular}
\caption{\small \label{tab:vGM} 
Gell-Mann vector components for all generators, with $F=F(n,a,b)$ and $G=G(d,r,s)$. Any generator $\mathcal{G}$ in Eq.~\eqref{eq:Gens} can be displayed as $\mathcal{G} = \exp(i \pi \vec{g}[\mathcal{G} ] \cdot \vec{T}) = \exp(i \pi \sum_{a=1}^8 g[\mathcal{G} ]_a  T_a)$. $T_a$ for $a = 1,..,8$ are the standard Gell-Mann matrices, which can be found in any textbook and in our package \name, satisfying the commutation relations as given in App.~\ref{app:csh} in Eq.~\eqref{eq:LA}. The generators $J$, $M$, $P$, and $Q$ can be obtained through Eq.~\eqref{eq:JMPQ}.}.
\end{table}

%%%%%%%%%%%%%%%%%%%%%%%%%%%%%%%%%%%%%%%%%%%%%%%%%%%%%%%%%%%%%%%%%%%%%%%%%%%%%%%
\subsection{The subgroup tree within the group database }
\label{app:group_tree}
%%%%%%%%%%%%%%%%%%%%%%%%%%%%%%%%%%%%%%%%%%%%%%%%%%%%%%%%%%%%%%%%%%%%%%%%%%%%%%%

\begin{figure}[!]
\vspace{-3.2cm}
    \includegraphics[angle=90,width=\textwidth]{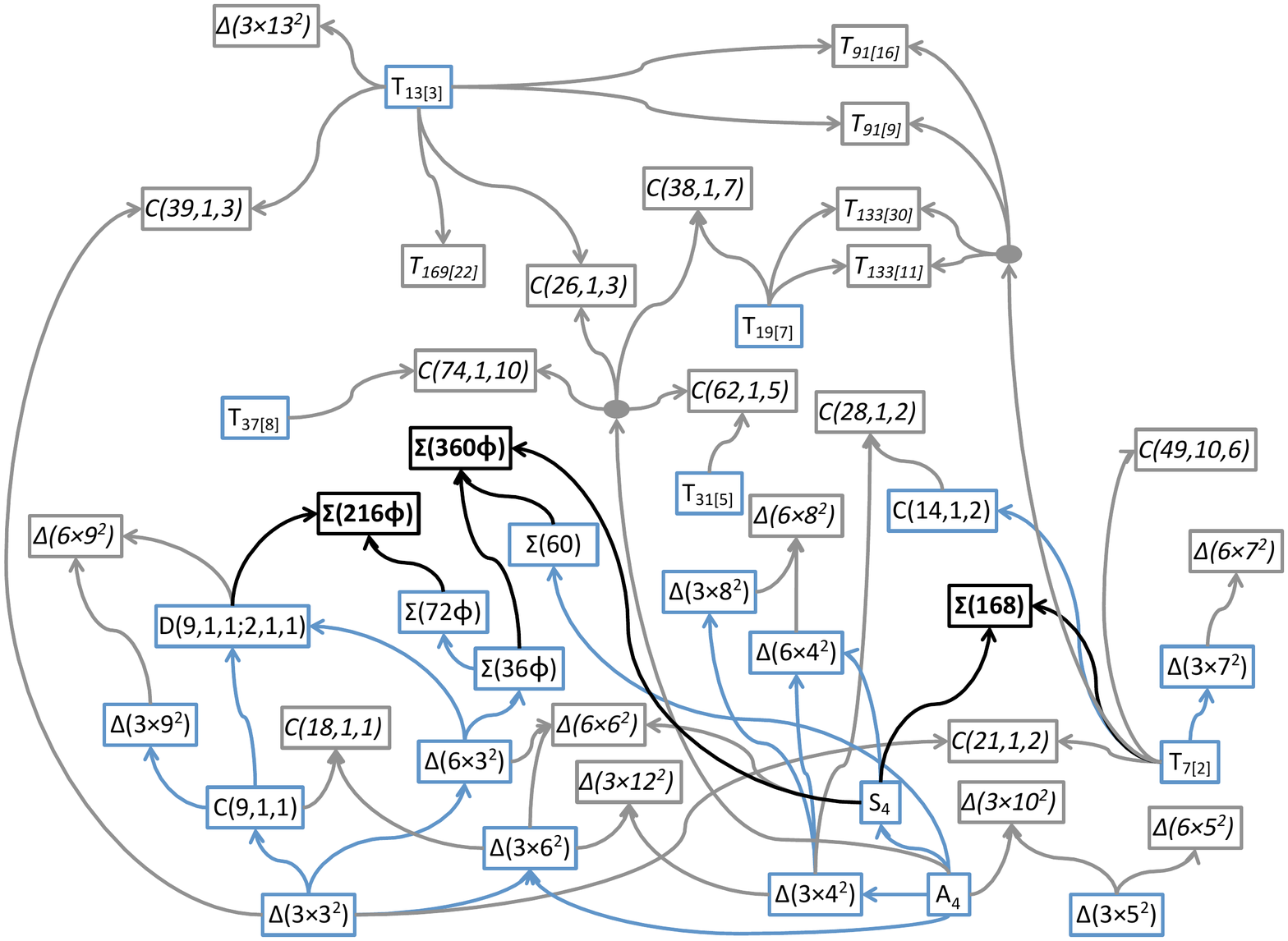}
    \vspace{-1cm}
  \caption{\small Subgroup tree within our database in Tab.~\ref{tab:GroupList}. Black (bold faced) groups denote
  maximal subgroups of $SU(3)$. Gray (Italic) groups denote largest groups within 
  our tree (within a branch). Note that ten subgroup relations do not follow 
  directly from the generators as in Tab.~\ref{tab:SU3_SubGen}. The similarity transformations relating the generators are described in App.~\ref{app:group_tree}.}
\label{fig:su3tree}
  \vspace{-5pt}
\end{figure} 

\begin{table}
\centering
\begin{tabular}{|c|c||l|}\hline
No. & Group & Nearest supergroups\\ \hline\hline
01 & $A_4=T=\Delta(3\cdot 2^2)$ & $S_4=O=\Delta(6\cdot 2^2)$, $\Delta(3\cdot 4^2)$, $A_5=\Sigma(60)=I=Y$,\\
 & & $\Delta(3\cdot 6^2)$, $C(26,1,3)$, $C(38,1,7)$, $\Delta(3\cdot 10^2)$, $C(62,1,5)$,\\
 & & $C(74,1,10)$\\ \hline
02 & $T_{7[2]}=C(7,1,2)$ & $C(14,1,2)$, $C(49,10,6)$, $\Delta(3\cdot 7^2)$, $PSL(2,7)=\Sigma(168)$,\\
 & & $C(21,1,2)$, $T_{91[16]}=C(91,1,16)$, $T_{91[9]}=C(91,1,9)$,\\
 & & $T_{133[11]}=C(133,1,11)$, $T_{133[30]}=C(133,1,30)$\\ \hline
03 & $S_4=O=\Delta(6\cdot 2^2)$ & $\Delta(6\cdot 4^2)$, $PSL(2,7)=\Sigma(168)$, $\Delta(6\cdot 6^2)$, $\Sigma(360\phi)$\\ \hline
04 & $\Delta(3\cdot 3^2)$ & $\Delta(6\cdot 3^2)$, $C(9,1,1)$, $\Delta(3\cdot 6^2)$, $C(21,1,2)$, $C(39,1,3)$\\ \hline
05 & $T_{13[3]}=C(13,1,3)$ & $C(26,1,3)$, $T_{91[16]}=C(91,1,16)$, $T_{91[9]}=C(91,1,9)$,\\ 
 & & $C(39,1,3)$, $T_{169[22]}=C(169,1,22)$,  $\Delta(3\cdot 13^2)$\\ \hline
06 & $\Delta(3\cdot 4^2)$ & $\Delta(6\cdot 4^2)$, $\Delta(3\cdot 8^2)$, $C(28,1,2)$, $\Delta(3\cdot 12^2)$\\ \hline
07 & $\Delta(6\cdot 3^2)$ & $\Sigma(36\phi)$, $D(9,1,1;2,1,1)$, $\Delta(6\cdot 6^2)$\\ \hline
08 & $T_{19[7]}=C(19,1,7)$ & $C(38,1,7)$, $T_{133[11]}=C(133,1,11)$, $T_{133[30]}=C(133,1,30)$\\ \hline
09 & $A_5=\Sigma(60)=I=Y$ & $\Sigma(360\phi)$\\ \hline
10 & $\Delta(3\cdot 5^2)$ & $\Delta(6\cdot 5^2)$, $\Delta(3\cdot 10^2)$\\ \hline
11 & $C(9,1,1)$ & $D(9,1,1;2,1,1)$, $\Delta(3\cdot 9^2)$, $C(18,1,1)$\\ \hline
12 & $C(14,1,2)$ & $C(28,1,2)$\\ \hline
13 & $T_{31[5]}=C(31,1,5)$ & $C(62,1,5)$\\ \hline
14 & $\Delta(6\cdot 4^2)$ & $\Delta(6\cdot 8^2)$\\ \hline
15 & $\Sigma(36\phi)$ & $\Sigma(72\phi)$, $\Sigma(360\phi)$\\ \hline
16 & $\Delta(3\cdot 6^2)$ & $\Delta(6\cdot 6^2)$, $C(18,1,1)$, $\Delta(3\cdot 12^2)$\\ \hline
17 & $T_{37[10]}=C(37,1,10)$ & $C(74,1,10)$\\ \hline
20 & $\Delta(3\cdot 7^2)$ & $\Delta(6\cdot 7^2)$\\ \hline
23 & $D(9,1,1;2,1,1)$ & $\Delta(6\cdot 9^2)$, $\Sigma(216\phi)$\\ \hline
27 & $\Delta(3\cdot 8^2)$ & $\Delta(6\cdot 8^2)$\\ \hline
29 & $\Sigma(72\phi)$ & $\Sigma(216\phi)$\\ \hline
34 & $\Delta(3\cdot 9^2)$ & $\Delta(6\cdot 9^2)$\\ \hline
\end{tabular}
\caption{\small \label{tab:group_tree} The subgroup structure among the 61 groups under consideration. Note that, for groups with numbers greater than 34, the number of elements is larger than $512\div 2 = 256$, so that they could never be subgroups of any group up to number 59, due to Lagrange's theorem, c.f.\ App.~\ref{app:facts}. It turns out that furthermore none of these groups is a subgroup of any of the two groups 60 and 61. We refer the reader to Eq.~\eqref{eq:CDsubgroups} for general subgroup relations among the $C$- and $D$-type groups. }
\end{table}

The subgroup structure within our choice of groups can be found in Tab.~\ref{tab:group_tree} and in Fig.~\ref{fig:su3tree}.
It has been obtained with the help of GAP~\cite{GAP,SmallGroups} and the  generator basis 
given in Tab.~\ref{tab:SU3_SubGen}. More precisely, we have first searched for the subgroup structure within our basis and then tested the 
remaining possibilities, allowed by Lagrange's theorem, with GAP. 
Note that the highest order groups that can be deduced from this table are not necessarily maximal groups, but they may only be so within our choice of groups, depending on the invariants chosen. In order to avoid confusion we shall refer to them as \emph{largest groups}.
As argued in Sec.~\ref{sec:crystal_break}, $\Sigma(216\phi)$, $\Sigma(360\phi)$, and $\Sigma(168)$ are the only maximal subgroups of $SU(3)$ in that list.

Not all subgroup relations in Fig.~\ref{fig:su3tree} and Tab.~\ref{tab:group_tree} follow
from the specific embedding in Tab.~\ref{tab:SU3_SubGen}. 
In fact there are ten cases, which we shall discuss to various degrees of detail according 
to importance and feasibility. For all cases there ought to be similarity transformations,
\begin{equation}
\label{eq:btrafo}
 g_i'  = A g_i A^{-1}  \;, \qquad A^{-1} = A^\dagger  \;,
\end{equation}
where $g_i$ are the group generators and 
$A$ can be written as  a unitary matrix as discussed in Sec.~\ref{sec:similar}.
There are a total of ten cases, which we make explicit below, in the group database which necessitate the transformation
\eqref{eq:btrafo} in order to make the subgroup relation apparent. 

\begin{itemize}
\item[4)] Four  cases include $T_{n[a]}$-relations:
\begin{eqnarray}
T_{7[2]} \subset T_{133[11]}   \supset T_{19[7]} \;, \qquad 
T_{91[9]} \supset T_{13[3]} \subset T_{169[22]} \;.
\end{eqnarray}
For the subgroups, the similarity transformation is given by:
\begin{equation}
F(n,a,1) = A F(n,1,a) A^{-1} \;,  \quad E  = A E A^{-1} \;, \quad A = G(2,1,1)\;.
\end{equation}
Essentially we are saying here that $T_{n[a]}$ can be generated either by 
$\{ E, F(n,1,a)\}$ or by $\{E,F(1,n,a)\}$.

\item[1)] \emph{ $\Sigma(36\phi)$ as a subgroup of $\Sigma(360\phi)$} 
is crucial for the criteria given in Sec.~\ref{sec:crystal_break}.
The embedding of $\Sigma(360\phi)$ is chosen such that 
the subgroup relation with $\Sigma(60)$ is most transparent.\footnote{The former
is equivalent to the central extension of the $A_6$~\cite{FFK}, and the latter is isomorphic 
to $A_5$.}
According to  Eq.~\eqref{eq:Gens}, $\Sigma(36\phi)$ is generated by $\{E,J,K\}$. 
First we note that  $E=K^3 J^2 K$. Furthermore we have verified that
$\Sigma(36\phi)$ is also generated by $\{F_1, F_2 \}$, where $F_1=E J E^2$ and $F_2 = K$, or equivalently, through the following algebraic relations:
\begin{equation}
\label{eq:P36}
\Sigma(36\phi) =\ \ll F_1, F_2 | \;\; F_1^3 = F_2^4 =  (F_2F_1)^4 =  F_1 F_2 F_1^2 F_2 F_1 F_2^3 F_1 F_2^3 = 1 \gg \;,
\end{equation}
where $\ll \dots \gg$ denotes the generating relations. Relations such as the one above are called \emph{presentations}. We are not aware of 
such a result in the literature. This result was achieved by working out a number of 
relations with explicit generators, and then the corresponding group was generated 
in GAP and found to be isomorphic to $\Sigma(36\phi)$ in the {\tt SmallGroups} library~\cite{GAP,SmallGroups}.
Moreover, coming back to the main point, we have
verified that, among the $\Sigma(360\phi)$-generators, $F_1' =F E$ and $F_2'=E^2 F Q H Q F H$ with $F=F(2,0,1)$ are related to $F_1$ and 
$F_2$ as quoted above by:
\begin{equation}
F_1' = A F_1 A^{-1} \;,  \quad F_2' = A F_2 A^{-1}\;,\; \; {\rm where}\nonumber
\end{equation}
\begin{equation}
A=a
\begin{pmatrix}
 \frac{2 \left(\sqrt{3}-3 i\right)}{\sqrt{3}+3 i \sqrt{5}} & \frac{1}{8} \left(1+3 \sqrt{5}-i
   \sqrt{3} \left(\sqrt{5}-1\right)\right) & \frac{1}{4} \left(1-i \sqrt{15}\right) \\
 -1 & -1 & 1 \\
 \frac{1}{2} \left(1+i \sqrt{3}\right)  & \frac{1}{2} \left(1-i \sqrt{3}\right)  & 1
\end{pmatrix}\; .
\end{equation}
The parameter $a$ is not constrained, unless that one should have $a\neq 0$ in order for $A$ to remain invertible, but the choice $a=\frac{1}{\sqrt{3}} e^{i \phi}$ with $\phi = -\frac{1}{3} {\rm arccot} \left( \sqrt{15} \right)$ leads to $A \in SU(3)$.
\item[5)] 
For the five remaining cases are $A_4,\;S_4\subset \Sigma(168)$ and $S_4,\; \Delta(3\cdot 3^2),\; \Delta(6\cdot 3^2) \subset \Sigma(360\phi)$.  
The generators $E$ remain the same in both representations. 
This fixes the basis only partly. 
In fact  $A$ from Eq.~\eqref{eq:btrafo} is a matrix that commutes with $E$. 
By going to a diagonal basis this matrix is readily found to be the two-parameter matrix:
\begin{equation}
 A \to A(a,b) = \begin{pmatrix}
  f_1(a,b) & f_2(a,b) & f_3(a,b) \\
  f_3(a,b) & f_1(a,b) & f_2(a,b) \\
  f_2(a,b) & f_3(a,b) & f_1(a,b)
 \end{pmatrix} \;, \quad  [E,A(a,b)] = 0\;,
 \label{eq:E-commutant_1}
\end{equation}
with
\begin{eqnarray}
 f_1(a,b) &=& \frac{1}{3} \left(e^{2 i \pi  (a+b)}+e^{-2 i \pi  a}+e^{-2 i \pi  b}\right)\;,\nonumber\\
 f_2(a,b) &=& \frac{1}{3} e^{2 i \pi (a+b)} \left(e^{-2 i \pi  (2 a+b)}+\rho_3^2 e^{-2 i \pi  (a+2 b)}-\rho_3\right)\;,\nonumber\\
 f_3(a,b) &=& \frac{1}{3} e^{2 i \pi  (a+b)} \left(e^{-2 i \pi  (2 a+b)}-\rho_3 e^{-2 i \pi  (a+2 b)}+\rho_3^2\right)\;,
 \label{eq:E-commutant_2}
\end{eqnarray}
where here and further below we  use the notation:
$\rho_x \equiv \exp(2\pi i /x)$.
We shall denote the generators of the subgroup by $F_i$ and the ones of the supergroup
by $F_i'$ and the common generator $E$ is chosen to be $F_1 = F_1' = E$.
In this notation \eqref{eq:btrafo} reads:
\begin{equation}
F'_i = A(a,b) F_i A(a,b)^{-1} \;, \qquad A(a,b)^{-1} \stackrel{\eqref{eq:btrafo}}{=} A(a,b) ^\dagger =  A(-a,-b) 
\end{equation}
We would like to add that no attempt is made to choose the optimal representation i.e. 
find the presentation where the transformation matrices are simplest.

\begin{itemize}
\item \emph{$A_4$ as subgroup of $\Sigma(168)$}: A presentation of $A_4$ is given by
\begin{equation}
\ll F_1, F_2 | F_1^3 = F_2^2 = (F_1 F_2)^3 = F_2 F_1^2 F_2 F_1 F_2 F_1^2 F_2 F_1 =  \Id{} \gg\;, \nonumber
\end{equation}
with generators as given in  Tab.~\ref{tab:SU3_SubGen}. For $A_4$ and $\Sigma(168)$, 
\begin{equation*}
 \{ F_2=F(2,0,1)\}_{A_4}    \;, \qquad   \{ F'_2= M N M^6\}_{\Sigma(168)}   \;.
\end{equation*}
We find
\begin{equation}
 f_1(a_0,b_0) =\alpha_1^{-1/3} \;,\ \ f_2(a_0,b_0) = \rho_3^2 \beta_4^{1/3} \;,\ \ f_3(a_0,b_0) = \beta_2^{1/3}\;,
 \label{eq:trafo_0}
\end{equation}
where $\alpha_1$ is the first root of $175616 - 219520  x + 86240  x^2 - 11368  x^3 + 756  x^4 -  28  x^5 +  x^6$ and $\beta_i$ is the $i$-th root of $1 - 28  x + 756  x^2 - 11368  x^3 + 86240  x^4 - 219520  x^5 + 175616  x^6$. The ordering of the root is proportional to the real part of the root.
Numerically, it turns out that $(a_0,b_0)=(2.08998,2.04843)$.

\item \emph{$S_4$ as subgroup of $\Sigma(168)$}: A presentation of $S_4$ is given by
\begin{eqnarray}
&& \ll F_1, F_2, F_3 | F_1^3 = F_2^2 = (F_1 F_3)^2 =  (F_1^2 F_3)^2 = (F_1 F_2)^3\nonumber\\
&& = F_2 F_1^2 F_2 F_1 F_2 F_1^2 F_2 F_1 = F_1 F_3 F_2 F_3 F_1 F_2 F_1 = F_1 F_3 F_1^2 F_2 F_1 F_3 F_1^2 F_2 = \Id{} \gg\;,\nonumber
\end{eqnarray}
with generators as given in  Tab.~\ref{tab:SU3_SubGen}. For $S_4$ and $\Sigma(168)$, 
\begin{equation*}
\{ F_2=F=F(2,0,1) \;,F_3=G = G(2,1,1) \}_{S_4}    \;, \qquad  \{ F_2=E F E^2 \;, 
F_3=G\}_{\Sigma(168)}   \;.
\end{equation*}
We find
\begin{eqnarray}
 && f_1(a_0,b_0) = \gamma\;,\ \ f_2(a_0,b_0) = \frac{2-\rho_7-2 \rho_7^3+\rho_7^4+\rho_7^6}{1+\rho_7} \gamma \;,\label{eq:trafo_1} \\
 && f_3(a_0,b_0) = \frac{1+3 \rho_7-\rho_7^2-3 \rho_7^3+5 \rho_7^4-\rho_7^5+\rho_7^6}{2 \left(1-2 \rho_7+2 \rho_7^2+2 \rho_7^4\right)} \gamma \;, \nonumber\\
 && \gamma = \frac{7^{-2/3} \left(1-\rho_7+2 \rho_7^3+2 \rho_7^4+2 \rho_7^5 \right) }{\left(339-337 \rho_7+351 \rho_7^2-347 \rho_7^3+333 \rho_7^4-345 \rho_7^5+349 \rho_7^6\right)^{1/3}}\;.\nonumber
\end{eqnarray}
Numerically, it turns out that $(a_0,b_0)=(1.91002,2.13841)$.

\item \emph{$S_4$ as subgroup of $\Sigma(360\phi)$}: The presentation of $S_4$ is given as above and with generators as before we find,
\begin{equation*}
 \{ F'_2= F Q E F H E F H E F Q \;, F'_3 = E F E F Q E^2 F E H Q H Q\}_{\Sigma(360\phi)}\;.
\end{equation*}
It then turns out that
\begin{equation}
 f_1(a_0,b_0) = 2\delta \;,\ \ f_2(a_0,b_0) = \rho_3 \delta \left(3+\sqrt{5}\right)\;,\ \ f_3(a_0,b_0) = \frac{1}{2} \;,
 \label{eq:trafo_2}
\end{equation}
where $\delta=\frac{1}{4} \sqrt[3]{2-\sqrt{5}} $. Numerically, we obtain $(a_0,b_0)=(1.20978,1.12355)$.

\item \emph{$\Delta(3 \cdot 3^2)$ as subgroup of $\Sigma(360\phi)$}: A presentation of $\Delta(3 \cdot 3^2)$ is given by 
\begin{equation}
\ll F_1, F_2 | F_1^3 = F_2^3 = (F_1 F_2)^3 = F_2 F_1^2 F_2 F_1 F_2^2 F_1^2 F_2^2 F_1 =  \Id{} \gg\;, \nonumber
\end{equation}
with generators as given in Tab.~\ref{tab:SU3_SubGen}. For $\Delta(3 \cdot 3^2)$ and $\Sigma(360\phi)$, 
\begin{equation*}
\{F_2=F(3,0,1)\}_{\Delta(3 \cdot 3^2)}    \;, \qquad \{ F'_2= QH| \}_{\Sigma(360\phi)}   \;.
\end{equation*}
We find 
\begin{equation}
 f_1(a_0,b_0) = \frac{4}{\sqrt[3]{117-3 i \sqrt{15}}} \;,\ \ f_2(a_0,b_0) = \sqrt[3]{\frac{1+ i \sqrt{\frac{5}{3}}}{24} } \;,\ \ f_3(a_0,b_0) = \rho_3^2 f_2(a_0,b_0)\;.
 \label{eq:trafo_3}
\end{equation}
Numerically, it turns out that $(a_0,b_0)=(2.06993,1.86014)$.

\item \emph{$\Delta(6 \cdot 3^2)$ as subgroup of $\Sigma(360\phi)$}: A presentation of $\Delta(6 \cdot 3^2)$ is given by 
\begin{equation}
\ll F_1, F_2 | F_1^3 = F_2^3 = (F_1 F_2)^3 = F_2 F_1^2 F_2 F_1 F_2^2 F_1^2 F_2^2 F_1 =  \Id{} \gg\;, \nonumber
\end{equation}
with generators as given in Tab.~\ref{tab:SU3_SubGen}. For $\Delta(6 \cdot 3^2)$ and $\Sigma(360\phi)$  and 
\begin{equation*}
\{ F_2= E F E^2 \;, F_3 = G) \}_{\Delta(6 \cdot 3^2)}    \;, \qquad \{ F'_2= (EH)^2 E^2 (HQ)^2 \;, F_3 = H )\}_{\Sigma(360\phi)}   \;.
\end{equation*}
We find:
\begin{eqnarray}
 && f_1(a_0,b_0) = \frac{1}{\sqrt[3]{\frac{3}{2} \left(9+11 i \sqrt{3}+3 \sqrt{5}+5 i \sqrt{15}\right)}}\;,\nonumber \\ 
 && f_2(a_0,b_0) = \frac{\rho_3^2}{2\cdot 3^{2/3}} \sqrt[3]{\frac{1}{2} \left(21-9 \sqrt{5}-i \sqrt{6 \left(3-\sqrt{5}\right)}\right)}\;,\nonumber \\
 && f_3(a_0,b_0) = \rho_3^2 \sqrt[3]{\frac{19}{48}+\frac{3 \sqrt{5}}{16}-\frac{1}{2} i \sqrt{\frac{1}{288}-\frac{\sqrt{5}}{864}}}\;.
  \label{eq:trafo_4}
\end{eqnarray}
Numerically, it turns out that $(a_0,b_0)=(2.29215,1.74903)$.

\end{itemize}

\end{itemize}

\section{Tensor generating function}
\label{app:generating}
\setcounter{equation}{0}
\renewcommand{\theequation}{C.\arabic{equation}}

The aim of this appendix is to present the generating function for 
counting covariant tensors in our language. From the latter the branching 
rules can be obtained as shown in Sec.~\ref{app:branching}. 
For a summary on the generating function related to other problems 
in group theory the reader is referred to~\cite{Patera:1978qx}.

The Molien function~\eqref{eq:Molien}, by virtue  of Molien's theorem, counts the number
of invariants of a group in a certain representation ${\cal R}_f(H)$ of a finite group $H$. 
It is a natural to ask whether this can be generalized to count the number of \emph{covariants}. 
By covariants we mean tensors under a certain representation ${\cal R}_c(H)$.

The answer is given by the \emph{(tensor-) generating function}~\cite{burnside}:\footnote{In this language the Molien function is the \emph{invariant-generating function}.}
\begin{equation}
\label{eq:generating}
M_H(\R{c},\R{f};P) = \frac{1}{|{\cal R}_f(h))|} \sum_{h \in H } \frac{\chi_{c}[h]^* }{{\rm det}(\Id{} - P {\cal R}_f(h))} = \sum_{n \geq 0}  c_n P^n\;,
\end{equation}
where $P$ is a real number and $\chi_{c}[h]$, given by
\begin{equation}
\chi_{c}[h] = {\rm tr}[ {\cal R}_c(h)]   \;, 
\end{equation}
is the character of $h$ in the representation $c$. 
It should be emphasized that $f$ and $c$ are irreps. The generating function ought to reduce 
to the Molien function~\eqref{eq:Molien} in the case where $c$ is the trivial irrep,
\begin{equation}
M_H(P) = M(\R{1},{\cal R}_f(h);P)  \;,
\end{equation}
and  does so since  $\chi_{1}[h] = 1$. 
The \emph{generalization of the Molien theorem} states that the 
positive coefficients $c_n$ count the number of linearly independent 
${\cal R}_c(H)$-tensors whose components transform under ${\cal R}_f(H)$. 
The generating function can be written in the following way,
\begin{equation}
M_H(\R{c},\R{f};P)  = \frac{ \sum_i a^{c}_{n_i} \! \cdot\! P^{n_i}}{(1-P^{m_1})(1-P^{m_2})(1-P^{m_3})}\;, 
\end{equation}
in analogy to the form of the Molien function~\eqref{eq:Molien_ex}.
We shall quote here a few facts, assuming that the reader has digested some
of the material on the Molien function presented in the main text: 
\begin{itemize}
\item
There are $a_{n_i}^c$ linearly independent  
${\cal R}_c$-tensors of degree $n_i$, denoted by $E^{(n_i)}(f,c)$.
The entire set $\{E^{(n_i)}  \}$, for all irreps $\R{c}$ and 
$f$,  is known as 
the \emph{integrity basis}.
\item The denominator is the same as for the Molien function~\eqref{eq:Molien_ex},
and thus corresponds to the degrees of the primary invariants rather than covariants.
To appreciate the latter statement, in connection with the generalization of 
the Molien theorem, one has to note that a tensor times
an invariant is  a tensor of the same degree, or that a tensor times a tensor corresponds 
to a tensor of a higher degree.
\item  Knowing the degrees of the tensors, 
one can compute the various tensors by a taking a suitable polynomial ansatz for 
the $\R{c}$-tensor and then demand that its component elements transform 
as $\R{f}$-tensors under the generators ~\cite{Patera:1978qx}.  An example 
of an integrity basis element is given in the next section for the sake of clarity.
\item The generalization of \eqref{eq:proposition} is \cite{analytic},
\begin{equation}
\label{eq:proposition2}
   \sum_i a^c_{n_i}  = 
|\R{c}| \cdot \frac{m_1\cdot m_2 \cdot m_3}{|H|} \;,
\end{equation}
where $|\R{c}|$ is the order of the irrep $\R{c}$. 
We note that the numerator has no $1$ since, for representations
other than the trivial one, the identity is not an ${\cal R}_c$-tensor. 
\item The coefficients $a_{n_i}^c$ satisfy the following symmetry property~\cite{analytic}:
\begin{equation}
a_{n_i}^{c} = a_{n_i'}^{\bar c} \;, \quad \text{for } n_i + n_i' = m_1 + m_2 +m_3 -|\R{f}|  \quad .
\end{equation}
\item The composition laws are as follows \cite{generating_point}:
\begin{eqnarray}
\label{eq:compo}
M( \R{c}_1 + \R{c}_2,\R{f};P) &=& M( \R{c}_1,\R{f};P) \! \cdot \! M( \R{c}_2,\R{f};P)\;, \nonumber \\[0.1cm]
M(\R{c}, \R{f}_1 + \R{f}_2;P)  & = &  \sum_{\R{i}\R{i}'} n_{\R{i}\R{i}'}^{\R{c}} \! \cdot \!
M(\R{i}, \R{f}_1 ;P) \!\cdot \! M(\R{i}', \R{f}_2 ;P)  \;,
 \end{eqnarray}
 where the sum runs over all irreps $\R{c}$ and $\R{c}'$ of the finite group, and 
 $n_{\R{i}\R{i}'}^{\R{c}}$ is the number of times the irrep $\R{c}$  appears in 
 the Kronecker product $\R{i} \times \R{i}'$, which is easily computed from 
 the character table, c.f.\ App.~\ref{app:facts}.
\end{itemize}

Being aware that all of this is rather heavy to digest for the reader we pass on 
to our guinea pig $S_4$ of Sec.~\ref{sec:mainideas}, where some of the properties
mentioned above can be verified explicitly.

\subsection{$S_4$ as an example}

The group $S_4 = \Delta(6 \cdot 2^2)$ has irreps denoted by
$\{ \R{1},\R{1}',\R{2},\R{3},\R{3}'\}$.
The tensor generating functions are easily computed, using formula Eq.~\eqref{eq:generating}:
\begin{eqnarray}
\label{eq:S4g}
M_{S_4}(\R{1},\R{1};P) &=& \frac{1}{1-P}\;, \nonumber \\
M_{S_4}(\R{1},\R{1}';P) &=& \frac{1}{1-P^2}\;, \nonumber \\
M_{S_4}(\R{1},\R{3};P) &=& \frac{1+P^9}{(1-P^2)(1-P^4)(1-P^6)}   = 
M_{S_4}(P)|_{\rm Eq.~\eqref{eq:MolienS4}} \;, \nonumber \\
M_{S_4}(\R{3},\R{3};P) &=& 
\frac{ P^1 +  P^3 +P^4 + P^5 + P^6 + P^8}{(1-P^2)(1-P^4)(1-P^6)}   \;.
\end{eqnarray}
The first two generating functions are concerned with invariants of a one-dimensional
representation space,
\begin{equation}
\I(\R{1},\R{1})[S_4]_1 = x \;, \qquad \I(\R{1},\R{1}')[S_4]_1 = x^2\;, 
\end{equation}
which we have taken to be $x \in \RR$. Note that $\R{1}'$ acts as $x \to -x$.
The third generating function is the Molien function~\eqref{eq:MolienS4} for $S_4$, as
discussed in the main text, and thus we do not need to repeat the discussion here. 
The fourth one is new and we in particular see that the degrees of 
the primary invariants remain the same, as previously stated. Furthermore, the 
property from Eq.~\eqref{eq:proposition2} is verified.
Somewhat arbitrarily we quote, from Ref.~\cite{generating_point}, out of the six the three $\R{3}$-tensors of lowest degree:
\begin{equation}
 E^{(1)}(\R{3},\R{3}) = \left( \begin{array}{c}
	x \\ y  \\ z
	\end{array} \right) \;, \quad 
	E^{(3)}(\R{3},\R{3}) = \left( \begin{array}{c}
	x^3 \\ y^3  \\ z^3
	\end{array} \right)\;,  \quad 
	E^{(4)}(\R{3},\R{3}) = \left( \begin{array}{c}
	(y^2-z^2)yz \\  (z^2-x^2)zx  \\  (x^2-y^2)xy
	\end{array} \right)  \;,
\end{equation}
for the sake of clarity through an  example.
As stated previously, it would be no problem to compute 
them with a suitable ansatz and the Reynolds operator.

\subsection{Branching rules for $SO(3) \to {\cal F}_3$ and $SU(3) \to {\cal F}_3 $ }
\label{app:branching}

The \emph{branching rules}, also known as \emph{correlation tables}, can be computed using 
the character generator~\cite{Patera:1978qx}, but here we shall use 
the method of tensor generating functions presented in~\cite{analytic}.
The problem is the following:
We would like to know how many times the irrep $\R{x}$ is contained in 
the representation $(l)$ or $(p,q)$, respectively, when restrained to
the subgroups $H_{SO(3)} \subset SO(3)$ and $H_{SU(3)} \subset SU(3)$, respectively,
\begin{equation}
\label{eq:model}
\text{\emph{Branching rules:}  } (l)_{SO(3)}  \to (r_l^{\R{x}} \R{x} + ...)_{H_{SO(3)}} \;, \qquad (p,q)_{SU(3)} \to (r_{p,q}^{\R{x}} \R{x} + ...)_{SU(3)}\;\;.
\end{equation}
This follows from the tensor generating functions, see e.g.~\cite{generating_point,analytic},\footnote{The function $B$ has been computed in 
the literature~\cite{analytic} for $\Sigma(168)$, $\Sigma(216\phi)$, 
 and $\Sigma(360\phi)$.}
\begin{eqnarray}
\label{eq:grand}
B(\R{x};l)    & =& (1-L^2) M(\R{x},\R{3},L)  = \sum_{l} r_{l}^{\R{x}}  L^l\;,  \nonumber \\[0.1cm]
B({\R{x}};P,Q) &=& (1-PQ) \sum_{\R{c},\R{c}'} n_{\R{c}\R{c}'}^{\R{m}} M(\R{c},\R{3},P) 
 M(\R{c}',\Rb{3},Q)  = \sum_{p,q} r_{p,q}^{\R{x}}  P^p Q^q \;,
\end{eqnarray}
where the prefactors $(1-L^2)$ and $(1-PQ)$ correspond to the  $O(3)$ 
and $U(3)$ conditions that $x^2+ y^2 + z^2 = \text{constant}$ and $xx^* + yy^* + zz^* = \text{constant}$, respectively. In the second equation in~\eqref{eq:grand}, use of the second composition law in \eqref{eq:compo} has been made. It is worth to note that $M(\R{c},\Rb{3},Q) = M(\Rb{c},\R{3},Q)$, since the generating function is real. Moreover, since the sum extends over all irreps, one may effectively replace $M(\R{c}',\Rb{3},Q) \to M(\R{c}',\R{3},Q)$ in the sum in  Eq.~\eqref{eq:grand}. The positive coefficients $r^{\R{x}}_l$ and $r^{\R{x}}_{p,q}$ give the numbers of linearly independent $\R{x}$-tensors\footnote{More precisely, here, linear independence is understood over the ring of denominator scalars \cite{analytic}.} whose components transform under $l_{SO(3)}$ and $(p,q)_{SU(3)}$ irreps, respectively. Thus they correspond to the multiplicity of the branching in Eq.~\eqref{eq:model}. 
All branching rules\footnote{Except for $SU(3)\to T_{163[58]}$ and $SU(3)\to T_{169[22]}$.} for the groups in our database can 
be obtained from our package \name. Below we shall illustrate the formalism, once more, through $SO(3) \to S_4$.
To this end we would like to add that the functions $B(\R{x};l)$ and  
$B({\R{x}};P,Q) $  can be brought into a form where there are two and five factors
in the denominator~\cite{analytic}, which corresponds the the two and five parameters
that characterize the corresponding representation vector, c.f.\ App.~\ref{app:csh}, Tab.~\ref{tab:SO3SU3}. 

\subsubsection{Examples of branching rules for $SO(3) \to S_4$}

The branching rules of $SO(3) \to S_4$ can be obtained by first identifying 
the $\R{3}_{SO(3)} \to \R{3}_{S_4}$.  The additional necessary generating functions to~\eqref{eq:S4g} are
\begin{eqnarray}
M_{S_4}(\R{1}',\R{3};P) &=& 
\frac{ P^3 +  P^6 }{(1-P^2)(1-P^4)(1-P^6)}\;,   \nonumber \\
M_{S_4}(\R{2},\R{3};P) &=& 
\frac{ P^2 +  P^4 +P^5 + P^7 }{(1-P^2)(1-P^4)(1-P^6)}\;,   \nonumber \\
M_{S_4}(\R{3}',\R{3};P) &=& 
\frac{ P^2 +  P^3 +P^4 + P^5 + P^6 + P^7}{(1-P^2)(1-P^4)(1-P^6)}\;. 
\end{eqnarray} 
Let us consider $l=2 \leftrightarrow \R{5}_{SO(3)}$. The only 
quadratic powers in the Taylor expansions of $(1-P^2) M_{S_4}(\R{c},\R{3};P)$ are $B(\R{2},l) = l^2 + ...$ and $B(\R{3}',l) = l^2 + ...$, 
and thus $5_{SO(3)} \to (\R{2} + \R{3}')_{S_4}$.
Let us quote a few more branching rules so that the reader can assure him- or herself:
\begin{alignat}{2}
& l = 1:  \quad & &   \R{3}_{SO(3)}  \to  \R{3}_{S_4}\;,   \nonumber \\ 
& l = 2: & &   \R{5}_{SO(3)}  \to  (\R{2} + \R{3}' )_{S_4}\;,  \nonumber \\ 
& l = 3: & &   \R{7}_{SO(3)}  \to  (\R{1}'+\R{3} + \R{3}' )_{S_4}\;,  \nonumber \\ 
& l = 4: & &   \R{9}_{SO(3)}  \to (\R{1}+ \R{2} + \R{3} + \R{3}' )_{S_4}\;. 
\end{alignat}
For the branching rules for $SU(3) \to S_4$ we refer the reader to our package \name.

%%%%%%%%%%%%%%%%%%%%%%%%%%%%%%%%%%%%%%%%%%%%%%%%%%%%%%%%%%%%%%%%%%%%%%%%%%%
\section{\label{app:subtle}From the Molien function to invariants in practice}
%%%%%%%%%%%%%%%%%%%%%%%%%%%%%%%%%%%%%%%%%%%%%%%%%%%%%%%%%%%%%%%%%%%%%%%%%%%
\setcounter{equation}{0}
\renewcommand{\theequation}{D.\arabic{equation}}

Ideally we would like infer from  the Molien 
function as given in~\eqref{eq:Molien} to the degrees of  primary and secondary invariants. 
Unfortunately this works only the other way around, as depicted in Eq.~\eqref{eq:Molien_ex}.
In the case where the degrees are not too degenerate, one can 
get the invariants, check their algebraic independence with the Jacobian criterion, 
and then determine the syzygies~\eqref{eq:syzygy}, to be certain that one has obtained the right
primary and secondary invariants. We shall discuss this in more detail below and 
first point towards an ambiguity of the Molien function.

\subsection{A manageable ambiguity of the Molien function }

We simply note that  a Molien function of the form~\eqref{eq:Molien_ex}, can be multiplied
by $(1+P^{m_1})/(1+P^{m_1})$, which leads to 
\begin{equation}
M_{\g{H}(\R{3})}(P) = \frac{1 + \sum_i a_{n_i} P^{n_i}}{(1-P^{m_1})(1-P^{m_2})(1-P^{m_3})} 
=  \frac{(1 + \sum_i a_{n_i} P^{n_i})(1+P^{m_1})}{(1-P^{2m_1})(1-P^{m_2})(1-P^{m_3})} \; ,
\label{eq:Molien_ex2} 
\end{equation}
from where we one could be tempted to infer  that the number of  secondary invariants changes by a factor 
of~$2$, and the product of degrees of primary invariants by a factor of $m_1$.
Supposing the first form was correct, then the second one would only satisfy the proposition 
\eqref{eq:proposition} in the case where $m_1 = 2$. So one has to pay special attention 
only to this case and for our list of subgroups the only invariant of degree two 
is the Euclidian distance~\eqref{eq:P2}. Indeed, a rather manageable ambiguity.

\subsection{Degeneracies}
\label{app:degeneracies}

Let us first note the rules for adding primary and secondary invariants, denoted by $\I$ and
$\Is$ respectively,
\begin{enumerate}
\item $\I_1 + \I_2$ is primary, 
\item $\Is_1 + \Is_2$ is \emph{not} secondary (not primary either),
\item $\Is_1 + \I_2$ is secondary.
\end{enumerate}
It is silently assumed that the degrees match. These rules follow from the definitions of the primary and secondary invariants, c.f.\ in particular~\eqref{eq:syzygy}. We further discuss two examples below to make
 these issues more transparent, of which \ref{app:exampleS4_reynolds} is
of the first type and \ref{app:degenerate_same} concerns types one and three mentioned 
in the list.

\subsubsection{Degeneracies of invariants of lower degrees. }
\label{app:exampleS4_reynolds}

From the Molien function~\eqref{eq:MolienS4} we know
that there is an invariant polynomial of degree four, for example $\I_4[S_4]$  given in~\eqref{eq:IS4}. 
The trial function $f(x,y,z) \in \{x^4, y^4, z^4\}$, using the Reynolds operator~\eqref{eq:genI}, will lead to this invariant.
A generic trial function leads to an invariant $\I_4' = a \I_2^{\,2} + b \I_4$. The choice
of $(a,b) \in \C^2$ corresponds to the choice of a basis and is arbitrary. We have made
the particular choice $(a,b)= (0,1)$.  
Thus whenever the sum of degrees of lower
invariants  equal the degree of an invariant in question there is an ambiguity in the choice.

\subsubsection{Degeneracies of invariants of the same degree }
\label{app:degenerate_same}

A prime example is the case of $\Delta(6\times 3^2)$. 
Note that in practice 
this example is doable as the degeneracies for $\Delta(6 n^2)|_{n \neq 3}$ are lifted, and we
may guess the primary and secondary invariants on ground of ``analytic continuation" in $n$, as discussed in Sec.~\ref{sec:Del36}. 

We shall discuss it without this trick for the sake of the example.
A Molien function of the following form can be found:
\begin{equation}
 M_{\Delta(6\times 3^2)}(P) = \frac{1+P^6 +P^9 + P^{15}}{(1-P^6)^3}\;.
 \label{eq:MolDel66new}
\end{equation}
Accordingly, we would expect 4 invariants of degree 6. Possible choices are:
\begin{equation}
 \I_{6a} =(x y z)^2,\ \I_{6b} = x^3 y^3 + y^3 z^3 + z^3 x^3, \I_{6c} = x^6  + y^6  + z^6,\ \I_{6d} =  x y z (x^3 + y^3 + z^3).
 \label{eq:MolDel66new}
\end{equation}
Using, e.g., the Jacobian criterion~\eqref{eq:Jac_Del3}, the algebraic independence of any three of them
is readily verified.
In order to find primary and secondary invariants, the syzygies~\eqref{eq:syzygy} have to be found.
For this question we can disregard the invariants of higher degree for the moment, since 
their degrees are too high to play a role,
 $9+6=15>2\cdot 6=12$.

As shown in Sec.~\ref{sec:Del36}, $\{ \I_{6a}, \I_{6b}, \I_{6c} \}$ are primary invariants. It may be instructive 
to see why or how  $\{ \I_{6a}, \I_{6b}, \I_{6d} \}$ fail to be primary invariants:
The left-hand side of the syzygy $\I_{6c}^2$ has got a term of the form $x^{12}$, but this term can never be obtained by multiplying any two of the invariants $\{ \I_{6a}, \I_{6b}, \I_{6d} \}$. We hope that this example is useful to the reader and the practitioner.

\section{Multiple representations }
\label{app:flavons}
\setcounter{equation}{0}
\renewcommand{\theequation}{E.\arabic{equation}}

Our setup could be generalized to include multiple spin-$0$ fields $\varphi_i$ and $\phi_j$, possibly carrying different representations $\rep{SU(3)}$. 
In models with flavour symmetries such fields are referred to as \emph{flavons}~\cite{Altarelli:2010gt}. In our view  two new features arise as opposed to a single spin-$0$ field. Consider the interaction of the SM with the flavon  sector,
\begin{equation}
\label{eq:Lflavon}
{\cal L} =   F_a O_{\rm SM}^a \;,
\end{equation}
where summation over repeated indices is understood, and 
$O_{\rm SM}^a$ consists of SM fields only.  The index $a$ is an index of a representation of the
flavour symmetry group $SU(3)$. In the case where we intended to be more complete we should
also sum over all irreps of the flavour group in the equation above.
In Eq.~\eqref{eq:Lflavon}, the complexity remains in the composite fields $F_b$, which can be written as follows:
\begin{equation}
F_a = \sum_{n,m} \frac{c_{mn}}{\Lambda^{n+m-4}} T_a^{i_1 ..i_n j_1  ..j_m} \varphi_{i_1} .. \varphi_{i_n} 
\phi_{j_1} .. \phi_{j_m} \;,
\end{equation}
where $\Lambda$ is some generic suppression scale and $c_{nm}$ are coefficients 
of order one. The new elements are first that more $a$-covariant objects in $F_b$ can 
be formed, since two antisymmetric indices do not vanish under contraction of 
$\varphi$ and $\phi$ and second that the relative direction of the VEV of the two fields does matter.
In connection with the latter, suppose the two fields were in  irreps which are complex conjugate to each other. 
Then ${\cal L} = m^2 \varphi^a  \phi_a = m^2 \varphi \cdot \phi$ is not invariant to 
separate rotations of the fields $\varphi$ and $\phi$.
One speaks of \emph{vacuum alignment}. 
Thus, by combining the two fields in  one potential, one can enforce rich patterns of
flavour symmetry breaking, which have the potential to shine light on the hierarchies in the flavour
sector.  We would like to add to this end that the generating function
as discussed in App.~\ref{app:generating} constitutes a powerful tool in tackling this problem in the most general way.

%%%%%%%%%%%%%%%%%%%%%%%%%%%%%%%%%%%%%%%%%%%%%%%%%%%%%%%%%%%%%%%%%%%%%%%%%%%
\section{\label{app:T-conj}Conjectures concerning the $\boldsymbol{T_{n[a]} }$-groups }
%%%%%%%%%%%%%%%%%%%%%%%%%%%%%%%%%%%%%%%%%%%%%%%%%%%%%%%%%%%%%%%%%%%%%%%%%%%
\setcounter{equation}{0}
\renewcommand{\theequation}{F.\arabic{equation}}

Contrary to $\Delta(3n^2)$ and $\Delta(6n^2)$, we have not been able to find the first and 
second primary invariants of the $T_{n[a]}$ groups in full generality. As for the latter we can start to guess the 
primary invariants on grounds of the examples in our database. Our guesses are:
\begin{eqnarray}
 \I_3 &=& x y z\;,\nonumber\\
 \I_{2a+1} &=& x^{a+1} y^{a} + y^{a+1} z^{a} + z^{a+1} x^{a}\;,\nonumber\\
 \I_n &=& x^n + y^n + z^n\;.
 \label{eq:Prim_Inv_Tn}
\end{eqnarray}
They are invariant under the $T_{n[a]}$ generators $E$ and $F(n,1,a),$\footnote{Recall that 
$n$ are the primes out of $3k+1$ where $k$ is an integer.} and they
are also algebraically independent. 

If we assume that~\eqref{eq:Prim_Inv_Tn} are the correct primary invariants then,
by virtue of proposition~\eqref{eq:proposition}, the number of secondary invariants is
given by:
\begin{equation}
\frac{3\cdot n \cdot (2a+1)}{3n} = 2a+1\;. 
\end{equation}
Furthermore, from the examples in our database we are led to conjecture 
the following patterns:
\begin{enumerate}

\item If the secondary invariants are put in ascending order of their respective degrees, then the degree of the $(2a)$-th invariant is
\begin{equation}
{\rm deg} \left( \overline{\I}^{(2a)} \right)=n+2a+1={\rm deg}(\I_n)+{\rm deg}(\I_{2a+1})\;.
\end{equation}
Note that we denote the $k$-th secondary invariant by $\overline{\I}^{(k)}$, starting with $\Is^{(0)} = 1$.

\item The degree of the $a$-th invariant is
\begin{equation}
{\rm deg} \left( \overline{\I}^{(a)} \right)=\frac{1}{2}(n+2a+1)=\frac{1}{2} [{\rm deg}(\I_n)+{\rm deg}(\I_{2a+1})]\;.
\end{equation}

\item If $m$ is the degree of the first invariant, then the degree of the $(2a-1)$-th invariant is given by $(n+2a+1-m)$.

\item For $k=1, 2, ..., a-1$ it holds that:
\begin{equation}
{\rm deg} \left( \overline{\I}^{(a)} \right) - {\rm deg} \left( \overline{\I}^{(k)} \right) = {\rm deg} \left( \overline{\I}^{(a + k)} \right) - {\rm deg} \left( \overline{\I}^{(a)} \right).
\end{equation}

\end{enumerate}
We hope that these observations may help to solve out this problem in future studies.

\section{$\boldsymbol{SU(3)}$}
\setcounter{equation}{0}
\renewcommand{\theequation}{G.\arabic{equation}}

\subsection{The complex spherical harmonics }
\label{app:csh}

A widely used method to construct irreps is the method of highest weights, as advocated 
in many textbooks~\cite{groups}. For our purposes it is more convenient to work in an explicit
basis. For $SO(3)$, explicit representations in terms of spherical harmonics 
$Y_{l,m}$ are well known as the representations of the Lie algebra elements directly relate to coordinate and momentum
representations in quantum mechanics. What are the spherical harmonics of $SU(3)$? It appears that this question was explicitly studied in the late sixties, in connection with 
the eightfold way~\cite{solidSU3}, and the corresponding representation functions are known as the \emph{complex spherical harmonics}. 
A thorough mathematical treatment of so-called \emph{solid $SU(n)$ harmonics} can be found 
in the book of Louck~\cite{Louck}. In this appendix we shall present the material in a 
rudimentary way, relying on the analogy to $SO(3)$ and  the spherical harmonics.

The spherical harmonics $Y_{l,m}$ 
are the solutions of the Laplace equation on the two-sphere 
$S_2$. This can be seen to originate from the quotient of $SO(3)$ with 
the stabilizer of a representive vector, which is $SO(2)$. In analogy one gets,
\begin{equation}
\label{eq:S5}
S_2 \simeq SO(3)/SO(2) \;, \qquad S_5 \simeq SU(3)/SU(2) \;,
\end{equation}
the group manifold for the complex spherical harmonics. 
The five-sphere can be embedded into $\C^3$. 
Thus there will be five parameters as opposed to two, $m$ and $l$, 
for $Y_{l,m}$. 

The same conclusions can be reached in a way which parallels 
the introduction of the spherical harmonics in quantum mechanics and possibly 
justifies the name complex spherical harmonics best~\cite{Ikeda}. Consider 
complex coordinates $(z_1,z_2,z_3) \in \C^3$, and the Laplace equation:\footnote{In this section only, honouring the standard notation of complex analysis, we use $\bar{\phantom{}}$ to denote 
the complex conjugate instead of the $^*$-symbol.}
\begin{equation}
\left( \frac{\partial^2}{\partial z_1  \partial  \bar z_1} +  \frac{\partial^2}{ \partial z_2\partial  \bar z_2}  +  \frac{\partial^2}{\partial  z_3 \partial  \bar z_3} \right) f = 0\;.
\end{equation}
Let $f_{(p,q)}$ be a polynomial solution of degree $p$ and $q$ in 
$z_i$ and $\bar z_i$, respectively, then 
$$f_{(p,q)}(z,\bar z) = \rho^{(p+q)} h_{(p,q)}(z,\bar z) \;,  \qquad 
\rho^2 \equiv \bar z_1 z_1 + \bar z_2 z_2 + \bar z_3 z_3 \;, $$
where $h_{(p,q)}(z,\bar z)$ is a complex spherical harmonic of order $(p,q)$.
The number of linearly independent $h_{(p,q)}$ is 
$\frac{1}{2} (p+1)(q+1)(p+q+2)$. All $SU(3)$ irreps can be generated in this way. 
If real coordinates are chosen,  $\vec{z} \in \mathbb{R}^3$, the discussion reduces
to the spherical harmonics of $SO(3)$.

An orthogonal basis can be obtained from the following generating function:
\begin{eqnarray}
\label{eq:SU(3)_GF}
G(a_1,a_2,b) &=& (\bar z_1- a_1 \bar z_2)^{-q-1} (\bar z_1-a_1 \bar z_2 -a_2
   \bar z_3)^{p+1}  \times  \nonumber \\
   & & (b (a_1 z_1 +z_2 ) (\bar z_1 -a_1
   \bar z_2-a_2 \bar z_3)+ z_3  (\bar z_1-a_1 \bar z_2)+a_2
   (z_1 \bar{z_1}+z_2 \bar{z_2}))^q   \nonumber  \\
  &=&  \sum_{r=0}^q \sum_{s=0}^{p+q+1} \sum_{t = 0}^\infty h^{rst}_{(p,q)} a_1^t a_2^s b^r  \;.
\end{eqnarray}
$ h^{rst}_{(p,q)}$ is an orthogonal basis for a $(p,q)$-representation whose states are characterized by the labels $(r,s,t)$ ranging from:
\begin{equation}
\label{eq:rst}
r = 0..q \;, \quad s=0..p \;,\quad  t = 0..(p+r-s) \;.
\end{equation}
It is readily verified that $r,s,t$ sums over $\frac{1}{2} (p+1)(q+1)(p+q+2)$ elements.
The parameters $p,q,r,s,t$ correspond to the five parameters of the five-sphere~\eqref{eq:S5}.  In our work we adapt the phase convention which follows from 
\eqref{eq:SU(3)_GF}. An alternative convention based on isospin has been suggested in Ref.~\cite{solidSU3}.
 For the readers convenience, we give a summary of some basic facts in Tab.~\ref{tab:SO3SU3}, in comparison of $SO(3)$ and $SU(3)$.
 \begin{table}[h]
\begin{center}
\begin{tabular}{l | l | l  }
group   & $SO(3)$ & $SU(3)$ \\
\hline
rank     & $1 \leftrightarrow l$ & $2 \leftrightarrow (p,q)$ \\
repres.\ fct.  & $Y_{l,m}$ & $h_{(p,q)}^{rst}$ \\
fct.\ on manifold & $SO(3)/SO(2) \simeq S_2$ & $SU(3)/SU(2) \simeq S_5$ \\
embedding   & $\emb \RR^3$ with $x^2+y^2 + z^2 = r^2$   & 
$\emb \C^3$ with $z_1 \bar z_1 + z_2 \bar z_2 + z_3  \bar z_3  = \rho^2$ \\
labelling irrep & $(l) \in \N_0$    &  $(p,q) \in \N_0^2$  \\ 
dim(irrep)  & $(2l+1)$ & $(p\!+\!1)(q\!+\!1)(p\!+\!q\!+\!2)/2$ \\
labelling states irrep & $m =  -l .. l$   & $r = 0..q \;, \; s=0..p \;,\;  t = 0..(p+r-s)$ \\
 \end{tabular}
\end{center}
\caption{\small Comparison of $SO(3)$ vs.\ $SU(3)$ data. The acronym ``fct'' stands for function.} 
\label{tab:SO3SU3}
\end{table}

\subsection{\label{app:pqBasis}Construction of explicit $(p,q)$ representations }

\subsubsection{Polyomial basis}
\label{app:Pbasis}

We have stated that $h_{(p,q)}$ are polynomials of degree $p$ and $q$ 
in the variables $z_i$ and $\bar z_i$. We shall denote such a space by 
${\cal H}_{(p,q)}$. As an example let us quote
\begin{equation}
x^2 y^3 z \in {\cal H}_{(6,0)}\;.
\end{equation}
The dual vector is given by
\begin{equation}
(x^2 y^3 z )^\dagger \equiv ( | 2 3 1 000 \rangle )^\dagger = 
\langle 2 3 1 000 | \equiv   \partial_x^2  \partial_y^3 \partial_z \;,
\end{equation}
where the association with bra and ket should be obvious.
The normalization then follows:
\begin{equation}
\langle 231 000 | 231000 \rangle = 2! 3! 1! \quad \Rightarrow \quad 
|abcdef \rangle_N = \frac{1}{\sqrt{a!b!c!d!e!f!}} |abcdef \rangle\;.
\end{equation}
The entire space is spanned by ${\cal H} = \oplus_{p \geq 0;,q \geq 0}  {\cal H}_{(p,q)}$,
and the identity  on ${\cal H}_{(p,q)}$ is represented as $\mathbf{1}_{(p,q)} 
=\frac{1}{(p+q)} (x \partial_x + y \partial_y 
+ z \partial_z+ \bar x \partial_{\bar{x}}
 + \bar y \partial_{\bar{y}} 
+ \bar{z} \partial_{\bar{z}})$.

\subsubsection{Gell-Mann basis on polynomial space}
Noting that the fundamental representation space $(1,0)$ 
in the polynomial basis is given by $\{x,y,z\}$, 
the Gell-Mann operators of the $SU(3)$ Lie-algebra are readily read off:
\begin{eqnarray}
 & & {\cal B}^{(1,0)}_{GM} = \{T_1, T_2, ..., T_8\}  =  \\[0.1cm]
 &   & \!\!\! \frac{1}{2} \{   
 ( y \partial_x + x \partial_y), i ( y \partial_x - x \partial_y), 
 2T_3, i ( z \partial_y - y \partial_z)
( z \partial_x + x \partial_z) ,  ( z \partial_y + y \partial_z), 
 i ( z \partial_x - x \partial_z),  2T_8 \nonumber
   \} \;,
\end{eqnarray}
with the Cartan sub algebra,
\begin{equation}
\label{eq:T3T8_2}
T_3 = \frac{1}{2} ( x\partial_x - y\partial_y)  \;, \quad T_8 =  \frac{1}{2} \frac{1}{\sqrt{3}} ( x\partial_x +  y\partial_y - 2 z\partial_z) \;.
\end{equation}
The Gell-Mann matrices satisfy the $SU(3)$ Lie-algebra relations:
\begin{equation}
\label{eq:LA}
[T_a,T_b] = i f_{abc} T_c\;,
\end{equation}
with $f_{123} = 1$ for example. The basis ${\cal B}^{(1,0)}_{GM}$ works 
on ${\cal H}_{(p,0)}$ space, but it does not act on ${\cal H}_{(0,q)}$ space.
We must therefore construct ${\cal B}^{(0,1)}_{GM}$. This follows by complex 
conjugation,
\begin{equation}
{\cal B}^{(0,1)}_{GM} =  - ({\cal B}^{(1,0)}_{GM})^* \;,
\end{equation}
where the extra minus sign stems from the fact that an extra factor of $i$ comes in 
when the representation is exponentiated, $\exp(i v^a T^a)$. Then, 
\begin{equation}
 {\cal B}_{GM}  =  {\cal B}^{(1,0)}_{GM} +  {\cal B}^{(0,1)}_{GM} 
\end{equation}
is a basis that gives all $(p,q)$ representations:
\begin{equation}
\label{eq:pq-irrep}
[(T_i)_{(p,q)}]_{kl} =    \langle k | T_i | l \rangle \;,
\end{equation}
where $T_i \in  {\cal B}_{GM}$, and $ | l \rangle$ corresponds 
to  $ |ab (p\!-\!a\!-\!b) de (q\!-\!d\!-\!e) \rangle_N \in {\cal H}_{(p,q)}$ and is understood
to be an orthonormal basis. 
Note that we have taken into account 
the degree of the polynomial state, which constrains the third and sixth entries with 
 $p$ and $q$. We have verified this construction for many examples, 
 and we have also verified the Dynkin index,
 \begin{equation}
 {\rm Tr}[ (T_a)_{(p,q)} (T_b)_{(p,q)} ] = k_{(p,q)} \delta_{ab} \;,
 \end{equation}
 which can be computed using Racah's formula~\cite{Slansky}. 
 A few examples are: $k_{(1,0)} = 1/2$, $k_{(2,0)} = 5/2$, $k_{(1,1)} = 3$, $k_{(3,0)} = 15/2$, $k_{(2,1)} = 10$, $k_{(4,0)} = 35/2$.  
 The symbol  $\delta_{ab}$ corresponds to the well-known Kronecker-symbol.

Thus, given a normalized basis which is not hard to obtain, the $(p,q)$-irreps
can be computed in an extremely efficient way. By virtue of the explicitness of 
the differential-polynomial representation the normalization factors, 
which are obtained in the abstract highest weight method by solving a set of equations~\cite{groups}, are quasi-free or result from simple differentation 
of polynomials.
We have implemented the construction \eqref{eq:pq-irrep} in our package \name.


\begin{thebibliography}{199}


\bibitem{tribi}
  P.~F.~Harrison, D.~H.~Perkins, W.~G.~Scott,
  ``Tri-bimaximal mixing and the neutrino oscillation data,''
  Phys.\ Lett.\  {\bf B530} (2002)  167 [hep-ph/0202074].

\bibitem{FlavourModels}

 \emph{An incomplete list:}
 F.~Caravaglios, S.~Morisi,
  %``Neutrino masses and mixings with an S(3) family permutation symmetry,''
  arXiv:hep-ph/0503234. 
  W.~Grimus, L.~Lavoura,
  %``S(3) x Z(2) model for neutrino mass matrices,''
  JHEP {\bf 0508} (2005) 013 [arXiv:hep-ph/0504153].
 G.~Altarelli, F.~Feruglio,
  %``Tri-bimaximal neutrino mixing from discrete symmetry in extra dimensions,''
  Nucl.\ Phys.\  B {\bf 720} (2005) 64 [arXiv:hep-ph/0504165].
 W.~Grimus, L.~Lavoura,
  %``A Model realizing the Harrison-Perkins-Scott lepton mixing matrix,''
  JHEP {\bf 0601} (2006) 018 [arXiv:hep-ph/0509239].
 G.~Altarelli, F.~Feruglio,
  %``Tri-bimaximal neutrino mixing, A(4) and the modular symmetry,''
   Nucl.\ Phys.\  B {\bf 741} (2006) 215 [arXiv:hep-ph/0512103].
 H.~K.~Dreiner, C.~Luhn, M.~Thormeier,
 %``What is the discrete gauge symmetry of the MSSM?,''
 Phys.\ Rev.\  D {\bf  06} 075007 [arXiv:hep-ph/0512163].
 I.~de Medeiros Varzielas, S.~F.~King, G.~G.~Ross,
  %``Tri-bimaximal neutrino mixing from discrete subgroups of SU(3) and SO(3)
  %family symmetry,''
  Phys.\ Lett.\  B {\bf 644} (2007) 153 [arXiv:hep-ph/0512313].
 C.~Hagedorn, M.~Lindner, R.~N.~Mohapatra,
  %``S(4) flavor symmetry and fermion masses: Towards a grand unified theory of
  %flavor,''
  JHEP {\bf 0606} (2006) 042 [arXiv:hep-ph/0602244].
 S.~F.~King, M.~Malinsky,
  %``A(4) family symmetry and quark-lepton unification,''
  Phys.\ Lett.\  B {\bf 645} (2007) 351 [arXiv:hep-ph/0610250].
 S.~Morisi, M.~Picariello, E.~Torrente-Lujan,
  %``Model for fermion masses and lepton mixing in SO(10) x A(4),''
  Phys.\ Rev.\  D {\bf 75} (2007) 075015 [arXiv:hep-ph/0702034].
 C.~Luhn, S.~Nasri, P.~Ramond,
  %``Tri-bimaximal neutrino mixing and the family symmetry semidirect product of
  %Z(7) and Z(3),''
  Phys.\ Lett.\  B {\bf 652} (2007) 27 [arXiv:0706.2341 [hep-ph]].
 F.~Bazzocchi, S.~Kaneko, S.~Morisi,
  %``A SUSY A(4) model for fermion masses and mixings,''
  JHEP {\bf 0803} (2008) 063 [arXiv:0707.3032 [hep-ph]].
 G.~Altarelli, F.~Feruglio, C.~Hagedorn,
  %``A SUSY SU(5) Grand Unified Model of Tri-Bimaximal Mixing from A(4),''
  JHEP {\bf 0803} (2008) 052 [arXiv:0802.0090 [hep-ph]].
 F.~Feruglio, C.~Hagedorn, Y.~Lin, L.~Merlo,
  %``Lepton Flavour Violation in Models with A(4) Flavour Symmetry,''
  Nucl.\ Phys.\  B {\bf 809} (2009) 218 [arXiv:0807.3160 [hep-ph]].
 F.~Bazzocchi, M.~Frigerio, S.~Morisi,
  %``Fermion masses and mixing in models with SO(10) x A(4) symmetry,''
  Phys.\ Rev.\  D {\bf 78} (2008) 116018 [arXiv:0809.3573 [hep-ph]].
 F.~Bazzocchi, S.~Morisi,
  %``S(4) as a natural flavor symmetry for lepton mixing,''
  Phys.\ Rev.\  D {\bf 80} (2009) 096005 [arXiv:0811.0345 [hep-ph]].
 F.~Bazzocchi, L.~Merlo, S.~Morisi,
  %``Fermion Masses and Mixings in a S(4)-based Model,''
  Nucl.\ Phys.\  B {\bf 816} (2009) 204 [arXiv:0901.2086 [hep-ph]].
 F.~Bazzocchi, L.~Merlo, S.~Morisi,
  %``Phenomenological Consequences of See-Saw in S(4) Based Models,''
  Phys.\ Rev.\  D {\bf 80} (2009) 053003 [arXiv:0902.2849 [hep-ph]].
 M.~C.~Chen, S.~F.~King,
  %``A4 See-Saw Models and Form Dominance,''
  JHEP {\bf 0906} (2009) 072 [arXiv:0903.0125 [hep-ph]].
 G.~Altarelli, F.~Feruglio, L.~Merlo,
  %``Revisiting Bimaximal Neutrino Mixing in a Model with S(4) Discrete
  %Symmetry,''
  JHEP {\bf 0905} (2009) 020 [arXiv:0903.1940 [hep-ph]].
 G.~Altarelli, D.~Meloni,
  %``A Simplest A4 Model for Tri-Bimaximal Neutrino Mixing,''
  J.\ Phys.\ G {\bf 36} (2009) 085005 [arXiv:0905.0620 [hep-ph]].
 W.~Grimus, L.~Lavoura, P.~O.~Ludl,
  %``Is S(4) the horizontal symmetry of tri-bimaximal lepton mixing?,''
  J.\ Phys.\ G {\bf 36} (2009) 115007 [arXiv:0906.2689 [hep-ph]].
 F.~Feruglio, C.~Hagedorn, L.~Merlo,
  %``Vacuum Alignment in SUSY A4 Models,''
  JHEP {\bf 1003} (2010) 084 [arXiv:0910.4058 [hep-ph]].
 S.~F.~King, C.~Luhn,
  %``A Supersymmetric Grand Unified Theory of Flavour with PSL(2)(7) x SO(10),''
  Nucl.\ Phys.\  B {\bf 832} (2010) 414 [arXiv:0912.1344 [hep-ph]].
 C.~Hagedorn, S.~F.~King, C.~Luhn,
  %``A SUSY GUT of Flavour with S4 x SU(5) to NLO,''
  JHEP {\bf 1006} (2010) 048 [arXiv:1003.4249 [hep-ph]].
 C.~Hagedorn, M.~Serone,
  %``Leptons in Holographic Composite Higgs Models with Non-Abelian Discrete
  %Symmetries,''
  arXiv:1106.4021 [hep-ph].
 R.~d.~A.~Toorop, F.~Feruglio, C.~Hagedorn,
  %``Discrete Flavour Symmetries in Light of T2K,''
  Phys.\ Lett.\  B {\bf 703} (2011) 447 [arXiv:1107.3486 [hep-ph]].
 S.~F.~King, C.~Luhn,
  %``Trimaximal neutrino mixing from vacuum alignment in A4 and S4 models,''
  JHEP {\bf 1109} (2011) 042 [arXiv:1107.5332 [hep-ph]].
  
  
%
\bibitem{Buchmuller:2008uq}
  W.~Buchm{\"u}ller, J.~Schmidt,
  %``Higgs versus Matter in the Heterotic Landscape,''
  Nucl.\ Phys.\  {\bf B807} (2009)  265-289.
  [arXiv:0807.1046 [hep-th]].
  
  \bibitem{Kobayashi:2006wq}
  T.~Kobayashi, H.~P.~Nilles, F.~Ploger, S.~Raby, M.~Ratz,
  %``Stringy origin of non-Abelian discrete flavor symmetries,''
  Nucl.\ Phys.\  B {\bf 768} (2007) 135
  [arXiv:hep-ph/0611020].
  %%CITATION = NUPHA,B768,135;%%
  

\bibitem{Adulpravitchai:2009kd}
  A.~Adulpravitchai, A.~Blum, M.~Lindner,
  %``Non-Abelian Discrete Groups from the Breaking of Continuous Flavor
  %Symmetries,''
  JHEP {\bf 0909} (2009) 018 [arXiv:0907.2332 [hep-ph]].
  %%CITATION = JHEPA,0909,018;%%
  
  
\bibitem{burnside}
W.~Burnside, ``Theory of groups of finite order,'' Cambridge University Press, second edition, 1897.

\bibitem{noether}
E.~Noether, ``Der Endlichkeitssatz der Invarianten endlicher Gruppen,'' Math. Ann. {\bf 77} (1916).

\bibitem{sturmfels}
B.~Sturmfels, 
``Algorithms in Invariant Theory,''
Texts and Monographs in Symbolic Computation, Springer.

\bibitem{Meyer}
 B.~Meyer, ``On the symmetries of spherical harmonics,'' Canad. J. Math. {\bf 6} (1954).

%\bibitem{McLellan}
%A.~G.~McLellan, J. Phys. C: Solid St. Phys. 7, 3326-40, 1974.


\bibitem{turkey}
  M.~Koca, M.~Al-Barwani and R.~Koc,
  ``Breaking SO(3) into its closed subgroups by Higgs mechanism,''
  J.\ Phys.\ A  {\bf 30} (1997) 2109.
  %%CITATION = JPAGB,A30,2109;%%




%\cite{Patera:1978qx}
\bibitem{Patera:1978qx}
  J.~Patera, R.~T.~Sharp,
  ``Generating Functions For Characters Of Group Representations And Their Applications''
  (Lecture Notes in Physics 94), New York: Springer, pp. 175Ð83.
  
  

  
\bibitem{Molien}
Th.~Molien, ``{\"U}ber die Invarianten der linearen Substitutionsgruppen,'' 
Sitzungber. Konig. Preuss. Akad. Wiss. (J. Berl. Ber.) {\bf 52} (1897): 1152-1156. 

\bibitem{permutationgroups}
J.~D.~Dixon, B.~Mortimer, ``Permutation Groups", Graduate Text in Mathematics, Springer, 1996.

%\cite{O'Raifeartaigh:1986vq}
\bibitem{O'Raifeartaigh:1986vq}
  L.~O'Raifeartaigh,
  ``Group Structure Of Gauge Theories,''
  Cambridge, UK: Univ. Pr. (1986) 172 P. (Cambridge Monographs On Mathematical Physics).
  
\bibitem{Michel}
  L.~Michel,
  ``Symmetry Defects And Broken Symmetry. Configurations - Hidden Symmetry,''
  Rev.\ Mod.\ Phys.\  {\bf 52} (1980) 617-651.
  %%CITATION = RMPHA,52,617;%%

\bibitem{counter}
M.~J.~Linehan, G.~E.~Stedman,
 ``Little groups of irreps of $O(3)$, $SO(3)$, and the infinite axial subgroups,''
Journal of Physics A: Mathematical and General, Volume 34, Issue 34, pp. 6663-6688 (2001).


  
\bibitem{Ludl:2010bj}
  P.~O.~Ludl,
  ``On the finite subgroups of $U(3)$ of order smaller than 512,''
  J.\ Phys.\ A  {\bf 43} (2010) 395204
  [Erratum-ibid.\  A {\bf 44} (2011) 139501]
  [arXiv:1006.1479 [math-ph]].
  %%CITATION = JPAGB,A43,395204;%%


\bibitem{Milleretal}
Chapter XII in 
G.~A.~Miller, H.~F.~Blichfeldt, L.~E.~Dickson,  ``Theory and Applications of Finite Groups", John 
Wiley \& Sons, New York, 1916, and Dover Edition, 1961.

\bibitem{MichelPR}
L.~Michel, B.~I.~Zhilinskii,
 ``Symmetry, Invariants, and Topology. I. Basic Tools". 
Phys. Rep. 341, 11-84 (2001).


%\cite{Parattu:2010cy}
\bibitem{Parattu:2010cy}
  K.~M.~Parattu, A.~Wingerter,
  %``Tribimaximal Mixing From Small Groups,''
  Phys.\ Rev.\  {\bf D84} (2011) 013011
  [arXiv:1012.2842 [hep-ph]].



%\cite{Ishimori:2010au}
\bibitem{Ishimori:2010au}
  H.~Ishimori, T.~Kobayashi, H.~Ohki, Y.~Shimizu, H.~Okada, M.~Tanimoto,
  ``Non-Abelian Discrete Symmetries in Particle Physics,''
  Prog.\ Theor.\ Phys.\ Suppl.\  {\bf 183 } (2010)  1-163
  [arXiv:1003.3552 [hep-th]].


\bibitem{response}
  W.~M.~Fairbairn, T.~Fulton,
  ``Some Comments On Finite Subgroups Of $SU(3)$,''
  J.\ Math.\ Phys.\  {\bf 23} (1982) 1747.
  %%CITATION = JMAPA,23,1747;%%




%\cite{Bovier:1980ga}
\bibitem{Bovier:1980ga}
  A.~Bovier, M.~Luling, D.~Wyler,
  ``Representations And Clebsch-gordan Coefficients of Z Metacyclic Groups,''
  J.\ Math.\ Phys.\  {\bf 22} (1981) 1536.
  
  

\bibitem{FFK}
W.~M.~Fairbairn, T.~Fulton, W.~H.~Klink,
``Finite and Disconnected Subgroups of
$SU(3)$ and their Application to the Elementary-Particle Spectrum,"
 J. Math. Phys. 5: 1038, 1964.



\bibitem{GAP}
The GAP~Group, \emph{GAP -- Groups, Algorithms, and Programming, Version 4.4.12}, 2008,
\\
\url{http://www.gap-system.org}

\bibitem{SmallGroups}
H.~U.~Besche, B.~Eick and E.~A.~O'Brien, \textit{SmallGroups - a GAP package}, 2002,
\\
\url{http://www.gap-system.org/Packages/sgl.html}
\\
\url{http://www-public.tu-bs.de:8080/~beick/soft/small/small.html}



\bibitem{Jacobi}
Chapter 3 in J.~E.~Humphreys, ``Reflection groups and Coxeter groups,'' Cambridge University Press, Cambridge, 1990.




\bibitem{Ludl:2011gn}
  P.~O.~Ludl,
  ``Comments on the classification of the finite subgroups of $SU(3)$,''
  J.\ Phys.\ A  {\bf 44} (2011) 255204
  [arXiv:1101.2308 [math-ph]].
  %%CITATION = JPAGB,A44,255204;%%
  
\bibitem{GL11}
  W.~Grimus and P.~O.~Ludl,
  ``Finite flavour groups of fermions,''
  arXiv:1110.6376 [hep-ph].
  %%CITATION = ARXIV:1110.6376;%%

\bibitem{Ludl_diplom}
  P.~O.~Ludl,
  ``Systematic analysis of finite family symmetry groups and their application
  to the lepton sector``,
  arXiv:0907.5587 [hep-ph].
  %%CITATION = ARXIV:0907.5587;%%

  
  
  
 \bibitem{DMFV}
  R.~Zwicky, T.~Fischbacher,
  ``On discrete Minimal Flavour Violation,''
  Phys.\ Rev.\  D {\bf 80} (2009) 076009
  [arXiv:0908.4182 [hep-ph]].
  %%CITATION = PHRVA,D80,076009;%%



\bibitem{Luhn:2011ip}
  C.~Luhn,
  ``Spontaneous breaking of $SU(3)$ to finite family symmetries: a pedestrian's
  approach,''
  JHEP {\bf 1103} (2011) 108
  [arXiv:1101.2417 [hep-ph]].
  %%CITATION = JHEPA,1103,108;%%

\bibitem{conformal}
  P.~Di Francesco, P.~Mathieu, D.~Senechal,
  ``Conformal field theory,''
  New York, USA: Springer (1997) 890 p.
  
 \bibitem{Ramond}
 P.~Ramond,``Group Theory in Physics - A physicists survey," CUP, 2010.
  
\bibitem{Cornwell}
J.~F.~Cornwell, ``Group Theory in Physics'' (Three volumes), Volume 1, Academic Press, New York (1997).

\bibitem{Specht}
W.~Specht, ``Zur Theorie der Gruppen linearer Substitutionen. II.'' (German), 
Jber. Deutsch. Math. Verein. 49 (1940) 207Ð215.

%\cite{Hanany:1999sp}
\bibitem{Hanany:1999sp}
  A.~Hanany, Y.~-H.~He,
  ``A Monograph on the classification of the discrete subgroups of $SU(4)$,''
  JHEP {\bf 0102 } (2001)  027
  [hep-th/9905212].

%\cite{Everett:2008et}
\bibitem{Everett:2008et}
  L.~L.~Everett, A.~J.~Stuart,
  ``Icosahedral $A_5$ Family Symmetry and the Golden Ratio Prediction for Solar Neutrino Mixing,''
  Phys.\ Rev.\  {\bf D79 } (2009)  085005
  [arXiv:0812.1057 [hep-ph]].


%\cite{Etesi:1997jv}
\bibitem{Etesi:1997jv}
  G.~Etesi,
  ``Spontaneous symmetry breaking in $SO(3)$ gauge theory to discrete subgroups,''
  J.\ Math.\ Phys.\  {\bf 37} (1996)  1596-1602
  [hep-th/9706029].


\bibitem{turkey2}
 M.~Koca, R.~Koc, H.~Tutunculer,
  ``Explicit Breaking of $SO(3)$ with Higgs Fields in the Representations $L=2$
  and $L =3$,''
  Int.\ J.\ Mod.\ Phys.\  A {\bf 18} (2003) 4817
  [arXiv:hep-ph/0410270].
  %%CITATION = IMPAE,A18,4817;%%


\bibitem{Holthausen:2011vd}
  M.~Holthausen, M.~A.~Schmidt,
  %``Natural Vacuum Alignment from Group Theory: The Minimal Case,''
  [arXiv:1111.1730 [hep-ph]].


\bibitem{Berger:2009tt}
  J.~Berger, Y.~Grossman,
  ``Model of leptons from $SO(3) \to A_4$,''
  JHEP {\bf 1002} (2010) 071
  [arXiv:0910.4392 [hep-ph]].
  %%CITATION = JHEPA,1002,071;%%


  


%\cite{Altarelli:2010gt}
\bibitem{Altarelli:2010gt}
  G.~Altarelli, F.~Feruglio,
  ``Discrete Flavor Symmetries and Models of Neutrino Mixing,''
  Rev.\ Mod.\ Phys.\  {\bf 82} (2010)  2701-2729
  [arXiv:1002.0211 [hep-ph]].


\bibitem{generating_point}
 J.~Patera, R.~T.~Sharp, P.~Winternitz, 
 ``Polynomial irreducible tensors for point groups,'' J. Math. Phys. {\bf 19} (11) 2362, 1978.
 

\bibitem{nice}
R.~King, J.~Patera, R.~T.~Sharp,
``On finite and continuous little groups of representations of semi-simple Lie groups,"
 J. Phys. A: Math. Gen., 15, 1143, 1982.
 
\bibitem{analytic} 
 P.~E.~Desmier, R.~T.~Sharp, J.~Patera,
  ``Analytic $SU(3)$ States in a Finite Subgroup Basis,''
  J.\ Math.\ Phys.\  {\bf 23} (1982) 1393.
  %%CITATION = JMAPA,23,1393;%%



\bibitem{Louck}
J.~D.~Louck, ``Unitary symmetry and combinatorics'', World Sci., Hackensack, NJ, 2008.

\bibitem{Ikeda}
M.~Ikeda, ``On Complex Spherical Harmonics,'' Prog. Theo. Phys. {\bf 32} (1964), p.178.

 \bibitem{solidSU3}
  T.~Kayama,
  ``On the normalization of solid harmonics for $U(3)$,''
  Prog.\ Theor.\ Phys.\  {\bf 39} (1968) 850.
  %%CITATION = PTPKA,39,850;%%

%\cite{Georgi:1982jb}
\bibitem{groups}
  H.~Georgi,
  ``Lie Algebras in Particle Physics. From Isospin to Unified Theories,''
  Front.\ Phys.\  {\bf 54} (1982) 1.
  %%CITATION = FRPHA,54,1;%%


\bibitem{Slansky}
  R.~Slansky,
  ``Group Theory for Unified Model Building,''
  Phys.\ Rept.\  {\bf 79} (1981) 1.
  %%CITATION = PRPLC,79,1;%%




%\bibitem{Berger:2009tt}
  %J.~Berger and Y.~Grossman,
  %``Model of leptons from SO(3) ---> A(4),''
  %JHEP {\bf 1002} (2010) 071
  %[arXiv:0910.4392 [hep-ph],.
  %%CITATION = JHEPA,1002,071;%%


  


%\bibitem{Ludl_June1}
 % W.~Grimus and P.~O.~Ludl,
  %``Principal series of finite subgroups of SU(3),''
  %J.\ Phys.\ A  {\bf 43} (2010) 445209
  % [arXiv:1006.0098 [hep-ph],.
  %%CITATION = JPAGB,A43,445209;%%

%\bibitem{SU(3)/Z3}
 % A.~J.~Macfarlane,
  %``Description Of The Symmetry Group SU(3)/Z(3) Of The Octet Model,''
 % Commun.\ Math.\ Phys.\  {\bf 11} (1968) 91.
  %%CITATION = CMPHA,11,91;%%



\end{thebibliography}
\end{document}